\pdfoutput=1
\documentclass[11pt,twoside,a4paper,cmspaper,final,collab]{cms-tdr}
\def\svnVersion{146a66f}\def\svnDate{2026/06/22}\def\cmsCernNoTag{CERN-EP-2025-269}\def\cmsCernDate{\today}\def\cmsMessage{Published in Machine Learning: Science and Technology as \href{http://dx.doi.org/10.1088/2632-2153/ae7d87}{\doi{10.1088/2632-2153/ae7d87}.}}
\begin{document}\cmsNoteHeader{MLG-23-002}

\newcommand{\doc}{paper\xspace}

\newcommand{\INTLUMI}{138\fbinv}

\newcommand{\px}{\ensuremath{p_x}\xspace}
\newcommand{\py}{\ensuremath{p_y}\xspace}
\newcommand{\pz}{\ensuremath{p_z}\xspace}

\newcommand{\DeepB}{\text{DeepB}\xspace}
\newcommand{\DeltaEta}{\ensuremath{\Delta\eta_\text{jj}}\xspace}
\newcommand{\DeltaMJJ}{\ensuremath{\Delta m_\mathrm{j1j2}}\xspace}
\newcommand{\LSF}{\ensuremath{\mathrm{LSF_3}}\xspace}
\newcommand{\mSD}{\ensuremath{m_\mathrm{SD}}\xspace}
\newcommand{\nPF}{\ensuremath{n_\mathrm{PF}}\xspace}

\newcommand{\mPBpr}{\ensuremath{m_{\PBpr}}}%
\newcommand{\mA}{\ensuremath{m_{\text{A}}}}%
\newcommand{\mWp}{\ensuremath{m_\PWpr}\xspace}%
\newcommand{\mX}{\ensuremath{m_{\PX}}\xspace}%
\newcommand{\mY}{\ensuremath{m_{\PY}}\xspace}%
\newcommand{\mYp}{\ensuremath{m_{\PYpr}}\xspace}%
\newcommand{\mjj}{\ensuremath{m_\mathrm{jj}}\xspace}%

\newcommand{\fs}{\ensuremath{f_\text{s}}\xspace}
\newcommand{\pbg}{\ensuremath{p_\text{bg}(x)}\xspace}
\newcommand{\pdata}{\ensuremath{p_\text{data}(x)}\xspace}
\newcommand{\psig}{\ensuremath{p_\text{sig}(x)}\xspace}
\newcommand{\Rx}{\ensuremath{R(x)}\xspace}

\newcommand{\KERAS} {{\textsc{Keras}}\xspace}
\newcommand{\TENSORFLOW} {{\textsc{TensorFlow}}\xspace}
\newcommand{\PyTorch} {{\textsc{PyTorch}}\xspace}
\newcommand{\Pyro} {{\textsc{Pyro}}\xspace}
\newcommand{\nflows} {\textsc{nflows}\xspace}

\newcommand{\cwola}{\textit{CWoLa Hunting}\xspace}
\newcommand{\cathode}{\textit{CATHODE}\xspace}
\newcommand{\cathodeb}{\textit{CATHODE-b}\xspace}
\newcommand{\vae}{\textit{VAE-QR}\xspace}
\newcommand{\TNT}{\textit{TNT}\xspace}
\newcommand{\quak}{\textit{QUAK}\xspace}

\newcommand{\PQQst}{\HepParticle{Q}{}{*}\Xspace}
\newcommand{\PQTpr}{\HepParticle{\PQT}{}{\prime}\Xspace}
\newcommand{\PY}{\HepParticle{Y}{}{}\Xspace}
\newcommand{\PYpr}{\HepParticle{\PY}{}{\prime}\Xspace}
\newcommand{\PBpr}{\HepParticle{B}{}{\prime}\Xspace}
\newcommand{\PRad}{\HepParticle{R}{}{}\Xspace}

\newcommand{\Qstar}{\ensuremath{\PQQst \to \PQq \PWpr \to 3 \PQq}\xspace}
\newcommand{\XtoYY}{\ensuremath{\PX\to \PY\PYpr \to 4 \PQq}\xspace}
\newcommand{\Wp}{\ensuremath{\PWpr \to \PBpr\PQt \to \PQb \PZ \PQt}\xspace}
\newcommand{\Wkk}{\ensuremath{\PW_\mathrm{KK} \to \PRad\PW \to 3 \PW}\xspace}
\newcommand{\Zp}{\ensuremath{\PZpr \to  \PQTpr \PQTpr \to \PQt\PZ \PQt \PZ}\xspace}
\newcommand{\GtoHH}{\ensuremath{\PXXG_\mathrm{KK} \to \PH\PH \to 4 \PQt}\xspace}

\ifthenelse{\boolean{cms@external}}{\providecommand{\cmsLeft}{upper\xspace}}{\providecommand{\cmsLeft}{left\xspace}}
\ifthenelse{\boolean{cms@external}}{\providecommand{\cmsRight}{lower\xspace}}{\providecommand{\cmsRight}{right\xspace}}
\ifthenelse{\boolean{cms@external}}{\newcommand{\cmsAppendix}{}}{\providecommand{\cmsAppendix}{Appendix~}}
\ifthenelse{\boolean{cms@external}}{\newcommand{\cmsTable}[1]{\resizebox{\textwidth}{!}{#1}}}{\providecommand{\cmsTable}[1]{#1}}

\newlength\cmsTabSkip\setlength{\cmsTabSkip}{1ex}
\newlength\cmsFigWidthv
\newlength\cmsFigWidthvi
\newlength\cmsFigWidthvii

\cmsNoteHeader{MLG-23-002}

\title{Machine-learning techniques for model-independent searches in dijet final states}

\date{\today}

\abstract{
   Anomaly detection methods used in a recent search for new phenomena
   by CMS at the CERN LHC are presented.
   The methods use machine learning to detect anomalous jets produced in
   the decay of new massive particles without depending on a specific theory model.
   The effectiveness of these approaches in enhancing sensitivity to various
   simulated signal samples is studied and compared using data collected in
   proton-proton collisions at a center-of-mass energy of 13\TeV.
   In an example analysis, the capabilities of anomaly detection methods are further
   demonstrated by identifying large-radius jets consistent with Lorentz-boosted hadronically
   decaying top quarks in a model-agnostic framework.
}

\hypersetup{%
pdfauthor={CMS Collaboration},%
pdftitle={Machine-learning techniques for model-independent searches in dijet final states},%
pdfsubject={CMS},%
pdfkeywords={CMS, machine learning, anomaly, dijet, resonance}}

\maketitle

\section{Introduction}

\ifthenelse{\boolean{cms@external}}{
  \setlength{\cmsFigWidthv}{0.5\textwidth}
  \setlength{\cmsFigWidthvi}{0.5\textwidth}
  \setlength{\cmsFigWidthvii}{0.5\textwidth}
}{
  \setlength{\cmsFigWidthv}{0.55\textwidth}
  \setlength{\cmsFigWidthvi}{0.6\textwidth}
  \setlength{\cmsFigWidthvii}{0.7\textwidth}
}

Modern physics is remarkably successful at describing the world around us, from
the largest to the smallest scales.
Yet, our understanding of the universe is known to be incomplete.
Important questions include, among others, the nature of dark matter and the
origin of dark energy, as well as the imbalance between matter and antimatter in
the universe.
From a different perspective, general relativity, the theory of the
gravitational force, is incompatible with the description of electromagnetism
and interactions given by the standard model (SM) of particle physics.
Many of the proposed solutions to these problems involve the existence of new
particles that could be produced at particle accelerators such as the CERN LHC.
However, searches conducted so far at the ATLAS~\cite{ATLAS} and
CMS~\cite{Chatrchyan:2008zzk}
experiments have yielded negative results.

By far the most common events produced in collisions at the LHC originate in
processes mediated by the strong interaction.
Together with other SM processes, they form an overwhelming
background from which the (typically few) signal events resulting from the
production of new particles have to be carefully separated.
Machine learning (ML) is commonly employed for this
task~\cite{CMS:2025kje,hepmllivingreview}, mainly in a supervised
approach: classifiers are trained to distinguish between examples of events
predicted by the SM and another set of events following a beyond-the-SM (BSM)
theory~\cite{PDG2024}, requiring an explicit formulation for each alternative
theory.

Machine learning has also been proposed~\cite{Kasieczka:2021xcg} as a tool to
suppress the background in a largely model-agnostic way---that is, without prior
knowledge about the kind of signal events that may be present in the data
(anomaly detection~\cite{Belis:2023mqs}).
This paper reports on the techniques used
in a recent search for BSM particles~\cite{CMS:2024nsz}.

\begin{figure}[ht]
    \centering
    \includegraphics[width=0.49\textwidth]{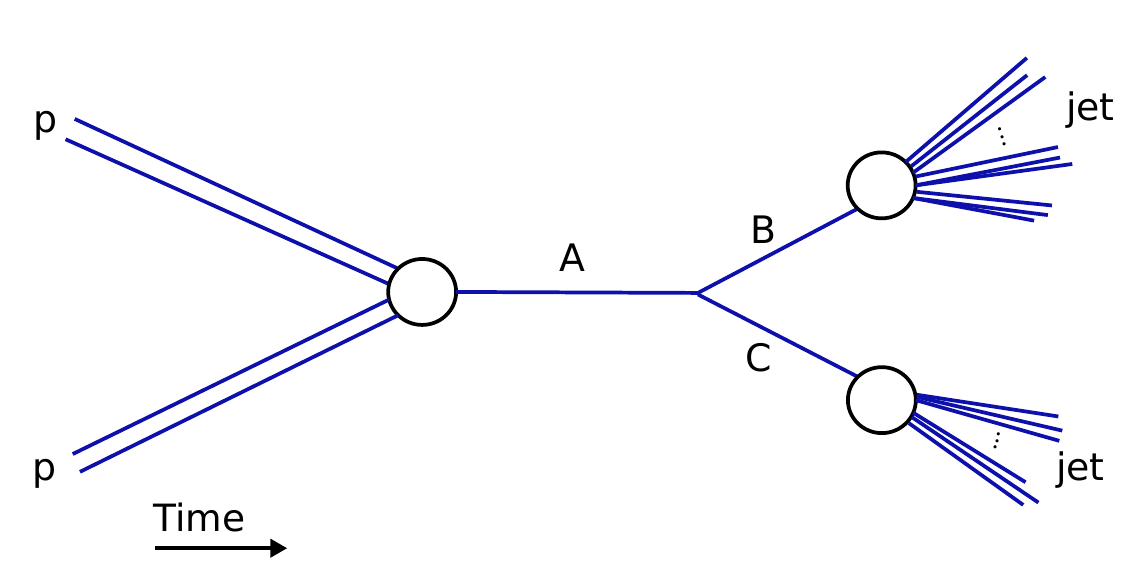}
    \caption{%
        Diagram of the $\text{A}\to\text{BC}\to\text{2 jets}$ signal topology
        targeted in this work.
        The particle A is produced in a collision between two protons, with time
        flowing from left to right.
        Adapted from Ref.~\cite{CMS:2024nsz}.
    }
    \label{fig:dijet-diagram}
\end{figure}

The BSM particles (noted A) considered in this \doc are not observed
directly in the detectors.
Instead, they decay almost instantly into a pair of lighter (less massive)
particles B and C, which in turn decay into SM particles, as shown in
Fig.~\ref{fig:dijet-diagram}.
The end result, as measured by the detector, consists of two collimated
highly-energetic sprays of SM particles, called jets.
The distribution of individual particles within the jets, also called the jet
``substructure'', depends on the properties of the intermediate
particles B and C: each quark or gluon appearing in the decay chain of B and C
typically leads to one ``subjet''.
The substructure can be thus used to identify the particles initiating the jets.
In addition to decaying to a pair of jets with nontrivial substructure, we
assume that particle A is heavy (2--6\TeV) and has a narrow intrinsic width
compared to the detector resolution, allowing to construct signal-depleted and
signal-enriched regions.

The results in Ref.~\cite{CMS:2024nsz} are derived using a ``bump hunt''
technique based on the invariant mass of the two jets, \mjj.
In the SM, \mjj is expected to follow a smoothly falling distribution, whereas
the new particle A would manifest itself as a sharp peak.
However, the amount of signal may be too small to produce a visible peak.
In this \doc, we consider five ML methods that use jet substructure
to detect signal events and separate them from the background.

The use of five very different methods raises the question of their relative and
absolute performance, as well as their complementarity in finding new particles.
We perform a detailed comparison in this paper, studying in particular the
impact of the set of input features used by each method and the performance when
using a normalized feature set.
We also assess the complementarity of the methods by studying the correlations
between the anomaly scores they assign to each event.

The usefulness of anomaly detection techniques is best demonstrated by using
them to find known particles in experimental data.
In this paper, we showcase anomaly detection in a real-world scenario
where we enhance the signal produced by hadronically-decaying Lorentz-boosted
top quarks in large-radius jets and analyze the selected events to recover key
properties of the top quark.

This \doc is structured as follows.
In Section~\ref{sec:relatedwork}, we review related prior work on anomaly detection.
We introduce the datasets used in this study in Section~\ref{sec:datasets} and
describe the five anomaly detection techniques in Section~\ref{sec:mlmethods}.
In Section~\ref{sec:performance_comparison}, we investigate their properties and the
correlation between their results.
The application to Lorentz-boosted top quarks is presented in
Section~\ref{sec:ttbar_val}.
Finally, a summary of the paper can be found in Section~\ref{sec:summary}.

\section{Related work}
\label{sec:relatedwork}

There is a long history of quasi-model-independent, or model-agnostic, searches
for new particles at colliders~\cite{
    D0:2000vuh, D0:2000dnz, D0:2001mmn, D0:2000dnz,
    H1:2008aak, H1:2004rlm,
    CDF:2007iou, CDF:2007ykt, CDF:2008voc,
    CMS:2020zjg,
    ATLAS:2018zdn, ATLAS:2020iwa, ATLAS:2025obc}.
Such searches have traditionally been performed at collider experiments by
comparing distributions measured in data to predictions of the SM.
Observing a significant difference in the number of events or the shape of the
distributions would be interpreted as a discovery.
This approach is limited by the accuracy of the background predictions, to which
large uncertainties are often assigned.
In addition, the number of bins grows exponentially with the dimensionality,
limiting comparisons to a relatively small number of observables.
Fine categorization of the data further reduces the sensitivity.

The new class of model-agnostic searches explored in this \doc uses machine
learning to address these issues.
In contrast to traditional model-agnostic searches, these anomaly detection
methods generally focus on a single final state (\eg, dijet events) and use
estimates of the background based on control samples in data.
Many methods have been demonstrated using simulated events~\cite{Kasieczka:2021xcg, darkmachines, antiqcd, Hallin:2021wme, Metodiev:2017vrx, Collins:2019jip,cwola_monojet, nachman2020anode, Raine:2022hht, klein2022flows, curtains2, sengupta2023improving, mikuni2023highdimensional, feta, Jawahar:2021vyu, tsan2021particle, Finke_2021, Vaslin:2023lig, Anzalone:2023ugq, Cerri_2019, Dillon_2021,Cheng_2023, dillon2023normalized,vanBeekveld:2020txa, Kuusela_2012, ad_nmi, PhysRevD.105.055006,refId0,Knapp:2020dde},
including by CMS~\cite{CMS:2025lmn}, but thus far there
have been few applications to collider data.

Among the proposed techniques,
autoencoders~\cite{Farina:2018fyg, Heimel:2018mkt}
were deployed by ATLAS in three searches for BSM
physics~\cite{ATLAS_HiggsX,ATLAS_twobody_unsupervised,ATLAS:2025kuz}.
The signature considered by the first search consisted of a resonance decaying
to a Higgs boson and an unspecified particle Y reconstructed as a large-radius
jet.
A recurrent neural network trained on SM backgrounds was used to select Y
candidates.
In the second search, ATLAS looked for BSM particles decaying to $\text{jet}+X$,
where X could be a jet, b-tagged jet, lepton, or photon.
Events were required to contain at least one lepton to satisfy trigger requirements.
An autoencoder based on event-level properties, such as object multiplicities
and momenta, was trained on a fraction of the data and subsequently used to
select potentially anomalous events. Finally, in Ref.~\cite{ATLAS:2025kuz},
a variational autoencoder~\cite{kingma2014autoencoding} is used to identify anomalous jets in a search for all-hadronic
BSM signatures.

In a slightly different context, autoencoders are used for data quality
monitoring of the CMS electromagnetic calorimeters~\cite{CMSECAL:2023fvz},
as proposed in Ref.~\cite{Bourilkov:2019yoi}.

A second class of algorithms is based on weak
supervision~\cite{Metodiev:2017vrx},
a training mode that does not use explicit examples of the targeted signal,
while still producing classifiers capable of dramatically reducing the SM
background.
Weak supervision was first applied to collider data in a measurement of the
$\PQt\bar{\PQt}\Pb\bar{\Pb}$ cross section by the CMS
experiment~\cite{CMS:2019eih} to estimate the multijet background.
The first search using this technique was performed by the ATLAS
experiment~\cite{ATLAS:2020iwa}, seeking the same dijet signal topology as we
consider in this \doc.
However, as a first-of-its-kind search, it only used a two-dimensional feature
space for anomaly detection, consisting of the masses of the two jets.
This limitation was lifted in a recent iteration~\cite{ATLAS:2025obc} of this
search, in which the feature set was expanded to a total of six substructure
variables.

An extensive comparison of anomaly detection methods on simulated events was
already performed in Ref.~\cite{Kasieczka:2021xcg}.
In contrast to the simulations used in Ref.~\cite{Kasieczka:2021xcg}, in this
work, all the methods have been used in a search using collider data.
We also compare, for the first time, methods developed after the publication of
Ref.~\cite{Kasieczka:2021xcg}.

An anomaly detection technique based on weak supervision was already
used~\cite{Gambhir:2025afb} to
``search'' for a known particle, namely the \PGU meson, based on CMS open data.
This work focuses on a different SM particle, the top quark, which most commonly
gets produced in pairs.
In addition, we consider the hadronic decay of the top quark, whereas
Ref.~\cite{Gambhir:2025afb} focuses on a leptonic decay channel---two vastly
different experimental signatures.

\section{Datasets}
\label{sec:datasets}

The analysis is performed using proton-proton collision data collected with the
CMS detector at $\sqrt s = 13\TeV$ in the 2016--2018 data-taking period.
The detector is a cylindrical apparatus built around the interaction point of
the LHC beams and designed to identify the particles produced in the collisions
by combining the signal they leave while crossing several specialized detector
subsystems.
The information collected by all subsystems is merged and processed to
reconstruct particle properties and cluster them into jets.
More detailed descriptions of CMS and the data processing used in this paper can
be found in Refs~\cite{Chatrchyan:2008zzk,CMS:2023gfb} and
\cmsAppendix\ref{sec:cms}.

The collision data used in this paper correspond to an integrated luminosity of
\INTLUMI~\cite{CMS-LUM-17-003,CMS-PAS-LUM-17-004,CMS-PAS-LUM-18-002}.
Events are selected using triggers based on the transverse momentum
\pt of the jets or the scalar sum of their transverse momenta, \HT, with
thresholds between 450 and 500\GeV for jet \pt and between 800 and 1050\GeV for \HT,
depending on the data-taking year.
The events used in the analysis are required to contain two jets with
$\abs{\eta}<2.5$, $\pt>300\GeV$, and a pseudorapidity difference $\abs{\DeltaEta}<1.3$.
Jet pairs with $\abs{\DeltaEta}>1.3$ are also used when defining various control regions.
These two jets are referred to collectively as the ``dijet system''.
The invariant mass \mjj of this system is required to be larger than 1455\GeV, which guarantees a trigger efficiency above 99\%.
We refer to this set of requirements as the ``baseline preselection''.

Since jets are composite objects containing many particles, it is natural to use the properties $(\pt, \eta, \phi)$ of their constituents as inputs for anomaly detection; this strategy is referred to as using ``low-level features''.
In practice, it is often desirable to use a small set of ``high-level'' variables derived from the constituents' properties and known to contain physically relevant information.
The following variables are used in this \doc: the number of jet constituents
\nPF; the mass \mSD of the jet calculated using the soft-drop
algorithm~\cite{Larkoski:2014wba}; the $N$-subjettiness variables $\tau_N$ and
their ratios $\tau_{NM}=\tau_N/\tau_M$~\cite{Thaler:2010tr}, where $N$ and $M$
stand for the number of subjets the variables are sensitive to; \DeepB, the maximum
DeepCSV~\cite{BTV-16-002} b tagging score of the two leading subjets of the
large-radius jet selected using the soft-drop algorithm;
and the lepton subjet fraction
\LSF~\cite{Brust:2014gia}, encoding how much of the jet \pt is
carried by the jet's leading lepton.

A second dataset is constructed by combining Monte Carlo samples for the main
physics processes contributing to the analysis phase space.
The dominant background process is the production of multiple jets through
processes governed by quantum chromodynamics (QCD).
QCD multijet events, as well as hadronically decaying \PW and \PZ bosons, are
simulated at leading order using \MGvATNLO version 2.6.5~\cite{Alwall:2014hca}
with the MLM
merging scheme~\cite{Alwall:2007fs}.
Processes involving top quarks are generated at next-to-leading order using the
\POWHEG \textsc{box} v2.0~\cite{Nason:2004rx,Frixione:2007vw,Alioli:2010xd}: top
quark
pair production~\cite{Frixione:2007nw}, single top quark production in the $t$
channel~\cite{Frederix:2012dh}, and associated production of a top quark and W
boson~\cite{Re:2010bp}.
Parton shower and hadronization are simulated using
\PYTHIA 8.240~\cite{Sjostrand:2014zea} with the CP5 tune~\cite{CMS:2019csb}.
Events are sampled according to their respective cross sections and generator
weights to obtain an unweighted sample and passed through a detailed \GEANTfour~\cite{GEANT4} simulation of the CMS detector before undergoing the selection described above.
The equivalent luminosity of this dataset is about 27\fbinv, limited by the size
of the QCD multijet datasets.

\begin{table}
    \topcaption{%
        Signal processes considered in the analysis, categorized according to
        the number of partons produced in the decay of each jet.
        In addition to the $\text{A}\to\text{BC}$ process of
        Fig.~\ref{fig:dijet-diagram}, the decay products of particles B and C
        are indicated.
        The notation follows Ref.~\cite{CMS:2024nsz}.
    }
    \centering
    \begin{tabular}{cl}
        Category & Process \\
        \hline
        $1+2$ & \Qstar \\
        $2+2$ & \XtoYY \\
        $2+4$ & \Wkk   \\
        $3+3$ & \Wp    \\
        $5+5$ & \Zp    \\
        $6+6$ & \GtoHH \\
    \end{tabular}
    \label{tab:samples}
\end{table}

A selection of Monte Carlo samples describing representative BSM models is also used in the analysis.
They all involve the production of a heavy resonance (masses of 2, 3, and 5\TeV are considered) decaying to two daughter particles each producing a large-radius jet.
They are categorized according to the number of partons produced in the decay of
intermediate particles: for instance, the process \XtoYY is categorized as
$2+2$ because each jet contains two partons.
The considered topologies are listed in Table~\ref{tab:samples}.
Where relevant, masses of 25, 80, 170, and 400\GeV are used for the intermediate
particles in the decay chains.
The samples are generated at leading order using \MADGRAPH and undergo the same showering, hadronization, detector simulation, and selection process as the background samples discussed earlier.
A more complete description of the signal samples can be found in Ref.~\cite{CMS:2024nsz}.

The simulated samples are made available on Zenodo~\cite{zenodo} in the data
format described in \cmsAppendix\ref{sec:dataset-release}.

\section{Machine-learning based anomaly detection}
\label{sec:mlmethods}

This section describes the five machine-learning techniques used in the
model-agnostic search, focusing on their architecture, training procedure, and
statistical validation.
We sort them roughly by the strength of the assumptions made about the signal.
The method with the least assumptions, \vae (Section~\ref{subsec:vae}), is based
on a variational autoencoder followed by an additional quantile regression step.
In Section~\ref{subsec:weak-supervision}, we then cover three methods, \cwola,
\TNT, and \cathode that use ``weakly supervised'' classifiers to deduce
properties of the signal directly from data.
The construction used to train the classifiers assumes that the signal \mjj
distribution has a sharp peak.
Finally, Section~\ref{subsec:quak} covers a method, \quak, that incorporates
information from simulated signal events to help focus on anomalies created by
new particles---as opposed to, for instance, defects in the detector.

The methods described in this section do not test directly for the existence of
signal; a two-step procedure is followed instead.
Each method is first used to select a set of events in which SM backgrounds are
suppressed compared to a putative unknown signal.
The \mjj distribution of this set of selected events is then used, in a second
step, to derive statistical statements about the presence or absence of a
signal.
Specifically, the distribution is fitted using the sum of two generic functional
forms representing the background and signal contributions (as discussed in
Ref.~\cite{CMS:2024nsz} in more detail).
For this to be possible, the selection performed by the anomaly detection
methods must not significantly alter the shape of the \mjj spectrum, which we
refer to as ``sculpting'' the mass distribution.
As detailed below, several methods implement explicit countermeasures to combat
sculpting.

A brief summary of the methods is shown in Table~\ref{tab:methods-summary}.
Before applying them to collision data, all methods were validated in simulation
and in a background-dominated control region.
More details about the validation strategy can be found in
Ref.~\cite{CMS:2024nsz}.

\begin{table*}
    \topcaption{
        Summary of the methods used in this \doc.
        The type column distinguishes between unsupervised (Uns.), weakly
        supervised (Weak.), and semi-supervised (Semi.) methods.
        Depending on the method, the ML models use single jets or complete dijet
        events as inputs.
        We also list the input features and types of ML models used, with the
        following abbreviations:
        ``AE'' for autoencoders,
        ``VAE'' for variational autoencoders,
        ``DNN'' for fully connected deep neural networks, and
        ``NF'' for normalizing flows.
        The two variants of \cathode are shown separately.
        Details are given in the text.
    }
    \centering
    \cmsTable{
    \begin{tabular}{lllll}
        Method & Type & Input & Features & Model types \\
        \hline
        \vae & Uns. & Jet & Particle \px, \py, \pz & VAE, DNN \\
        \cwola & Weak. & Jet
            & $\mSD, \tau_{21}, \tau_{32}, \tau_{43}, \nPF, \LSF, \DeepB$
            & DNN \\
        \TNT & Weak. & Jet
            & $\mSD, \tau_{21}, \tau_{32}, \tau_{43}, \nPF, \LSF, \DeepB$
            & AE, DNN \\
        \cathode & Weak. & Event
            & $m_\mathrm{j1}, \DeltaMJJ, \tau_\mathrm{41,j1}, \tau_\mathrm{41,j2}$
            & NF, DNN \\
        \cathodeb & Weak. & Event
            & $m_\mathrm{j1}, \DeltaMJJ, \tau_\mathrm{41,j1}, \tau_\mathrm{41,j2}, \DeepB_\mathrm{j1}, \DeepB_\mathrm{j2}$
            & NF, DNN \\
        \quak & Semi. & Event
            & $\tau_s, \tau_{21}, \tau_{32}, \tau_{43}, \nPF, \rho, \DeepB$
                for both jets
            & NF \\
    \end{tabular}
    }
    \label{tab:methods-summary}
\end{table*}

\subsection{Variational autoencoder}
\label{subsec:vae}

The first method explored in this search makes use of a variational autoencoder
(VAE) and quantile regression (QR), and we call it \vae.
Autoencoders are a class of neural network architectures that utilize dimensionality reduction
and are commonly applied to low-level input data.
An autoencoder model is composed of two parts, an encoder and a decoder.
The encoder takes the input data and compresses it into a lower-dimensional representation called the latent space.
The decoder then attempts to decompress the latent representation to recover the original data.
The two components are trained simultaneously to minimize the difference between the decoder output and the original data.
The use of autoencoders to detect anomalous jets was first proposed in Refs.~\cite{Farina:2018fyg, Heimel:2018mkt}.
This approach relies on the observation that reconstruction quality is generally poorer for events that
were either not encountered during training, or seen in insufficient numbers.

Variational autoencoders extend the basic autoencoder concept to a probabilistic
model, increasing the model expressivity and adding generative capabilities.
Compared to an autoencoder, a VAE differs by turning the latent space into a
probabilistic distribution, typically a multivariate Gaussian, whose mean $\mu$
and standard deviation $\sigma$ are controlled by the encoder.
The decoder inputs are sampled from the latent-space distribution.

\begin{figure*}
\centering
\includegraphics[width=\textwidth]{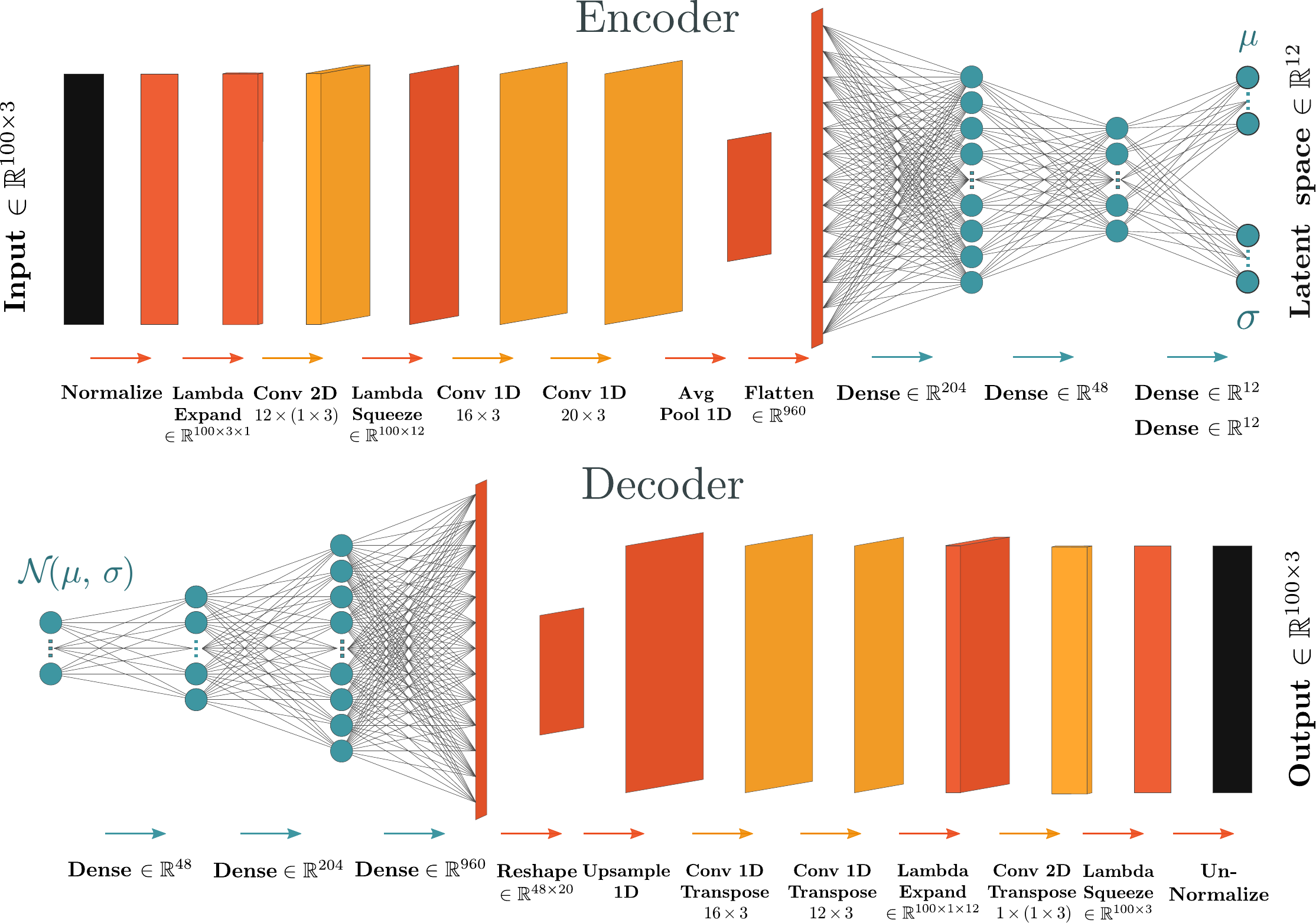}
\caption{%
  The VAE architecture used for jet anomaly detection.
  The output dimensions of layers are indicated with the $\in$ symbol.
  For convolution layers, the numbers of filters $f$ and kernel sizes $s$ are
  written as $f\times s$, with multidimensional $s$ for 2D layers.
}
\label{fig:VAEarch}
\end{figure*}

In our application, the VAE is trained to encode and decode the momentum of the
100 highest-\pt constituents of individual jets, encoded in Cartesian coordinates
as a zero-padded $100\times3$ matrix.
The network architecture is shown in Fig.~\ref{fig:VAEarch}.
Processing starts by normalizing the inputs to a mean of zero and a standard
deviation of one, which is followed by a series of convolutional layers.
The first convolution, with 12 two-dimensional (2D) filters of size $(1, 3)$,
generates an embedding for individual particles.
Correlations between particles are learned by two subsequent one-dimensional
(1D) convolution layers acting on three particles at a time, with 16 and 20
filters respectively.
After these convolutions, the number of latent-space dimensions is reduced to
12 by two fully connected layers.
The decoder uses the same operations in reverse order.

\begin{figure}
    \centering
    \includegraphics[width=0.49\textwidth]{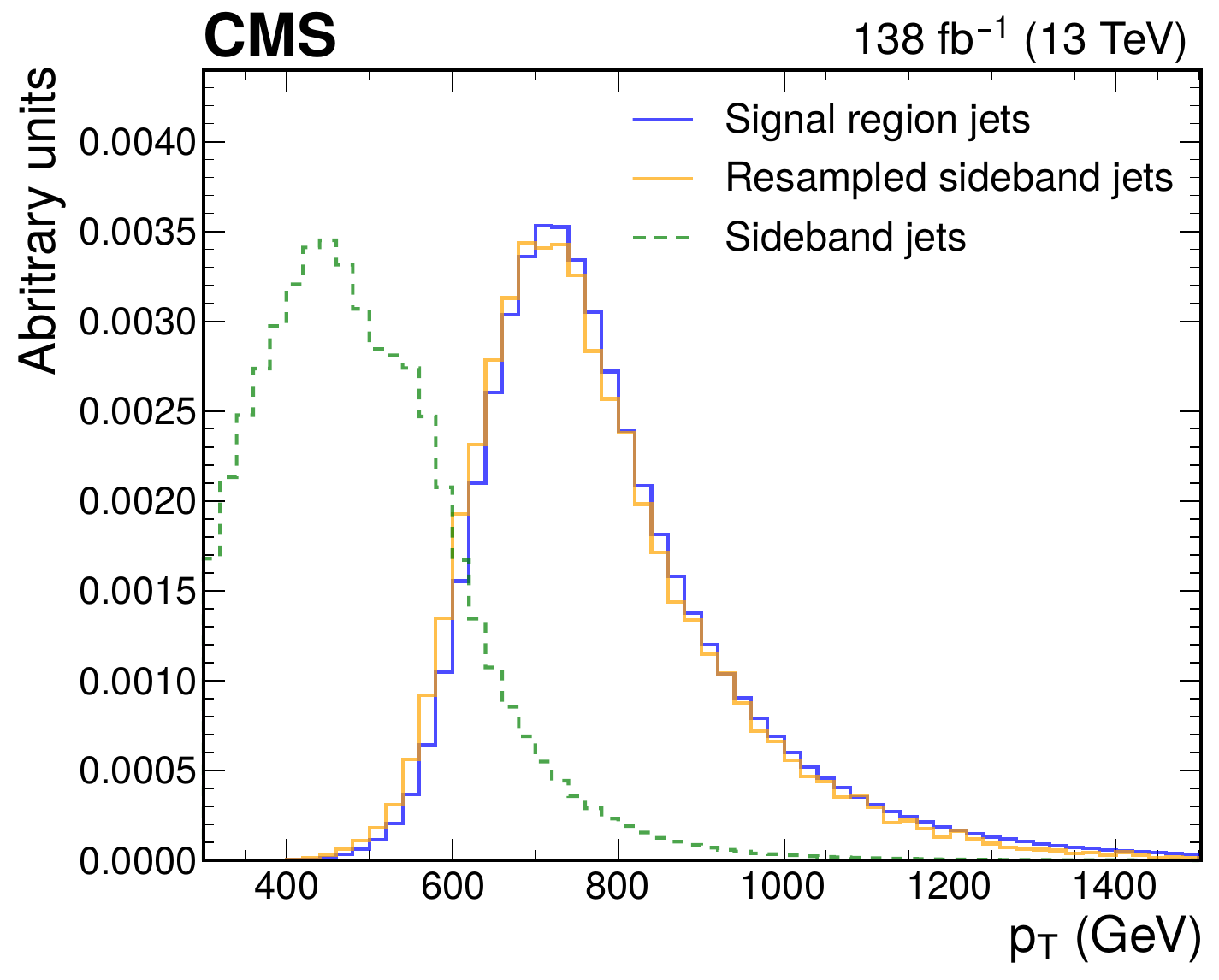}
    \includegraphics[width=0.46\textwidth]{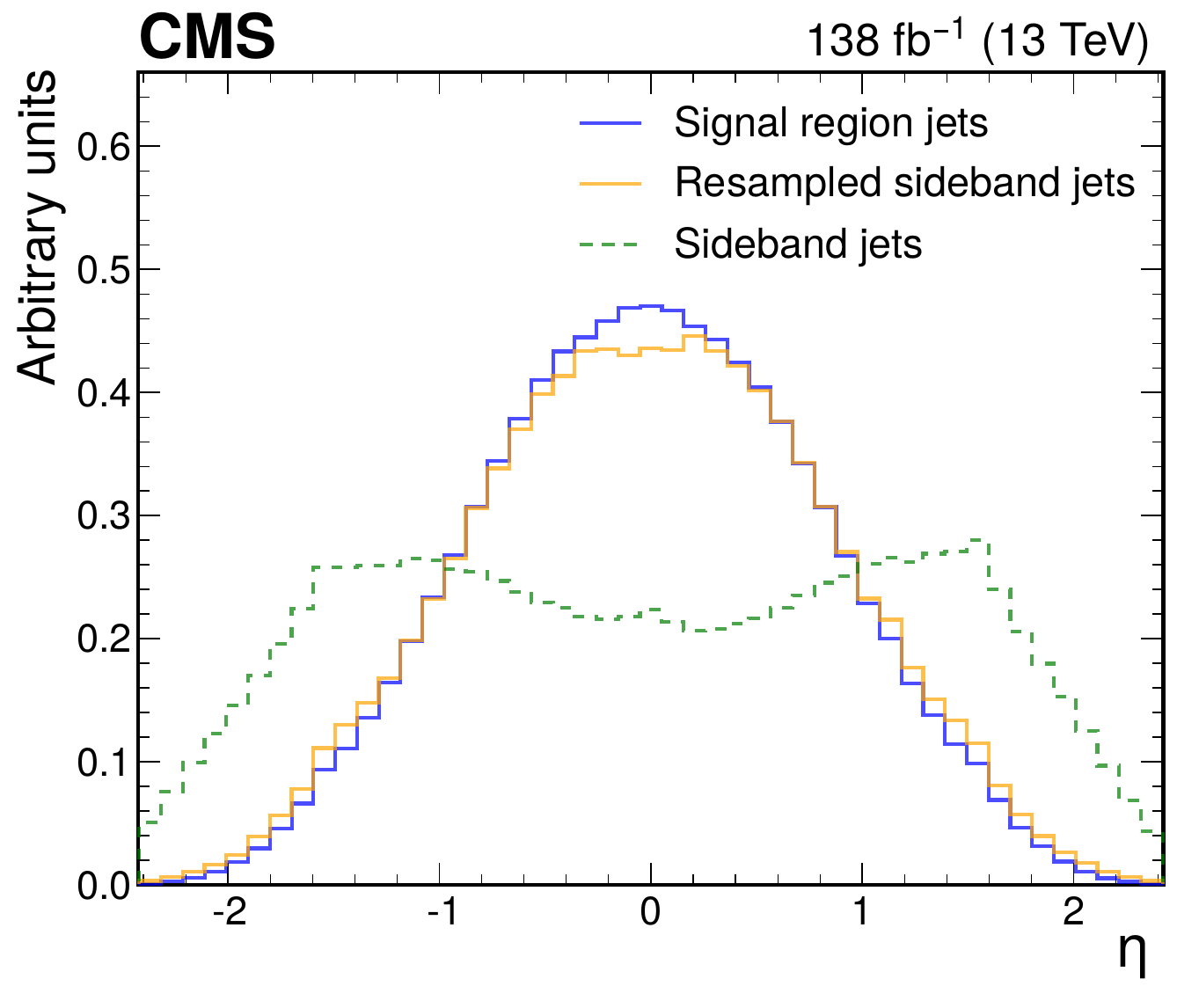}
    \caption{%
        Jet \pt (\cmsLeft) and $\eta$ (\cmsRight) distributions in data before and after
        resampling jets from the $2<\abs{\DeltaEta}<2.5$ sideband (green) to match
        the distribution in the signal region (blue).
        The distributions of resampled jets are shown in orange.
        Histograms are normalized to unity.
        The details are given in the text.
    }
    \label{fig:vae_jetmixing}
\end{figure}

The VAE is trained using jets from a sideband region in \DeltaEta in which signal
is expected to be suppressed compared to the SM background.
We select events by requiring
$2<\abs{\DeltaEta}<2.5$, a veto on additional jets with $\pt>300\GeV$, and \pt
imbalance between the two jets.
The two highest-$\pt$ jets in each event in this region are kept for training,
leading to a set of 7.4 million jets.
This collection is resampled to correct the three-dimensional $(\pt, \eta, \phi)$
distribution, which would otherwise differ significantly from that in the search
region, leading to a reduced set of 1.9 million jets.
This ensures that the VAE learns the distribution appropriate for jets in the
signal region (SR; for instance, jets in the \DeltaEta sideband region have,
on average, a smaller \pt).
The sampling uses simulated background events in the SR.
Each simulated jet is replaced by the closest jet from the sideband, selected
using a $k$-dimensional tree based on the jets' \pt, $\eta$, and $\phi$.
The resulting $\eta$ and \pt distributions are shown in
Fig.~\ref{fig:vae_jetmixing}.

Training is performed by minimizing a two-component loss function.
The first term is the permutation-invariant Chamfer distance~\cite{fan2016point}
calculated between the input and the output jets, rewarding the network for
correctly encoding and decoding its input.
The latent-space distribution is enforced by adding a weighted regularization term to the
loss function,
computed as the (closed-form) Kullback--Leibler
divergence~\cite{Kullback:1951zyt} between the $\mu$ and $\sigma$
vectors of the latent space and a unit Gaussian prior ($\mu=0$, $\sigma=1$).
The latter term rewards if the Gaussians in the latent space (the bottleneck of the architecture)
are not too disjoint and provide a good compromise between representing the different modes found
in the data and overlapping with one another.
The weight of the regularization term~\cite{higgins2017betavae} is set to 0.5,
which allows the VAE to focus more on the reconstruction quality while still
enforcing a well-behaved latent space.

The network is implemented using \TENSORFLOW/\KERAS 2.11.0 and trained on
batches of 256 jets with the Adam optimizer~\cite{Kingma:2014vow} and an initial
learning rate of $10^{-3}$.
The learning rate is multiplied by 0.3 if the improvement in the validation loss over 4 epochs is
smaller than 0.03.
Training is ended once this threshold is not reached over 8 epochs, typically
after 30 epochs.

\begin{figure*}
\centering
\includegraphics[width=0.333\textwidth]{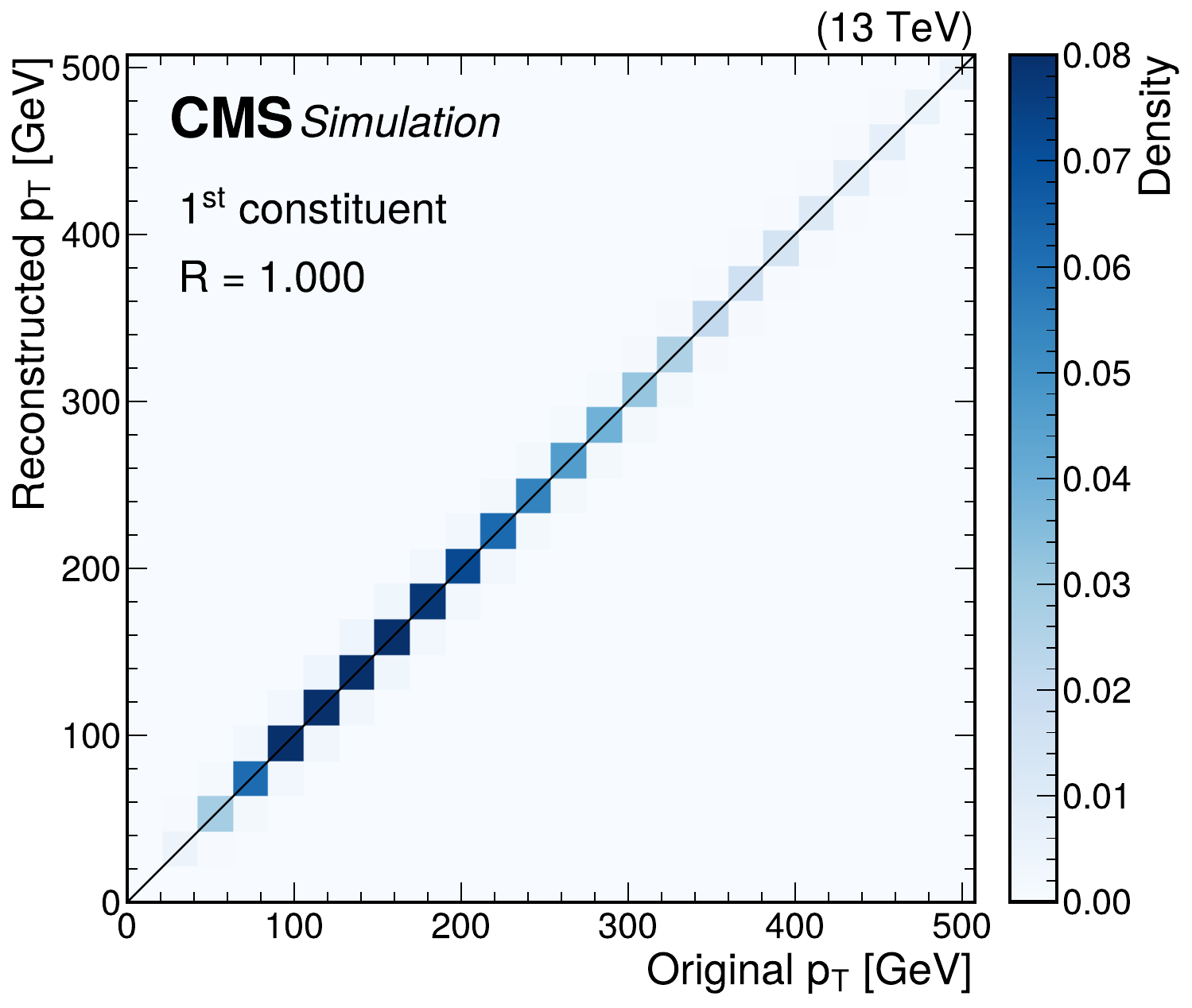}%
\includegraphics[width=0.333\textwidth]{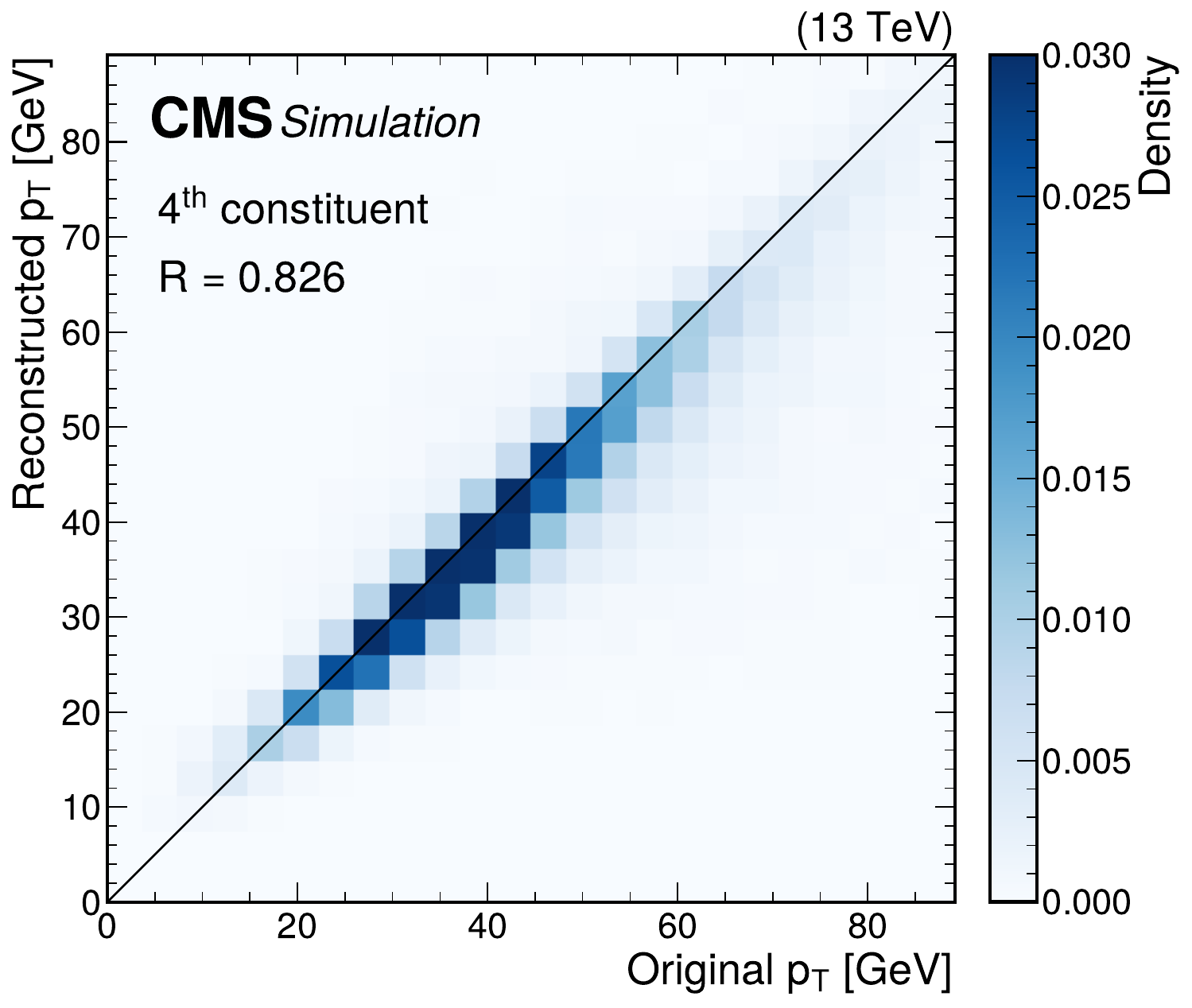}%
\includegraphics[width=0.333\textwidth]{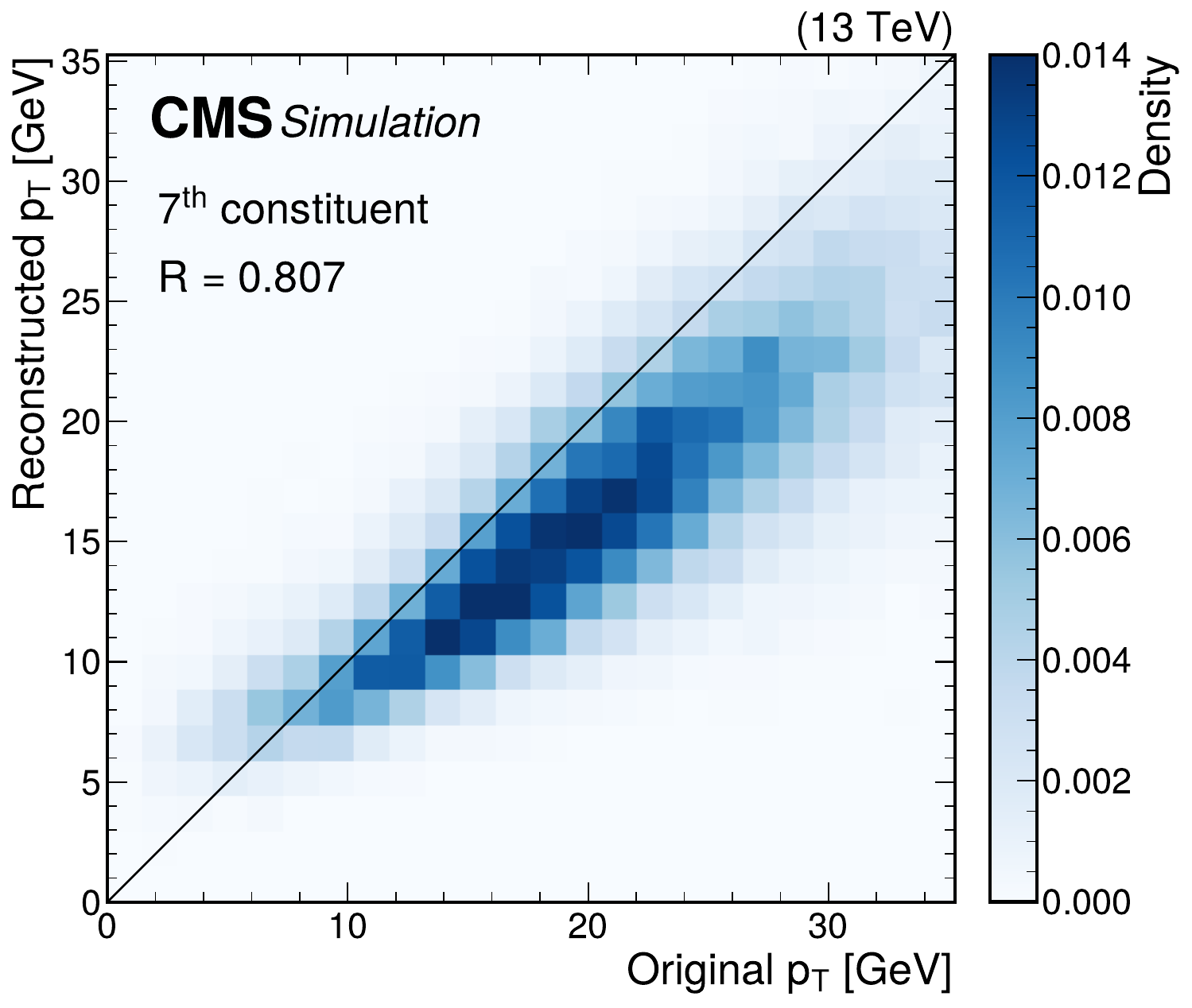}\\
\includegraphics[width=0.333\textwidth]{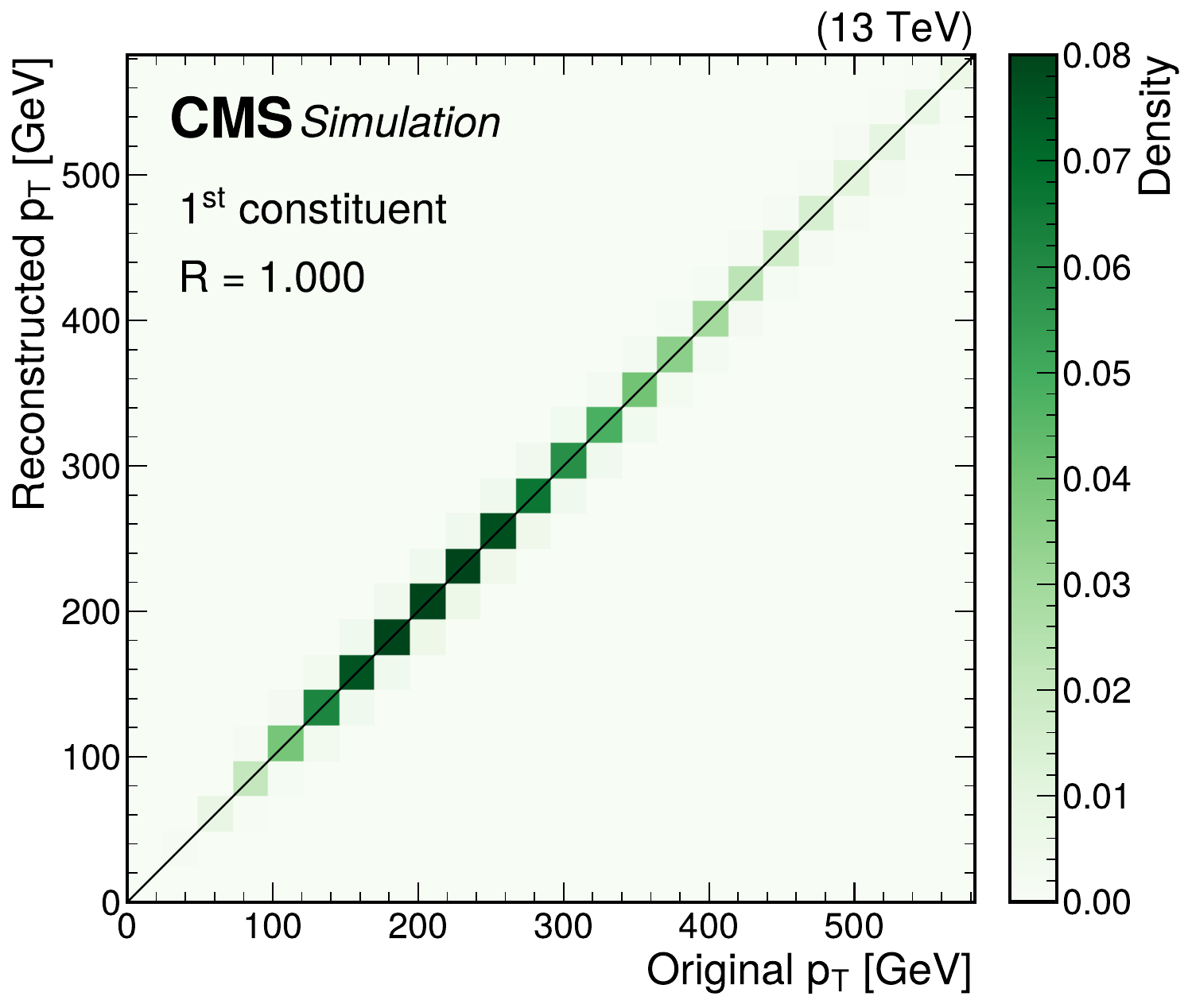}%
\includegraphics[width=0.333\textwidth]{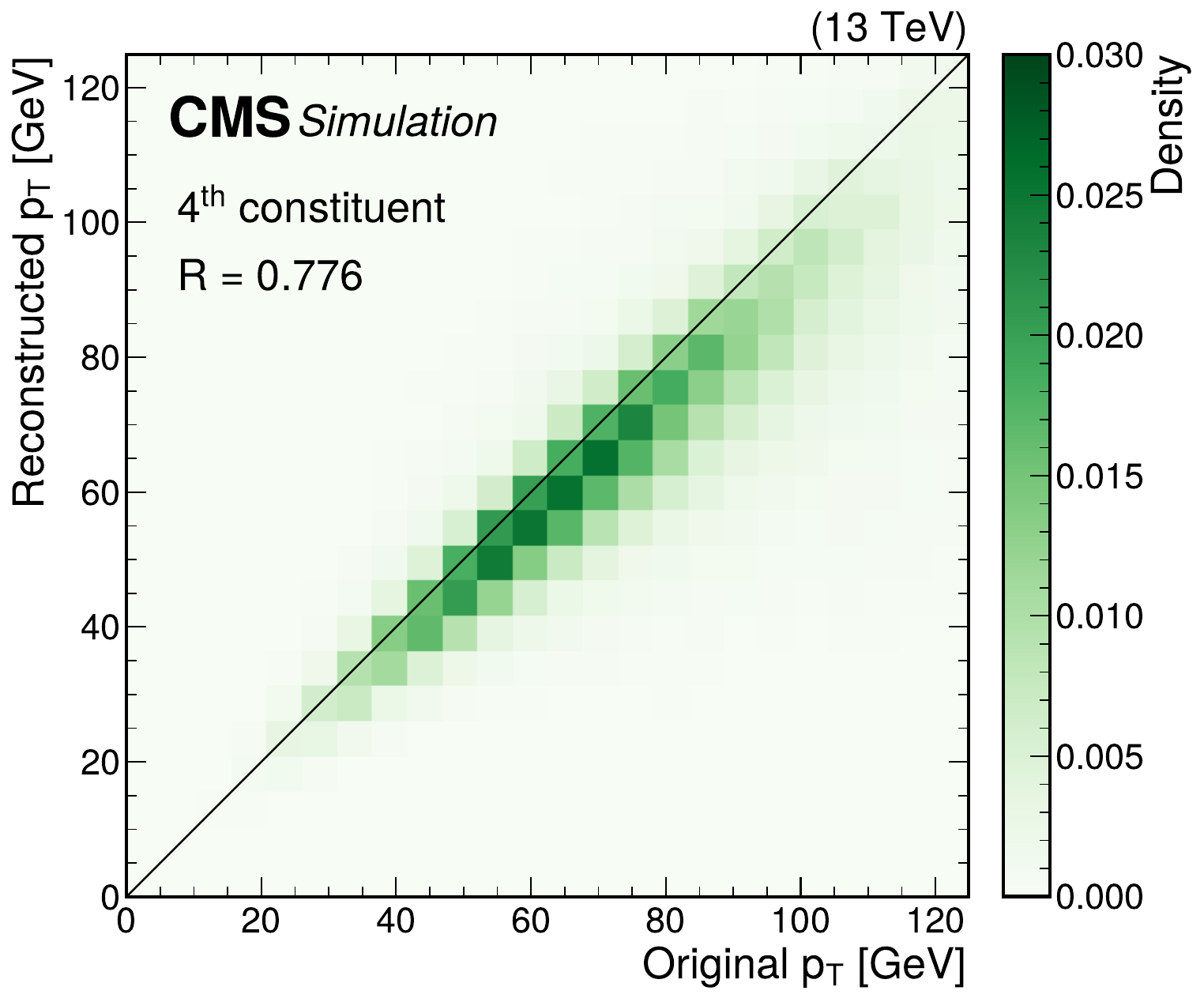}%
\includegraphics[width=0.333\textwidth]{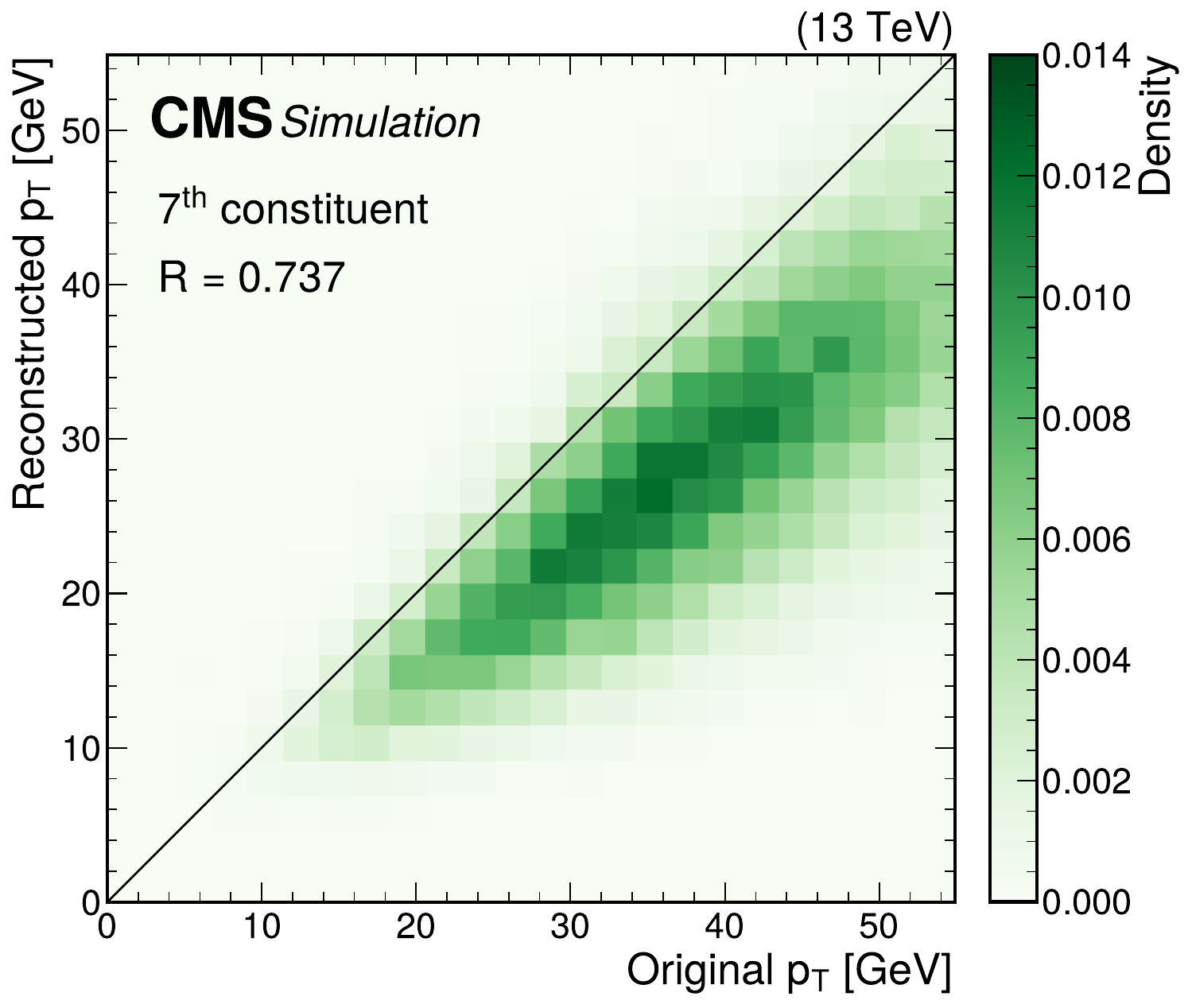}\\
\includegraphics[width=0.333\textwidth]{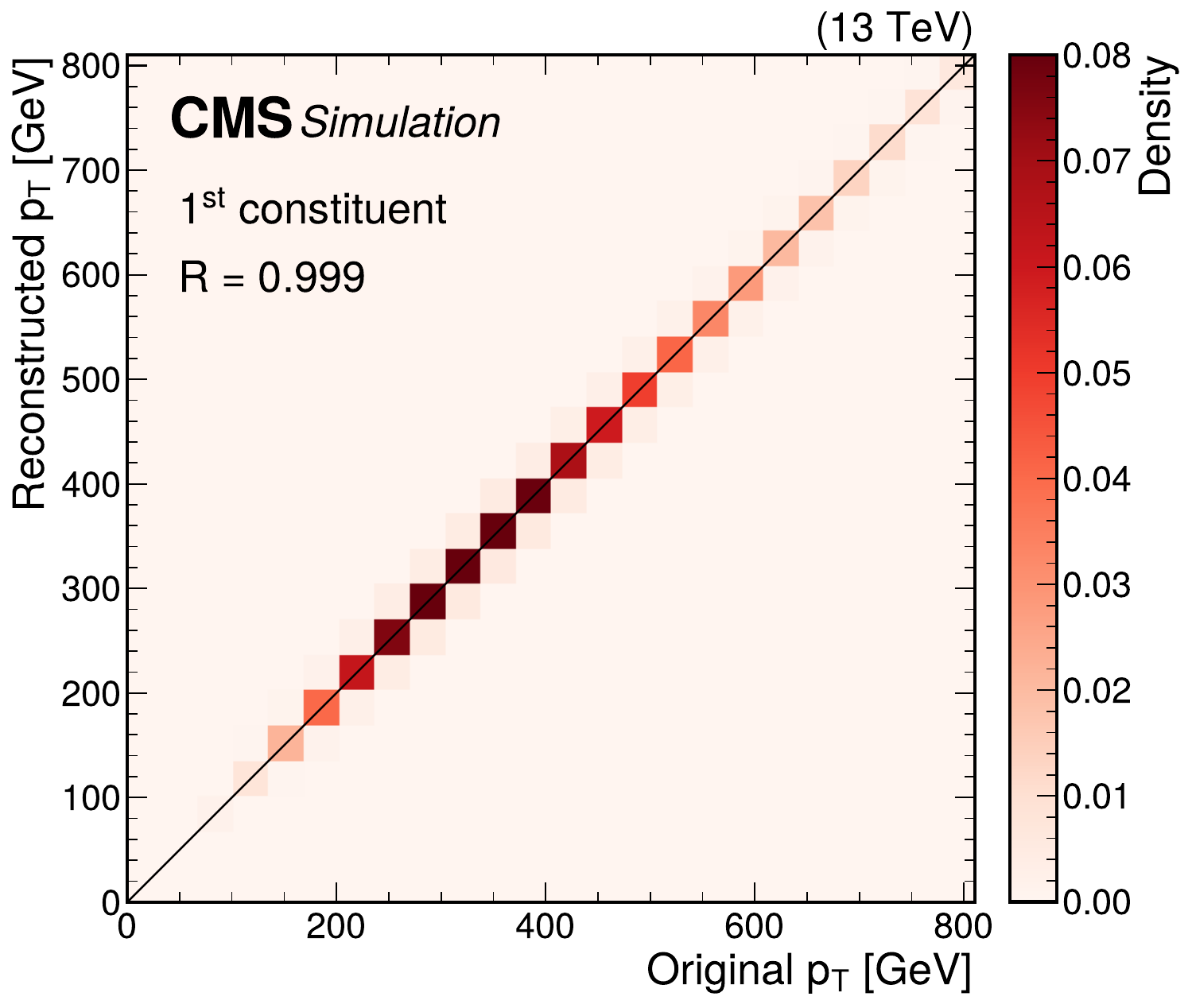}%
\includegraphics[width=0.333\textwidth]{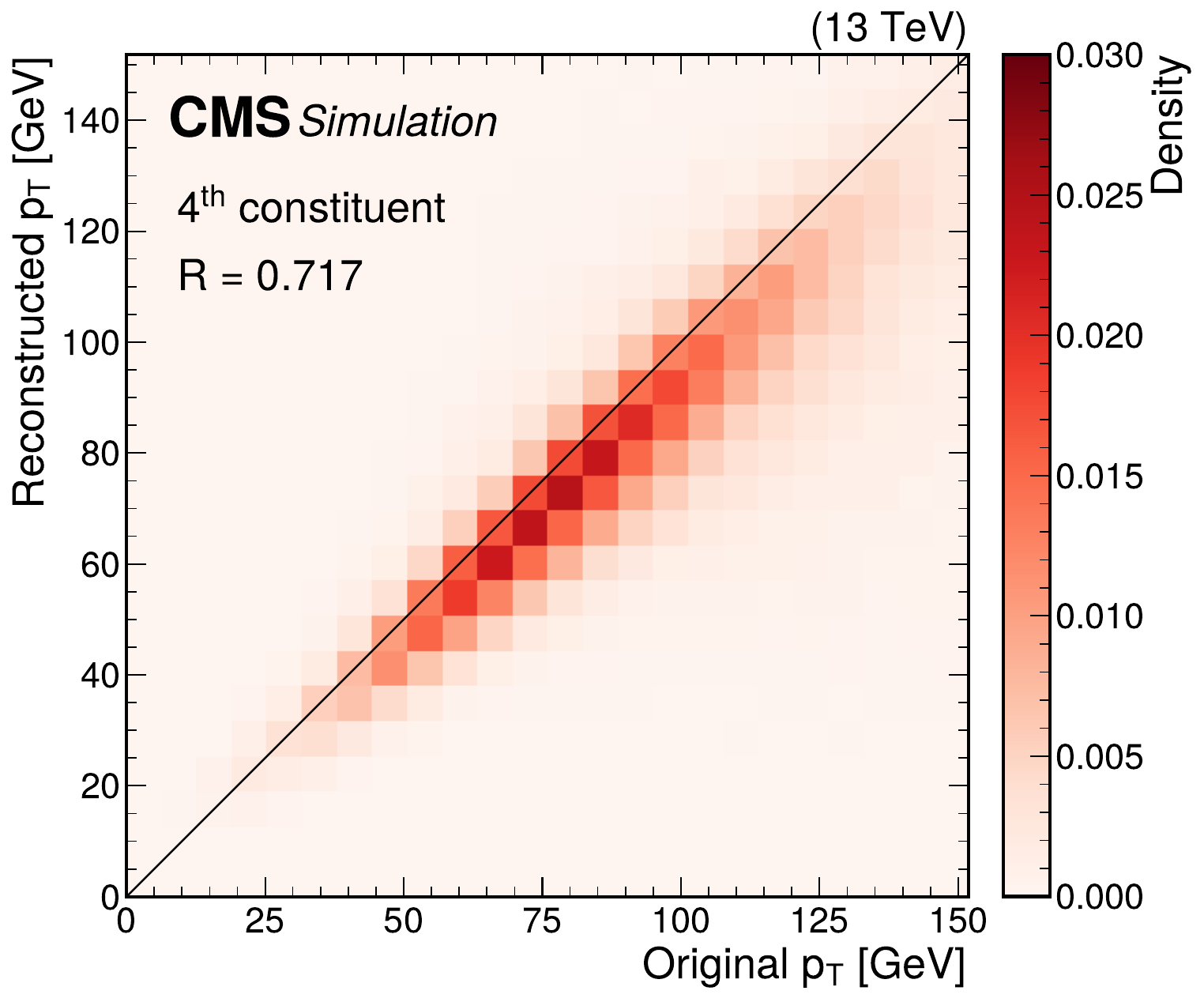}%
\includegraphics[width=0.333\textwidth]{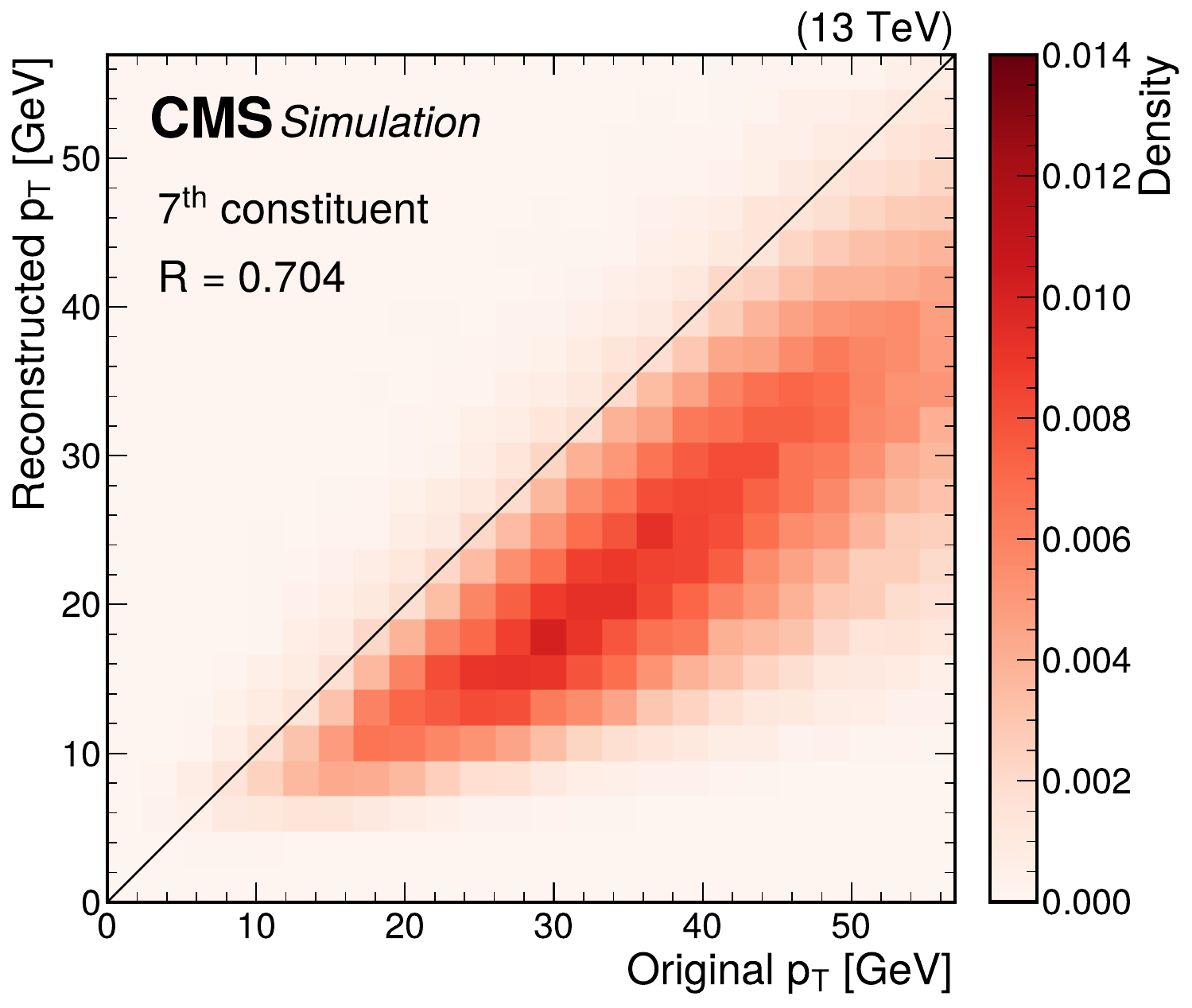}\\
\caption{%
    Quality of jet constituent $\pt$ reconstruction for simulated SM background
    jets
    (upper row), jets from a \Wp decay (middle row), and jets from an \XtoYY
    decay (lower row).
    The first (left column), fourth (middle column), and seventh (right column)
    constituents are shown.
    The Pearson correlation coefficient $R$ between the input and output is also
    shown.
}
\label{fig:SMj1Pt}
\end{figure*}

In Fig.~\ref{fig:SMj1Pt}, we show the performance of the VAE in reconstructing
the \pt of particles inside the leading jet from an SM background event, a \Wp
event, and an \XtoYY event.
For this figure, \pt is calculated from the $x$ and $y$ components of the
momentum vector, requiring the network to learn two variables and their
correlation.
The quality of the reconstruction, as measured by the Pearson correlation
coefficient $R$ between the input and output \pt, is better for background jets
than for signal jets.
To reduce sensitivity to outliers, the 2\% lowest-\pt and the 2\% highest-\pt
jets are excluded from the calculation of $R$.
Focusing first on background jets, we observe that the network reconstructs the
leading particle well.
The reconstructed \pt still matches the original \pt for the fourth particle,
the last one for
which an exact representation would fit in the 12-dimensional latent space.
For background jets, the network is also capable of recovering the seventh
particle, indicating that it successfully learned correlations between its
inputs.
This knowledge is, however, only useful for background, and reconstruction
quality is visibly worse for signal jets.
This property enables the use of the autoencoder as an anomaly detector.

Once trained, the VAE is used to compute the anomaly score of SR
jets, for which we exploit both reconstruction quality and latent-space
information~\cite{Dillon_2021} by reusing the loss function described above.
The SR jets with a large loss are considered anomalous.
The anomaly scores for each of the two jets of every event are considered, and
the smallest is taken.
This is equivalent to considering that, for an event to be anomalous,
both jets have to pass the same anomaly threshold.
Other choices, such as the maximum and the average were considered and shown not to
perform as well in conjunction with the quantile regression described below.

The construction outlined in the previous paragraph leads to a strong
correlation between the anomaly score and \mjj.
This is not desired, as a selection based on the anomaly score then results in
an arbitrarily modified \mjj spectrum even in the absence of signal, hindering
the statistical analysis.
We solve this problem using quantile regression.
That is, we construct the anomaly threshold $s$ used in the event selection to
have dependence on \mjj such that
we select a constant fraction $q$ of all events regardless of \mjj.
This is achieved by training a neural network to approximate the function
$s_q(\mjj)$ having this property for values of $q=10$, 5, and 1\%.

The QR models are fully connected networks with five hidden layers and 30 nodes per
layer using the exponential linear unit activation function, implemented in \KERAS.
They have one output node per value of $q$.
Network weights are initialized following a uniform distribution with variance
according to Ref.~\cite{HE_uniform}.
The networks are trained using events from the SR.
To avoid absorbing a signal into the regression, the dataset is split in four
subsets, each of which is used to train one model.
When evaluating the selection for an event from a given subset, we use the
average of the regressions derived in the three other subsets, so that no event
is used twice.
To further reduce the risk of absorbing a signal, the average is smoothed by
fitting it with a third-order polynomial.

\begin{figure}
    \centering
    \includegraphics[width=0.5\textwidth]{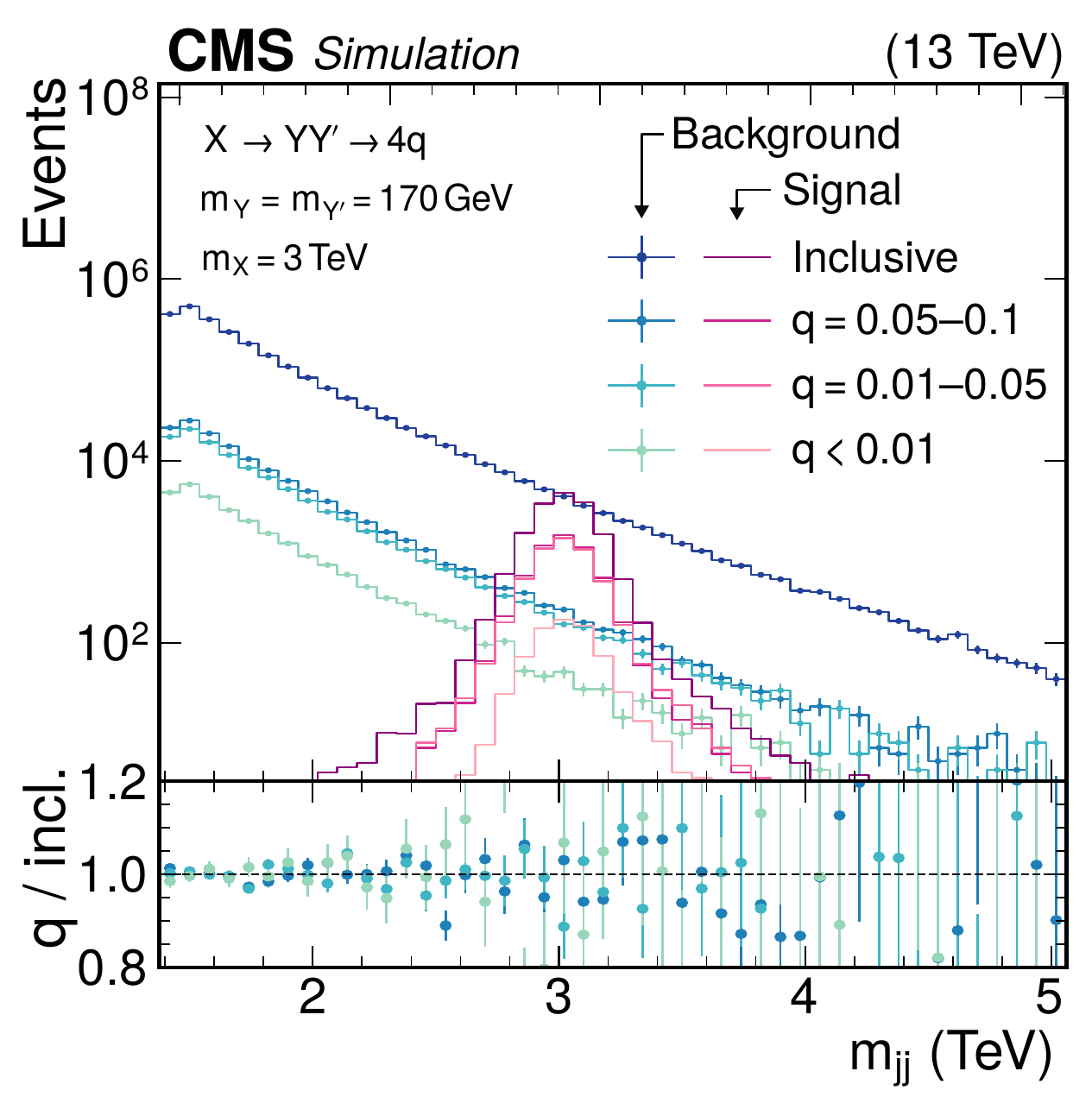}
    \caption{
        Simulated dijet invariant mass spectrum after selecting events based on
        the score of the \vae method.
        The distributions are shown in the upper panel for SM background
        (blue to green) and the \XtoYY signal model with $\mX=3\TeV$ and
        $\mY=\mYp=170\GeV$ (purple to pink).
        The inclusive selection is compared to the three quantile ranges used in
        the statistical analysis.
        The signal-to-background ratio increases when applying a tighter
        selection.
        No significant difference is observed between the shapes, showing that
        the quantile regression performs as expected.
        The lower panel shows the ratio of the quantile ranges to the inclusive
        distribution for the SM background, scaled by the respective quantile
        probabilities.
        The error bars shown for background represent the statistical
        uncertainty.
    }
    \label{fig:category_split}
\end{figure}

After the QR procedure, the \mjj distributions in each quantile are smooth and
their shapes closely follow the inclusive distribution.
Figure~\ref{fig:category_split} shows the \mjj distributions obtained after
the \mjj-dependent selection corresponding
 to each of the quantiles in simulated QCD multijet background and for an \XtoYY benchmark model:
one can observe a good agreement among these distributions within statistical
fluctuations, and in particular with the inclusive distribution.

The final results use events selected with $q<10\%$.
For model-independent results such as $p$-values~\cite{pvalue}, a single fit of
the \mjj
distribution is performed for all events meeting this criterion.
A different approach is followed when the signal shape is known, such as when
setting limits.
In these cases, the three orthogonal categories defined by $q<1\%$,
$1\le q<5\%$, and $5\le q<10\%$ are fitted simultaneously, using the \mjj shape and
yield distribution among the categories for the considered signal.
This more sensitive method cannot be applied in a model-agnostic way because the
shapes and yields were found to differ between signals.

\subsection{Weakly supervised methods}
\label{subsec:weak-supervision}

Three anomaly detection methods used in this \doc (\cwola, \TNT, and \cathode)
use weakly supervised learning~\cite{Metodiev:2017vrx}.
The goal of all of these methods is to learn, directly from data, with no input from simulations, an approximation to the likelihood ratio (LR) \Rx between the underlying probability densities of background, \pbg, and data (possibly including signal), \pdata, as a function of some input variables $x$:
\begin{equation}\label{eq:weak-supervision-likelihood-ratio}
\Rx=\frac{\pdata}{\pbg}.
\end{equation}
This LR, if it could be learned exactly, would be the most powerful model-agnostic anomaly detector.
Indeed, if data contains a fraction \fs of signal, its probability density can be rewritten as:
\begin{equation}
\pdata=(1-\fs)\pbg+\fs \psig,
\end{equation}
where \psig is the probability density of signal.
Injecting this expression into Eq.~\eqref{eq:weak-supervision-likelihood-ratio},
we find that \Rx is monotonically related to the signal-to-background LR
$\psig/\pbg$.
By learning \Rx, weakly supervised methods obtain an equivalent classifier
without prior knowledge of \psig.

The strategy for learning a good approximation to \Rx followed by all weakly
supervised approaches is to train a classifier between data and samples drawn
from the background model fully based on control samples in data.
If the background model is accurate and the classifier is well-trained, the classifier will approach \Rx asymptotically.
In practice, the performance typically degrades with smaller samples sizes available for training and lower fractions of signal events in the data sample.

All three weakly supervised methods train the classifier in an interval in \mjj
called the SR and use events from outside the interval to model the background
distribution \pbg in the SR.
In order to cover all possible signal masses, the training is repeated in twelve
partially overlapping SRs, listed in
Table~\ref{tab:weakly_supervised_massbins}.
The regions cover the whole available \mjj range with approximately uniform logarithmic spacing and are arranged in a staggered manner so that signals located at the boundary between two regions would not be missed.
The SRs are required to have sideband mass regions on either side for reliable background estimation, meaning the upper and lower extremes of the \mjj distribution are used solely as sideband control regions.

\begin{table}
    \topcaption{
        Mass regions used by the weakly supervised methods and corresponding
        observed number of events.
        Signal regions are required to have sideband mass regions on either side for reliable background estimation.
        This means only the A1--A6 and B1--B6 regions are used to seek signals,
        with the A0, B0, A7, and B7 regions used solely as sideband control
        regions.
    }
    \centering
    \begin{tabular}{c cc}
        Name & Mass region (\GeVns) & Num. of events \\
        \hline
        A0 & 1350--1650 & $1.4\times10^7$ \\
        A1 & 1650--2017 & $4.5\times10^6$ \\
        A2 & 2017--2465 & $1.4\times10^6$ \\
        A3 & 2465--3013 & $4.0\times10^5$ \\
        A4 & 3013--3682 & $1.0\times10^5$ \\
        A5 & 3682--4500 & $2.2\times10^4$ \\
        A6 & 4500--5500 & $3.9\times10^3$ \\
        A7 & 5500--8000 & 479 \\[\cmsTabSkip]

        B0 & 1492--1824 & $6.6\times10^6$ \\
        B1 & 1824--2230 & $2.1\times10^6$ \\
        B2 & 2230--2725 & $6.3\times10^5$ \\
        B3 & 2725--3331 & $1.7\times10^5$ \\
        B4 & 3331--4071 & $4.2\times10^4$ \\
        B5 & 4071--4975 & $8.5\times10^3$ \\
        B6 & 4975--6081 & $1.3\times10^3$ \\
        B7 & 6081--8000 & 144 \\
    \end{tabular}
    \label{tab:weakly_supervised_massbins}
\end{table}

The three methods rely on \mjj sideband regions but differ from each other with
respect to which events are used to train the classifier and in the modeling of
\pbg.
As we will review further in the sections below, \cwola and \TNT use \mjj
sideband regions directly as the background samples; these can serve as a good
model for the background if the $x$ features are statistically independent
(decorrelated) from \mjj. Meanwhile, \cathode first trains a generative model
on the sidebands and interpolates this into the SR;
samples from this model are used as the background model and correctly take into
account any correlation between $x$ and \mjj.

In order to make use of all available data events, the weakly supervised methods implement cross validation.
The data and the method-specific background samples are split into five equally
populated $k$-folds, one of which is set apart as the holdout set,
in which the trained model will be applied to data to select the most anomalous events.
Of the remaining four $k$-folds, one is used for validation and the other three
for training; training is repeated for each possible choice of validation fold.
When evaluating the models on the holdout set, we use the average prediction of
the four models trained with varied validation sets.

After training, the classifiers trained in each SR are used to select events from their respective holdout $k$-folds, without restriction on \mjj.
The events selected from all holdout $k$-folds are merged together and the resulting \mjj distribution is used as input to the fit.

This procedure is repeated for every mass region from
Table~\ref{tab:weakly_supervised_massbins}, yielding a different \mjj spectrum
depending on the signal mass hypothesis.
This results in as many smoothly falling \mjj spectra as there are SRs. A resonant signal identified by the classifiers would manifest itself as an excess of events localized in the corresponding SR.

\subsubsection{Per-jet CWoLa Hunting}
\label{subsec:cwola}

Classification without labels (\cwola)~\cite{Collins:2018epr,Collins:2019jip}
was the first proposed anomaly detection method based on weak supervision.
A classifier is trained to learn the LR $R(x)$ from Eq.~\eqref{eq:weak-supervision-likelihood-ratio}, using data selected from an interval in \mjj for $p_\text{data}$ and a reweighted sample of events from
sidebands on each side of the interval as a model of \pbg.
If a resonant signal is present in the chosen interval and not in the sidebands, the classifier will learn to tag it.
However, if there is no signal present, the classifier scores will not be meaningful.
The procedure is repeated for different choices of the signal mass region in order to cover the full \mjj range.

One very important aspect of this paradigm is that the features used for classification should not be correlated with \mjj.
Otherwise, the classifier will be able to differentiate background events in the SR from those in the sidebands.
It will then learn to consider background events within the SR as signal-like, which could introduce artificial bumps in the \mjj distribution even in the absence of a signal.
If, for example, the \pt of each jet were used as input features to the network,
then it would be able to identify events in the SR based on the \pt
difference between the sidebands and SR, leading to sculpting.
Due to the power of neural networks, subtle correlations between multiple features and \mjj can potentially lead to sculpting.
In the present work, the training strategy is modified to reduce sculpting.

In the original \cwola proposal, a single classifier is used for both jets in the event, allowing the classifier to capture correlations between the jets.
In this work, two distinct classifiers are used instead, trained on the heaviest and lightest of the two jets in each event, respectively.
This allows additional measures to be taken to mitigate background sculpting.
Each classifier uses seven input features:
\[ \mSD, \quad \tau_{21}, \quad \tau_{32}, \quad \tau_{43}, \quad \nPF, \quad \LSF, \quad \DeepB. \]

The sidebands used to model $p_\text{bg}(x)$ are the two mass regions
adjacent to the one used for training.
Three reweightings of the jets are performed to enhance the fidelity of this
model and to minimize correlations between the learned anomaly score and \mjj.
The relative weight of jets from the upper sideband is first increased to balance their contribution with that of the lower sideband.
This is necessary because the \mjj spectrum is steeply falling, and thus the upper sidebands are much less populated than the lower sidebands.
Additionally, to prevent class imbalance, jets in the SR are reweighted so that their total weight is the same as the total sideband weight.
Finally, jets in the SR are reweighted to ensure that their \pt distribution corresponds to the one observed in the sidebands after the previous reweighting steps.
This reweighting is accomplished by constructing a histogram of jet \pt's with
20 bins and then taking the ratio of the signal window and sideband jets.
To remove the effect of outliers from bins with low numbers of jets, the \pt reweighting factors are clipped to the range $[0.1,5.0]$.

\begin{figure}
    \centering
    \includegraphics[width=0.5\textwidth]{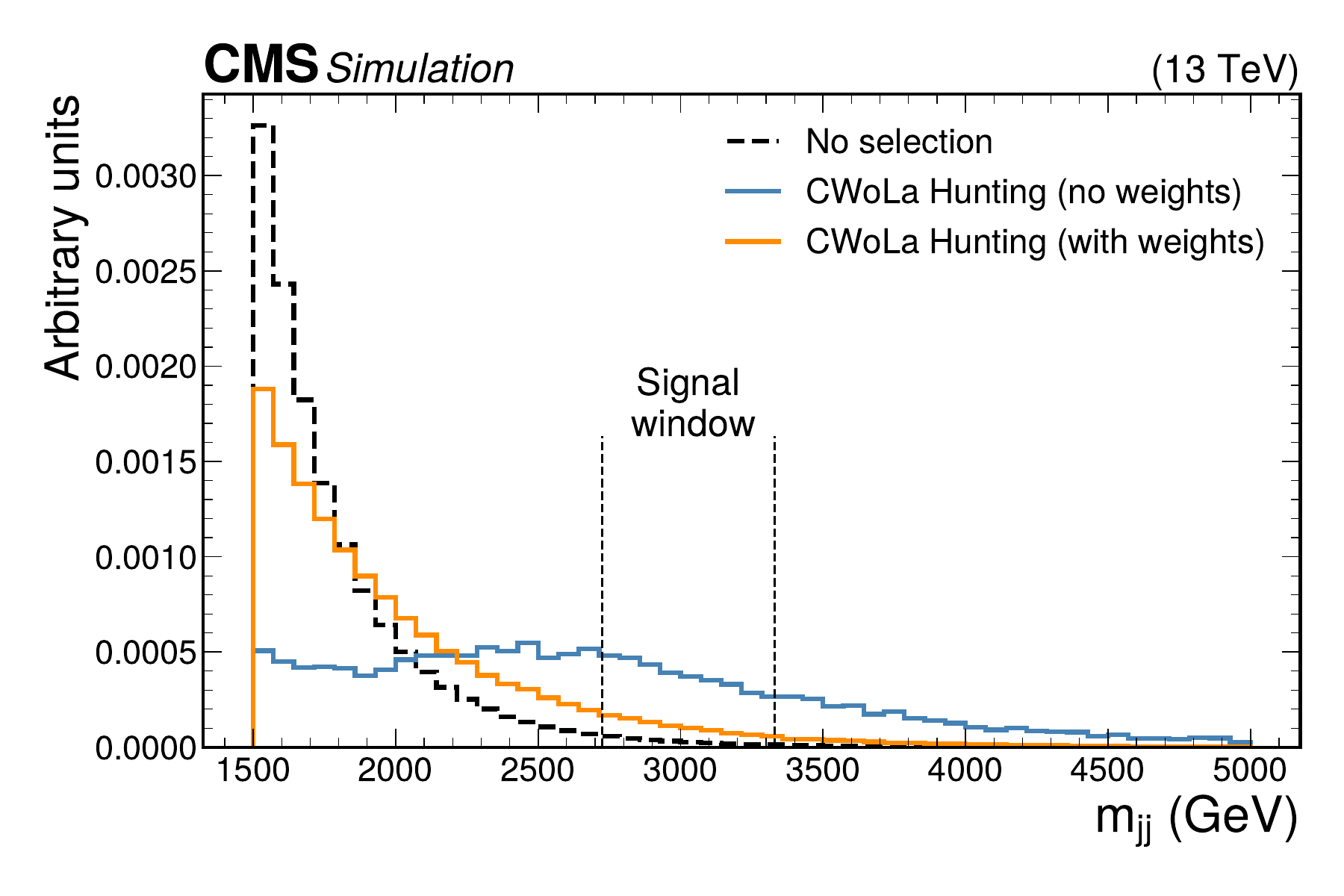}
    \caption{
        A comparison of the dijet mass distribution of simulated background events selected by the \cwola
        algorithm, performed with and without the reweighting procedure described in the text.
        The events selected without the reweighting procedure (solid blue) have a significantly distorted mass distribution
        as compared to the distribution before selection (dashed black).
        In contrast, the events selected when the reweighting procedure is employed (solid orange) retain a smoothly falling 
        mass distribution which allows a reliable signal extraction.
    }
    \label{fig:cwola_mjj_sculpt}
\end{figure}

The reduced correlation between the learned anomaly score and jet \pt then
mitigates mass sculpting effects in simulation studies.
This is shown in Fig.~\ref{fig:cwola_mjj_sculpt}, where the mass distribution of
events selected by the \cwola
algorithm is compared with and without the reweighting procedure.
Without the reweighting procedure, the algorithm learns a significant mass bias,
strongly distorting the mass shape.
With the reweighting procedure, the smoothly falling nature of the mass
distribution is retained, allowing a reliable signal extraction.

The classifiers are simple fully connected networks using the exponential linear
unit activation function~\cite{elu} and dropout~\cite{dropout} with a drop rate
of 0.2 between each layer to mitigate overtraining.
Because the amount of data used in the training for different SRs varies considerably, two network architectures are used for the classifiers: a network with 6 hidden layers of sizes 64, 128, 128, 32, 16, and 8 is used for SRs A1-A4 and B1-B4, and a smaller 4-layer network with 32, 54, 16, and 8 nodes is used for the rest.

The networks are trained in \TENSORFLOW~2.12.0 and \KERAS~2.11.0 using the Adam
optimizer with a learning rate of $10^{-4}$ and the default hyperparameters.
The binary cross-entropy loss function is used, with jets from the sidebands labeled as 0 and jets from the SR labeled as 1.
The batch size is 256 and up to 100 epochs are performed, with early stopping if the validation loss does not improve after 10 epochs.
Training is repeated three times with different random initialization of the
weights.
We keep the network that selects the largest number of jets in the SR of the
validation $k$-fold at the selection efficiency later used to select events, and
discard the other two\footnote{A similar idea has since been proposed in Ref.~\cite{Hein:2025uhj}.}.
Within each $k$-fold, the validation set used to select the best-performing
network is disjoint from the holdout set used to select anomalous events for
final statistical analysis.
Fluctuations are uncorrelated between $k$-folds, ensuring the model selection
criterion does not amplify statistical fluctuations in the final results.
Additionally, any bias to preferentially select events from the SR based only on the mass is mitigated by the reweighting procedure described above,
and is validated to not introduce significant mass bias in simulation and control region studies.

Once the networks for each jet are trained, their scores need to be combined to
select events.
Jet scores are first converted to percentiles separately for the heavier and
lighter jet in each event, and the per-event score is taken as the larger of
the two.
This method was chosen as it yielded the most robust performance
across the benchmark signals as compared to alternatives, such as multiplying,
adding, or taking the minimum of the jet scores.
The classifiers trained with different validation sets are combined after this
step.

Once the event-level anomaly score is defined, top-scoring events in each
$k$-fold are selected across the whole \mjj range for use in the fitting
procedure.
The selection threshold is chosen to reach a target efficiency in the two \mjj
sidebands,
with reweighting applied to compensate for the difference in the number of events.
This efficiency is chosen to ensure sufficient event counts for the fitting
procedure. It is set to 1\% up to SR B3, 3\% for SRs A4--B5, and 5\% above.

\subsubsection{Tag N' Train}
\label{subsec:tagntrain}

\begin{figure*}
    \centering
    \includegraphics[width = 0.9\textwidth]{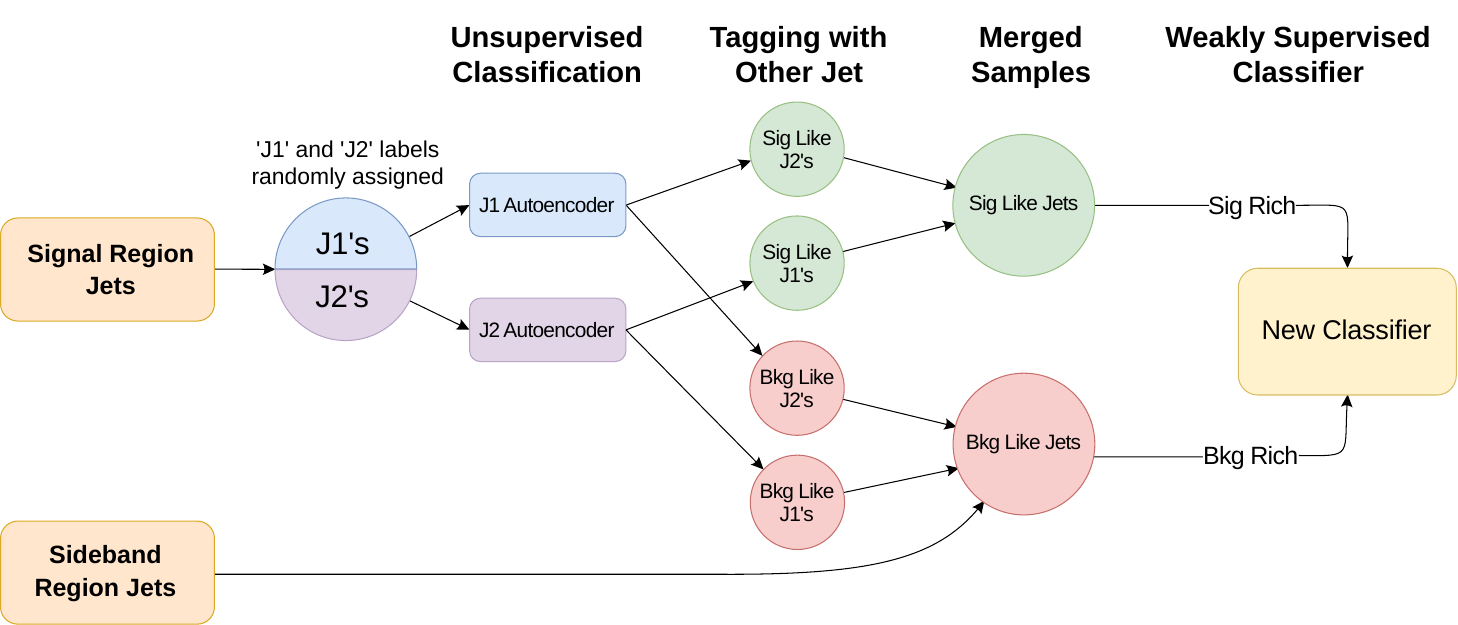}
    \caption{
        A schematic showing the training algorithm for \TNT.
        The two jets in the event are randomly assigned to subsets J1 and J2.
        The dijet invariant masses and the autoencoder scores are used to
        construct signal- and background-like subsets of J1 and J2.
        These subsets are then merged, and a single classifier is trained to
        distinguish between signal- and background-like jets.
        Adapted from Ref.~\cite{CMS:2024nsz}.
    }
    \label{fig:TNT_diagram}
\end{figure*}

The {\itshape Tag N' Train} (\TNT)~\cite{Amram:2020ykb} approach to anomaly
detection combines weak supervision with an initial selection based on an autoencoder.
The algorithm specifically targets signals in which both jets are anomalous,
using the autoencoder anomaly score of one jet to obtain a signal-enriched
sample of the other jet.
They are then combined for the weakly supervised training.
A schematic of the algorithm is shown in Fig.~\ref{fig:TNT_diagram}.

The version of \TNT implemented in this \doc starts by grouping the jets in two
sets J1 and J2, randomly assigning one jet in every event to J1 or J2 and the
other jet to the other set.
This contrasts with the original proposal, in which one sample was made of the heavier jet in each event and the other of the lighter jet.
This change reduces correlation between the two samples, improving the
performance of the \TNT algorithm.

The autoencoders used by \TNT are based on low-level features formatted as jet images, as proposed in Refs.~\cite{Farina:2018fyg,Heimel:2018mkt}.
The images are constructed as in  Ref.~\cite{Macaluso:2018tck}, with pre-processing applied before pixelating, and centered on the \pt-weighted centers of the jet constituents.
Constituents within an $\eta$ and $\phi$ range of $-0.6$ to $0.6$ around the
center of the jets are cast as images of 32 by 32 pixels.
In order to reduce dependence on the \pt of the jet, each image is
normalized so that the sum of all the pixel intensities is equal to one.
The autoencoders are based on convolutional networks with filter sizes of $3\times3$, where the image's dimensionality is reduced through max-pooling layers after each convolutional
layer. Three convolutional layers with 32, 24, and 16 filters are used.
The output is then fed through two dense layers of sizes 128 and 32, before reaching a latent space dimension of 6. The ReLU activation function is used after every convolutional and dense layer.
The architecture is then mirrored, with two fully connected layers and then three convolutional layers using 2D sampling layers in place of the max-pooling layers to produce
an image of the same dimensions as the input.

For each considered SR and test $k$-fold, an autoencoder is trained to reconstruct jets in the two \mjj sidebands.
Both jets in each event are used for training the autoencoder.
The network is trained for 100 epochs using the mean square error loss and the Adam optimizer with a learning rate of $10^{-4}$ and the default hyperparameters.

Once trained, the loss of the autoencoder is used in conjuction with \mjj to
construct the mixed samples for the weakly supervised training.
Within the SR, the loss of the J1 jet is used to tag the J2 jet as background- or signal-like, and vice-versa.
Jets with a loss in the top 20\% result in a signal-like tag for the other jet, and jets with a loss in the lowest 40\% result in a background-like tag.
Jets tagged in this way are then collected into a signal-like and a background-like sample, with all jets from the sidebands added to the background-like sample to increase its statistical power.
The remaining 40\% of events in the SR are not used further.

After the background-like and signal-like samples have been constructed, a weakly supervised classifier is trained to distinguish between them using the same architecture, input variables, and hyper-parameters as for \cwola.
Reweighting during training is slightly different: background-like jets from the SR and upper sideband are each reweighted to match the weight of the lower sideband.
As in  \cwola, the background- and signal-like classes are balanced with each other
and the \pt distribution of the signal-like class is adjusted to match the
background-like class in order to mitigate any potential \mjj sculpting when
selecting events.

Like \cwola, the distribution of anomaly scores for each jet is used to convert the raw \TNT anomaly score into a percentile.
The final event-level anomaly score is obtained as the product of the
percentiles of the two jets.
Compared to the maximum used for \cwola, this combination requires both jets to
be anomalous.
The rest of the procedure is identical to \cwola: top-scoring events are
selected across the whole \mjj range and combined to obtain the \mjj
distribution used in the fitting procedure.

\subsubsection{CATHODE and CATHODE-b}
\label{subsec:cathode}

\begin{figure*}
    \centering
    \includegraphics[width=\textwidth]{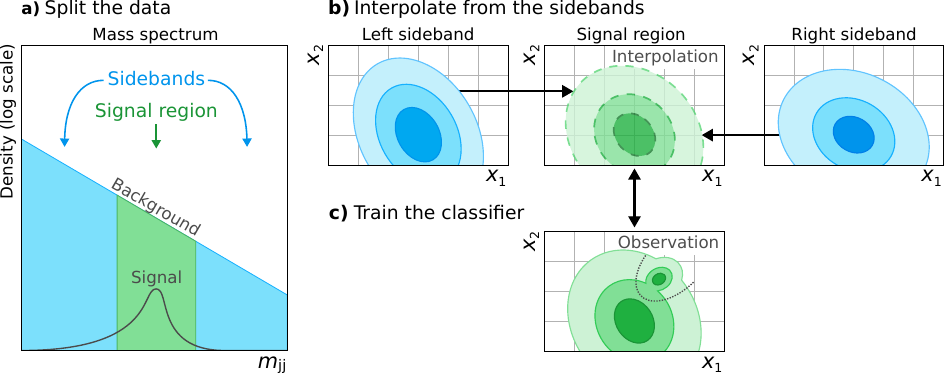}
    \caption{
        The three main steps of \cathode.
        a) Given a signal mass hypothesis, \mjj is used to define the signal
        region in which the signal would be localized.
        b) A background estimate for the input features $x$ is obtained from
        events outside the signal region by training a generative model and
        interpolating it.
        c) A weakly supervised classifier is trained to distinguish between the
        interpolated background and the observed signal region data, singling
        out any difference between them, \eg, caused by signal events.
    }
    \label{fig:cathode_diagram}
\end{figure*}

Classifying anomalies through outer density estimation
(\cathode)~\cite{Hallin:2021wme} also follows the weak supervision paradigm.
An overview of the three main steps of the method is shown in
Fig.~\ref{fig:cathode_diagram}.
First, the SR in which the LR $R(x)$ from
Eq.~\eqref{eq:weak-supervision-likelihood-ratio} will be learned is defined as
an interval in \mjj, similar to \cwola.
The second step is to estimate the background in the SR, which is done by
learning the background density in the sidebands ($\mjj\notin\text{SR}$) with a
conditional generative model, which is then sampled for $\mjj\in\text{SR}$.
Finally, the anomaly score is obtained by training a weakly supervised
classifier between the data and sampled background events.

Modeling \pbg with a generative architecture is a significant upgrade over
\cwola, to which \cathode is otherwise similar.
It allows the modeling of all correlations between \mjj and the features $x$
used for classification.
This allows the training of a single event-level classifier instead of the
jet-wise tagging done in \cwola.
The input features used in this paper are inspired by
Ref.~\cite{Hallin:2021wme}.
With j1 designating the heavier of the two jets and j2 the lighter, they are:
\begin{equation}
  m_\mathrm{j1},
  \quad \DeltaMJJ = m_\mathrm{j1} - m_\mathrm{j2},
  \quad \tau_\mathrm{41,j1},
  \quad \tau_\mathrm{41,j2}.
  \nonumber
\end{equation}
The $\tau_\mathrm{41}$ ratio was chosen instead of $\tau_\mathrm{21}$ as in
Ref.~\cite{Hallin:2021wme} because it was found to provide sensitivity to a broader set of signals.
Compared to \cwola, this smaller feature set is better suited for training
normalizing flows and minimizes performance degradation caused by the
presence of irrelevant features~\cite{Finke:2023ltw}.

A second advantage of the generative model is the possibility to use the
complete \mjj interval to constrain the background distribution, instead of the
comparatively narrow sidebands used in \cwola.
This results in a better estimate of \pbg.
The generative model is also not limited in the number of background samples it
can produce, which is used to reduce statistical fluctuations in the classifier
training.
Based on studies from the original \cathode paper, we sample four times as many
background events as there are data events in the SR.

Our implementation of \cathode closely follows Ref.~\cite{Hallin:2021wme}, with
slightly larger networks owing to the larger dataset.
The generative models are inverse autoregressive normalizing
flows~\cite{Kingma:2016wtg} with 15 MADE~\cite{germain2015made} blocks,
implemented in \PyTorch version 2.0.0~\cite{Ansel_PyTorch_2_Faster_2024} with
\Pyro version 1.8.4~\cite{JMLR:v20:18-403,Phan:2019elc}.
Each block has a single hidden layer with 128 nodes and uses ReLU activation.
Batch normalization layers~\cite{ioffe2015batchnorm}, with the momentum
parameter set to 0.1, are inserted between blocks to stabilize the training.
Optimization is performed using the Adam algorithm with a
learning rate of $10^{-4}$ and a weight decay (L2 regularization) parameter of
$10^{-6}$.
The loss function is the negative log-likelihood (NLL) of the transformed samples
under a standard Gaussian distribution.
The network is trained for 100 epochs and an ensemble of the five best epochs is
used for sampling.

The flows are conditioned on \mjj and learn $p(x|\mjj)$ for \mjj in the
sidebands.
To sample background events, we sample \mjj values from a kernel density
estimate~\cite{KDE} of the \mjj distribution in the SR and pass
them to the flows to obtain $x$.
The sampled events are then used to train the weakly supervised classifiers that
predict the final anomaly score.

The classifiers are simple fully connected networks with three hidden layers of 128 nodes using ReLU activation and a single output node with sigmoid activation,
implemented in the same \PyTorch release as the flows.
The binary cross entropy loss function is used with sampled events having label 0 and data events having label 1.
The two classes are weighted to compensate for the oversampling.
Training uses the Adam optimizer with a learning rate of $10^{-3}$ for a maximum of 100 epochs.
The results are stabilized by taking the average of the network output over the 10 epochs with the smallest validation loss.

The full training procedure is repeated for each possible choice of holdout and
validation folds.
The final anomaly score in a given holdout fold is obtained by averaging the
output of the four classifiers trained with different validation $k$-folds.
This is evaluated on all events from the SR and the sidebands, of which we keep
the 1\% most anomalous.
As in \cwola and \TNT, events selected from the five test $k$-folds are merged
together and used in the fit.
The complete procedure is repeated for SRs A1 to A6 and B1 to B5.
Region B6 is omitted because the number of events in region B7 alone is
insufficient to train the flows.

This \doc also features a second instance of \cathode, called \cathodeb, that differs from \cathode only in the input variables, adding the \DeepB scores of each jet to the previous list.
This modification is motivated by the many new physics models involving couplings to third-generation quarks and the additional sensitivity to these models brought by b tagging.
\cathodeb shares its methodology with \cathode but is otherwise treated as a
separate method.

\subsubsection{Signal selection efficiency}
\label{sec:limits_sys_uncs}

The classifiers obtained from weakly supervised methods are only approximations
of the ideal case, Eq.~\eqref{eq:weak-supervision-likelihood-ratio}, because of,
among others, the finite number of signal events in the training sample.
This can be understood from two extreme cases.
When the data contains no signal, the classifier cannot learn anything
meaningful and its score is a random function of the input variables.
If, on the other hand, the data contains only signal, we recover the maximally
powerful case of the supervised classifier.
Any intermediate signal contamination, such as the one that would originate from
a BSM signal, results in a classification power between these two
extremes.
In practice, a signal fraction of the order of 1\% or less is often sufficient
to achieve maximal performance.

The fraction $\epsilon$ of signal events that survives the event selection
is directly related to the classification power of the classifier, and therefore
also depends on the total number $N_\text{sig}$ of signal events present in the
data.
This dependence is illustrated in Fig.~\ref{fig:limit_diagram}, which shows
how efficiency increases as a function of $N_\text{sig}$.
As discussed in Appendix~C of Ref.~\cite{CMS:2024nsz}, obtaining this
$\epsilon(N_\text{sig})$ curve and estimating the associated uncertainties
is essential to set a limit on the signal cross section, one of the core results
of searches for new particles.
This section details the procedure followed in Ref.~\cite{CMS:2024nsz}.

\begin{figure}
\centering
\includegraphics[width=\cmsFigWidthv]{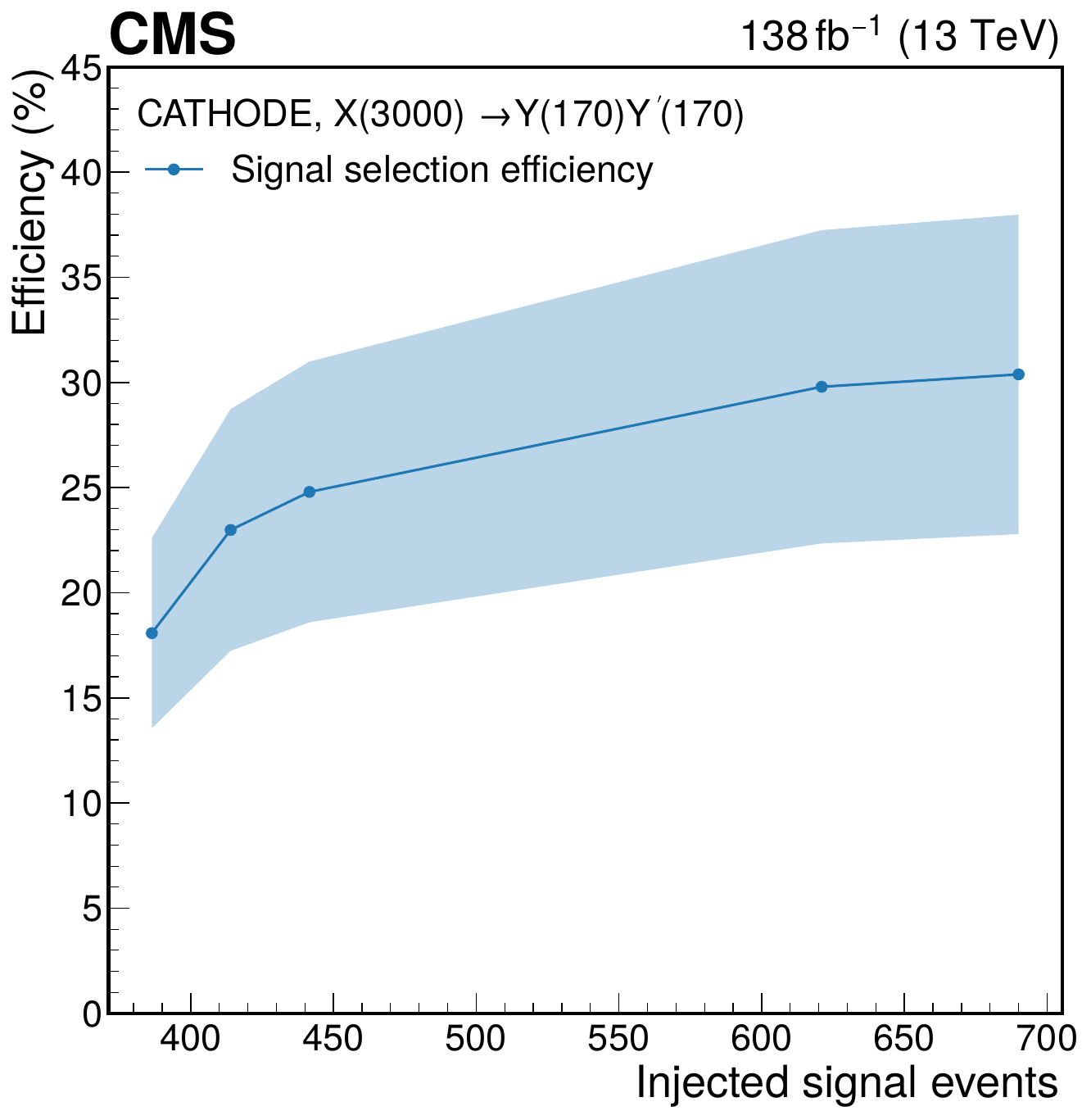}
\caption{
    Signal selection efficiency of the weakly supervised classifier in the
    \cathode method, evaluated by injecting simulated signal events in data for
    the \XtoYY signal at 3\TeV in SR B3.
    The shaded region represents the total statistical and systematic
    uncertainty evaluated as described in the text.
    The uncertainty is largely correlated between the different signal injection points,
    such that the rise in the signal efficiency is statistically meaningful despite the change
    over the considered range being smaller than the uncertainty band.
  }
  \label{fig:limit_diagram}
\end{figure}

The $N_\text{sig}$ values at which the efficiency $\epsilon$ needs to be
estimated are determined by the limit setting procedure and correspond to the
region where the search starts being sensitive to the chosen signal.
Assuming a given $N_\text{sig}$, we start by randomly sampling this number of
events from the signal sample of interest and mixing them with data from the
SR.
Using data directly instead of simulated background is necessary because of the
limited size of the background samples.
It also removes the need to consider systematic uncertainties related to
background simulation, which can be sizeable.
On the other hand, injecting signal into the data may induce a bias in case of
signal presence in the data.
We direct the interested reader to Ref.~\cite{CMS:2024nsz} for more details on
this effect, whose magnitude was deemed acceptable.

Having injected signal events into the data, we repeat the training of the
classifiers with the mixed datasets.
The trained classifiers and the remaining signal events are used to compute
$\epsilon(N_\text{sig})$.
This estimate is, however, noisy: the classifier performance depends on the
specific set of signal events injected and, to a smaller extent, on the random
initialization of network parameters.
To obtain stable results, the procedure is repeated five times with different
injected events and the results are averaged.
The envelope of the five runs is used as a systematic uncertainty.

As discussed in Ref.~\cite{CMS:2024nsz}, the distribution of signal events
is affected by systematic uncertainties related to the imperfect modeling of
the underlying physics process and the detector.
Significant changes to the features of signal events would affect the training
of the weakly supervised classifiers and may result in a change in $\epsilon$.
In principle, this can be accounted for by repeating the injection and retraining
procedure described above with versions of the signal adjusted for the
uncertainties.
However, repeating the training for many sources of systematic uncertainty
quickly becomes computationally prohibitive.
For small uncertainties, the residual stochastic fluctuations caused by
retraining can also be larger than the uncertainty.
A selection procedure is therefore used to perform the retraining only for the
most impactful sources.

The selection of these sources is based on the distributions of the
classifier input features for signal events.
A histogram is first constructed for every variable under consideration,
dividing the range in ten equally sized bins, excluding the upper and lower 1\%
of the distributions for stability.
Then, the histograms corresponding to variations of the systematic uncertainties
by $\pm1$ standard deviations are obtained within the same binning.
A source of uncertainty is selected for retraining if, for at least one bin in
any of the input features, the number of events that would be injected for the
up or down variation deviates from the nominal count $n$ by more than one fifth
of its statistical uncertainty $\sqrt n$.
This is a conservative criterion and the impact on the classifier is found to
still be small for several of the selected sources.

The final estimate of the uncertainty on $\epsilon(N_\text{sig})$ is obtained by
summing in quadrature the contributions from all sources:
the injection and retraining uncertainty,
and the systematic uncertainties in the signal description,
whether or not they were selected for retraining.
The uncertainty band shown in Fig.~\ref{fig:limit_diagram} is obtained using
this procedure.

\subsection{QUAK}
\label{subsec:quak}

The \textit{quasi anomalous knowledge} (\quak) method~\cite{Park:2020pak} is
the most model-dependent method employed in this analysis.
\quak uses a set of example signals as a prior to help enhance sensitivity to physical anomalies.
The method seeks a balance between the purely model-agnostic approaches of the prior methods and
fully supervised methods in which the model is optimized for a specific chosen signal.
The use of signal priors could help filter out physically irrelevant anomalies,
which could arise from bugs in the event reconstruction.
\quak uses two unsupervised models.
The first is trained on simulated background events and the second on the signal prior.
The final event selection makes use of both models.
Events which are dissimilar to background events and similar to the signal prior are selected as anomalous.

Our application of \quak uses autoregressive rational quadratic spline
(RQS) normalizing flows~\cite{durkan2019splineflows} implemented in \PyTorch~1.9
with \nflows~\cite{nflows}~0.14 as the unsupervised model.
We use six RQS transformations with 80 control points.
The parameters controlling the splines are obtained from a network comprising
five fully connected layers with 160 nodes and ReLU activation, evaluated on the
output of the previous layer.
The flows are trained with the Adam optimizer with a learning rate of $10^{-3}$ and
a batch size of 10,000.
Training stops when the loss does not improve over 10 consecutive epochs, with a
maximum of 100 epochs.
The NLL predicted by the flows for every event is
used to construct one ``signal'' and one ``background'' score.
This architecture takes advantage of the progress in generative models since the
publication of Ref.~\cite{Park:2020pak} and was found to improve performance
over the original autoencoder-based proposal.

The benchmark signals discussed in Section~\ref{sec:datasets} are used as the signal prior.
To ensure sensitivity to a wide range of signals,
the signal prior is split into six separate groups based on the masses of the B and C particles.
The groups are listed in Table~\ref{tab:quak-flows}.
A separate flow model is trained for each group.
The final signal score is the combination of the NLLs obtained from the six different flow models, calculated as the signed $L^5$ norm
$||\vec x||_{5} = (x_1^5 + x_2^5 + \dots + x_6^5)^{1/5}$.
This relatively large power was chosen so that the final signal score is sensitive to high signal-like scores from any of the six flows.
The background-like score is simply the NLL obtained from the corresponding flow.

Each flow estimates the joint probability density of 14 event-level input
variables, seven for each of the two jets, in its training dataset.
The two jets in each event are sorted by mass.
The following jet features are used:
\begin{equation*}
  \tau_s, \tau_{21}, \tau_{32}, \tau_{43}, \nPF, \rho, \DeepB,
\end{equation*}
where the modified $N$-subjettiness variable $\tau_s$ is defined as
$\tau_s=\sqrt{\tau_{21}}/\tau_1$ and $\rho$ is the ratio of the jet mass and
transverse momentum, $\rho = \mSD/\pt$.

\begin{table}
  \topcaption{%
    Signal models used to train the six signal normalizing flows used in the
    \quak method.
    Each flow uses signals with specific masses for the daughter particles B and
    C.
    For signals decaying to an SM and a BSM particle, the BSM particle is always
    the heavier of the two.
  }
  \centering
  \begin{tabular}{cccc}
    Daughter masses (\GeVns) & Processes \\
    \hline
               80, 80  & \XtoYY \\[\cmsTabSkip]
    \phantom{0}80, 170 & \Wkk   \\
                       & \Wp    \\
                       & \XtoYY \\[\cmsTabSkip]
    \phantom{0}80, 400 & \Wkk   \\
                       & \XtoYY \\[\cmsTabSkip]
              170, 170 & \Wp    \\
                       & \XtoYY \\[\cmsTabSkip]
              170, 400 & \Wp    \\
                       & \XtoYY \\[\cmsTabSkip]
              400, 400 & \GtoHH \\
                       & \Zp    \\
  \end{tabular}
  \label{tab:quak-flows}
\end{table}

A two-dimensional ``\quak~space'' is constructed and used to define the event selection.
The two dimensions of the space encode an event's background-like and signal-like scores, respectively.
The most anomalous events are expected to appear in the region characterized by a low background-like score and a high signal-like score.
As the exact location of an anomaly in the \quak~space could vary, the event selection procedure is varied for
different resonance mass hypotheses, $m_\mathrm{A}$.
First a signal region is defined as $m_\mathrm{A} - 400\GeV < \mjj < m_\mathrm{A} +
200\GeV$.
The width is motivated by the \mjj resolution of the detector, and a slight
shift towards low \mjj is added to account for longer tails observed on the
left side of signal \mjj distributions.
Two 500\GeV-wide sidebands, on each side of the SR, are used to estimate the background distribution in the \quak space.
These events are used to populate a 2D histogram representation of the \quak~space with 500 bins.
Bins exhibiting an excess of SR events relative to the sidebands are selected.
Events from the full dijet mass spectrum that fall within these chosen \quak~bins are then used in the resonance signal fit.

A broad search is conducted using \quak~with a signal prior consisting of a
combination of all benchmark signal samples, combining six flows as described
above.
Additionally, targeted searches for each benchmark signal are performed using
\quak with a single signal flow, trained solely on the specific signal being
tested.
Since this approach is optimized for detecting each benchmark signal individually, it is expected to perform better for known signals but generalize less effectively to unknown signals.
This method serves as a point of comparison when assessing the performance of anomaly detection techniques.

\section{Performance analysis}
\label{sec:performance_comparison}

The anomaly detection methods show varying performance across signal types.
Understanding these variations can guide the development of more generic and
performant methods that combine different approaches.
In this section, we explore the performance characteristics of our methods
using the benchmark signals.
First, Section~\ref{sec:tagging-performance} examines the ability of all methods
to identify the benchmark signals.
In Section~\ref{sec:correlations}, we then investigate whether the same events
are found by every method.
Finally, Section~\ref{sec:supervised} focuses on the impact of the choice of
input features on the performance and genericity of the methods.

A widely used metric to assess the performance of anomaly detection algorithms
in high-energy physics is the significance improvement characteristic
(SIC)~\cite{Gallicchio:2010dq},
defined from the true and false positive rates (TPR and FPR) as $\text{SIC}=
\text{TPR}/\sqrt{\text{FPR}}$.
The SIC quantifies how much the statistical significance ($z$-score) of an excess
is enhanced after applying a selection based on an anomaly detection algorithm, as
compared to the baseline selection only.
For instance, a SIC of 3 means that an initial excess of two standard deviations
becomes a six standard deviation excess after event selection.

\subsection{Anomaly tagging performance}
\label{sec:tagging-performance}

Unlike traditional classification problems, it is nontrivial to compare the
different anomaly detection techniques.
For the weakly supervised methods the tagging performance changes as a function of the amount of signal present in the data.
Therefore, comparing models on a SIC curve has limited utility as one
will have to pick an arbitrary signal cross section on which to evaluate the
performance.
Because the tagging performance can vary substantially across different cross
section ranges it is difficult to draw conclusions from such a comparison.

A more appropriate performance comparison evaluates the search sensitivity
achieved by each algorithm.
For this study, we construct a mock dataset by injecting signal at a chosen
cross section into the simulated background.
The complete version of each anomaly detection algorithm, including training, selection and signal extraction, is run on this mock dataset, and the final expected signal significance (or, equivalently,
$p$-value) is reported.
Better anomaly detection algorithms are those that can achieve a higher signal
significance (smaller $p$-value) at a given cross section.

Two signals are used as benchmarks for this study.
The first signal used as a benchmark is the $2+2$ prong \XtoYY signal with $\mX = 3\TeV$ and $\mY = \mYp = 170\GeV$.
This is a relatively straightforward signal in which the two large-area jets are both
2-pronged.
The second signal used as a benchmark is the $3+3$ prong \Wp signal with $\mWp = 3\TeV$ and $\mPBpr = 400\GeV$.
This signal has more complicated 3-prong jets on both sides ($\PQb\PZ$ and \PQt),
each of which also contains a \PQb quark jet.

To contextualize the performance of the anomaly detection methods,
their sensitivities are compared to those of several standard methods.
The standard comparison methods are chosen to differ only in their event selections,
and utilize the same preselection criteria,
fitting procedure, and statistical analysis as employed by the anomaly detection
methods.
The first comparison method is an inclusive search, which does not use any substructure-based selection criteria.
By not using any requirements on the jet substructure, this strategy is maximally inclusive to all signals.
However, its sensitivity is limited by the lack of a strategy to reduce the background yield.
Two model-specific event selections based on typical selections targeting jet
substructure are also used.
The first (second) selection targets 2-prong (3-prong) signals by requiring both jets in the
event to have $\tau_{21} < 0.4$ ($\tau_{32} < 0.65$) and $\mSD > 50\GeV$.
Finally, model-specific selections using \quak are employed to maximize
sensitivity, with the signal prior set to the specific signal model being
tested.

\begin{figure*}
   \centering
   \includegraphics[width = 0.9\textwidth]{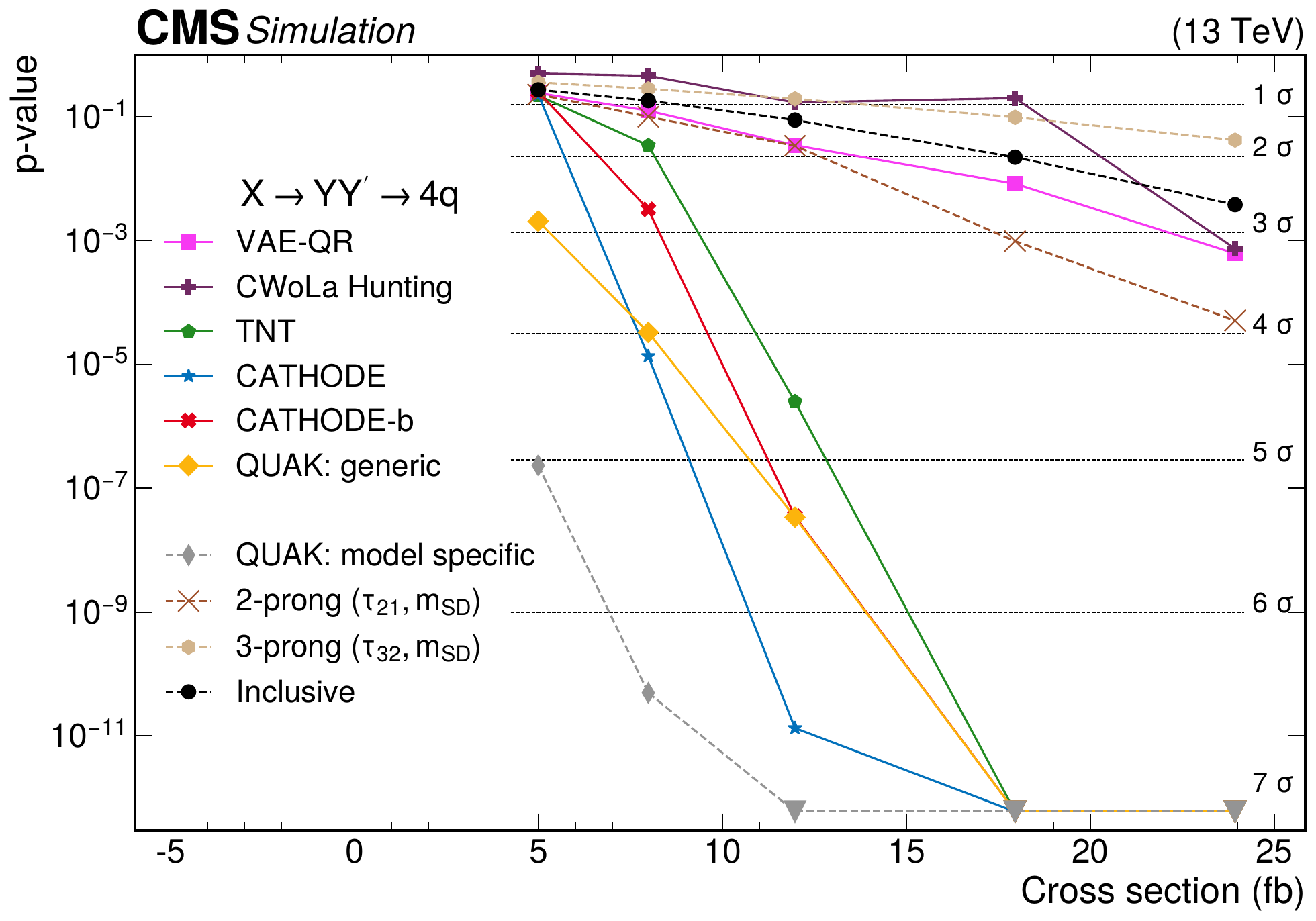}
   \includegraphics[width = 0.9\textwidth]{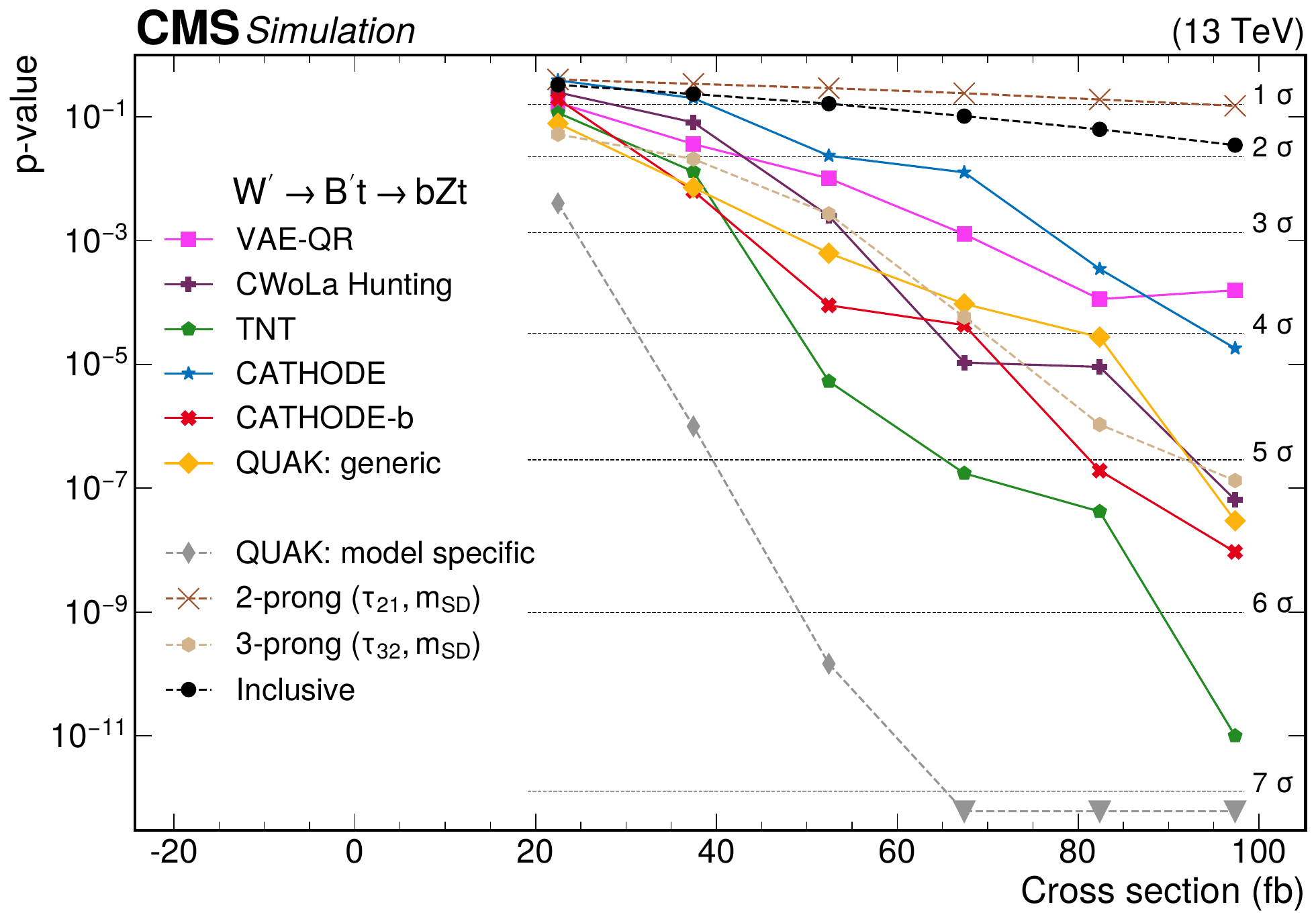}
   \caption{
    The $p$-values as a function of the injected signal cross sections for the
    different analysis procedures for two different signals, using different
    input features for each method as listed in Table~\ref{tab:methods-summary}
    and in the legend.
    The upper panel shows results for the 2-prong \XtoYY signal with
    $\mX=3\TeV$ and $\mY=\mYp=170\GeV$, while the lower panel shows
    results for the 3-prong \Wp signal with $\mWp=3\TeV$ and
    $\mPBpr=400\GeV$.
    Significance values larger than 7 standard deviations ($\sigma$) are denoted
    with downwards facing triangles.
    Reproduced from Ref.~\cite{CMS:2024nsz}.
   }
   \label{fig:pvals}
\end{figure*}

Figure~\ref{fig:pvals} shows the sensitivity of the search methods in
simulation, evaluated as explained above.
We compare the expected $p$-value extracted from the \mjj fit as a function of
the signal cross section for two benchmark signals.
This figure is reproduced from Ref.~\cite{CMS:2024nsz} and included here for
completeness.

As expected, the inclusive search is sensitive to both models, but reaches only low significances at the considered signal cross sections.
The model-specific selections targeting 2- and 3-prong signals have better sensitivity than
the inclusive search for the matching signal model, but significantly worse sensitivity for the other one.
This underscores the main drawback of traditional selection criteria: they introduce model dependence.
We thus use the inclusive baseline preselection as the model-agnostic baseline when
evaluating the performance of anomaly detection methods.
This choice is, for instance, present in the definition of the SIC.

In contrast to the 2- and 3-prong selections,
all anomaly detection methods demonstrate increased sensitivity
compared to the inclusive search, for both signals.
The relative performance of the anomaly detection methods differs between the two
signals and no single method is optimal for both.
As expected, the model-specific \quak search yields the best sensitivity for
both signals, reaching the discovery level (larger than five standard deviations) at significantly lower cross sections
than other methods.

\begin{figure*}
   \centering
   \includegraphics[width = 0.9\textwidth]{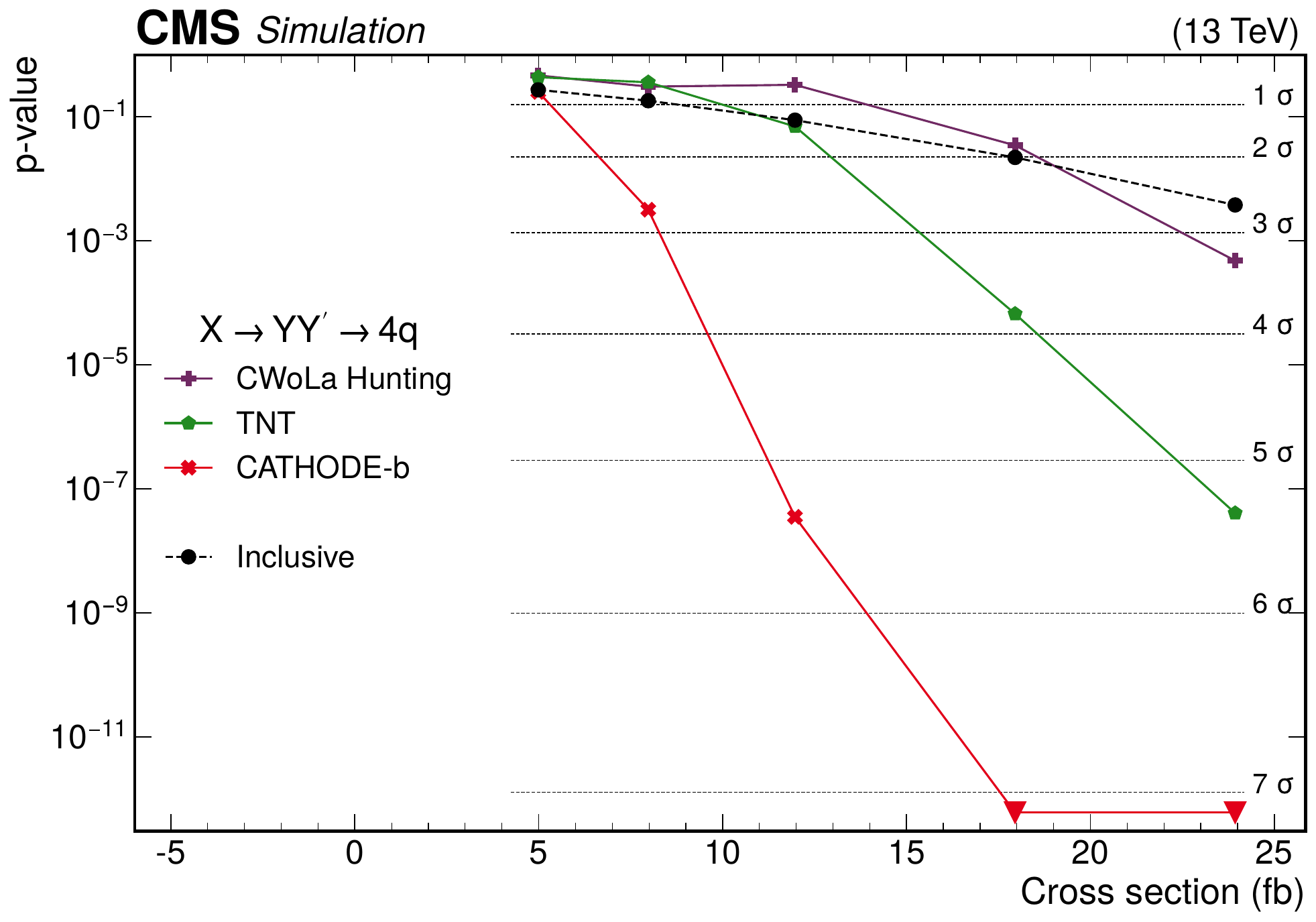}
   \includegraphics[width = 0.9\textwidth]{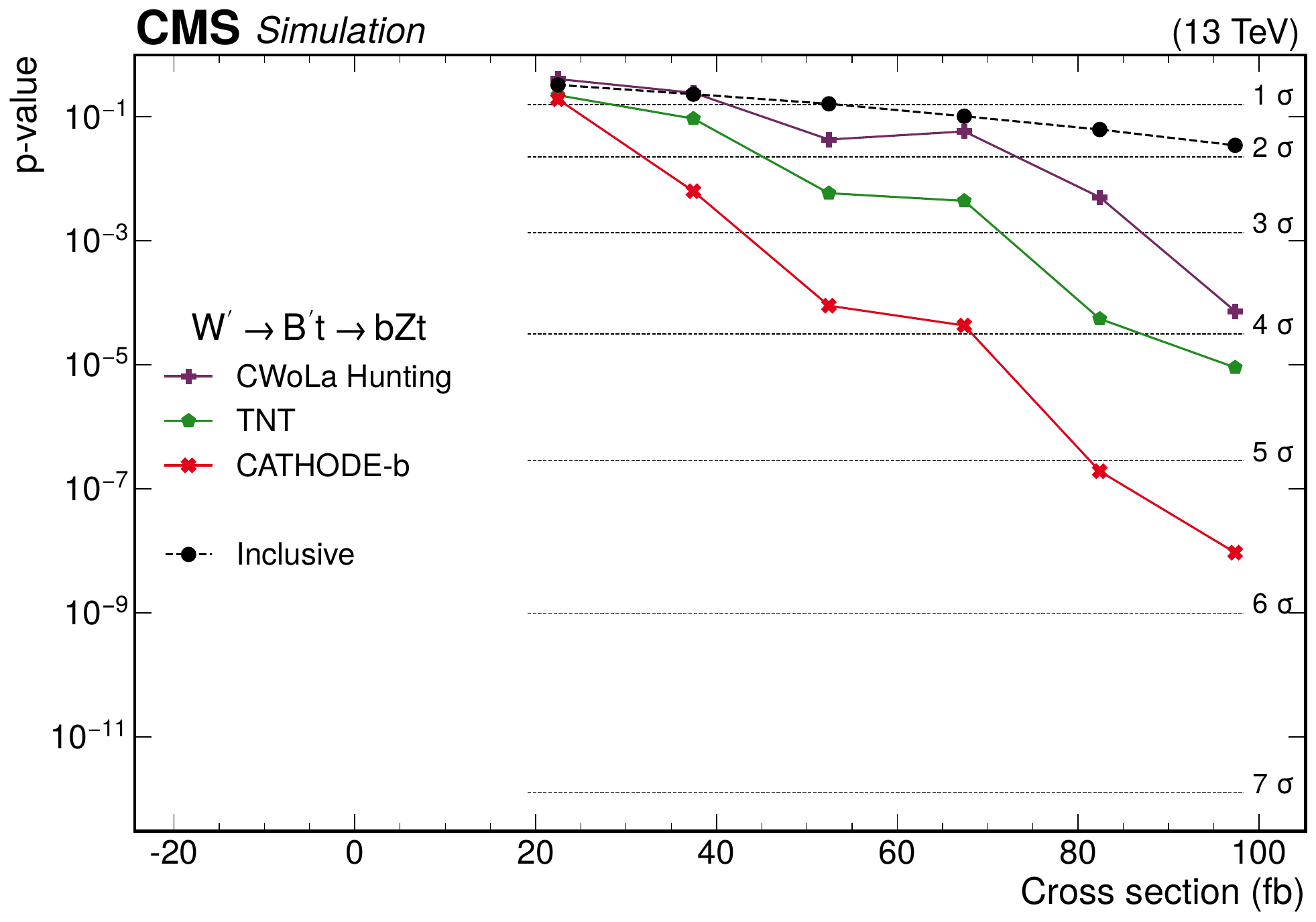}
   \caption{The $p$-values as a function of the injected signal cross sections for the different analysis procedures modified to use a common set of input features, for two different signals: (upper) the 2-prong \XtoYY signal with $\mX=3\TeV$ and $\mY=\mYp=170\GeV$, and (lower) 3-prong \Wp signal with $\mWp=3\TeV$ and $\mPBpr=400\GeV$.
   Significance values larger than 7 standard deviations ($\sigma$) are denoted
   with downwards facing triangles.}
   \label{fig:pvals_same_feats}
\end{figure*}

Because the methods use different feature sets, comparing them with common input
features provides a fairer assessment of algorithmic performance.
We therefore repeat this study, with each method using a common feature set,
that of \cathodeb.
The \vae method was not included in this study because it is based on
high-dimensional low-level input features and cannot readily be adapted to use a
small set of high level features.
For \TNT, we only modify the inputs of the classifier without changing the
unsupervised preselection.
Timeline constraints prevented the inclusion of \quak.
The results are shown in Fig.~\ref{fig:pvals_same_feats}.
\cathodeb achieves the best results because, unlike \TNT and \cwola, it
combines information from both jets to tag anomalous events.
Among the per-jet methods, \TNT outperforms \cwola because of its additional
preselection criteria~\cite{Amram:2020ykb}.

\subsection{Anomaly score correlations}
\label{sec:correlations}

We also explore the correlation between the anomaly scores of the different
methods on these two benchmark signals as well as background events.
We train the methods in simulation and compare their results on identical
events.
For the weakly supervised methods, an amount of signal equivalent to the limit
obtained from the inclusive search performed on the same dataset is used in the
comparison.

\begin{figure*}
    \centering
    \includegraphics[width=0.5\textwidth]{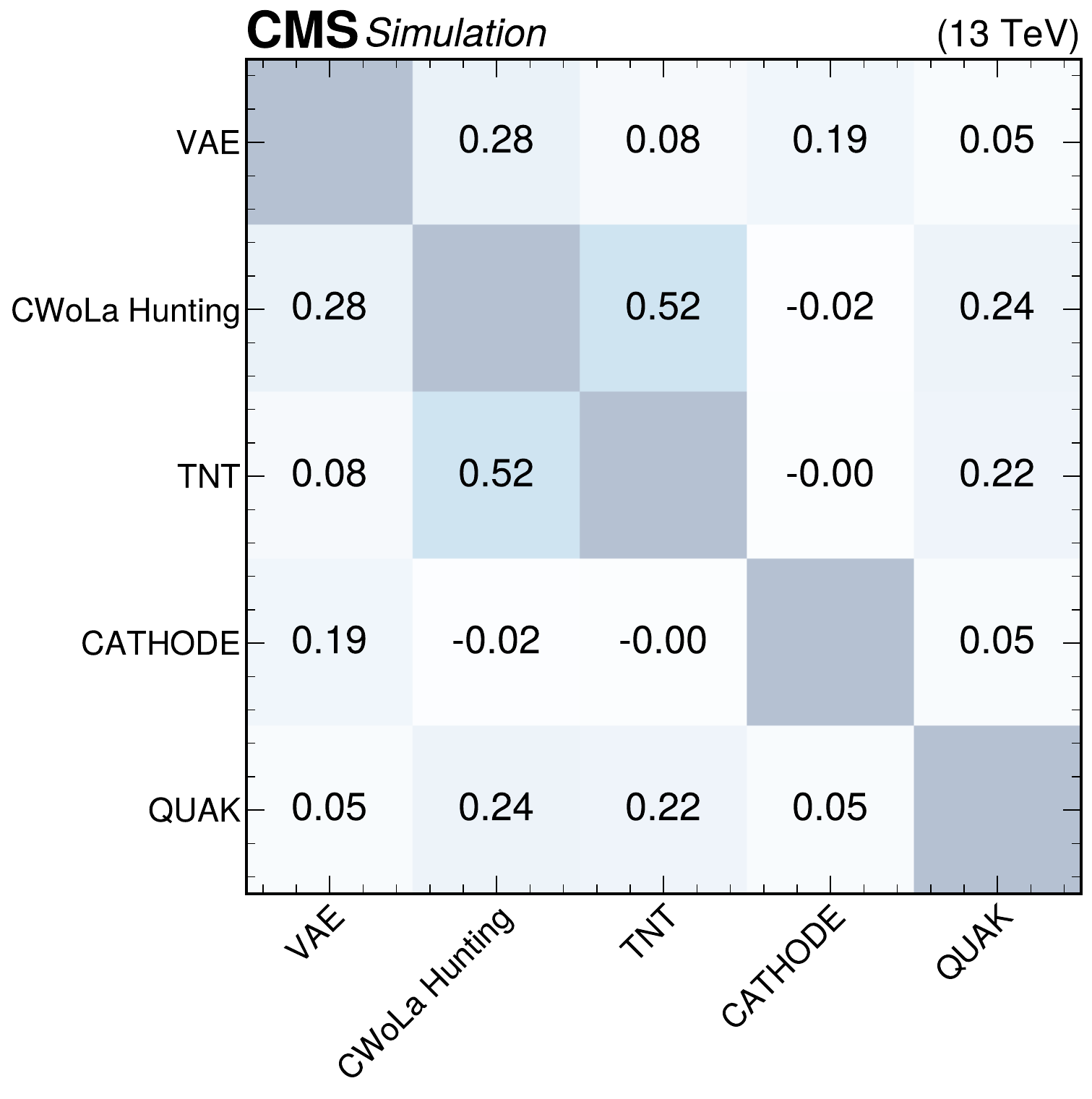}\\
    \includegraphics[width=0.5\textwidth]{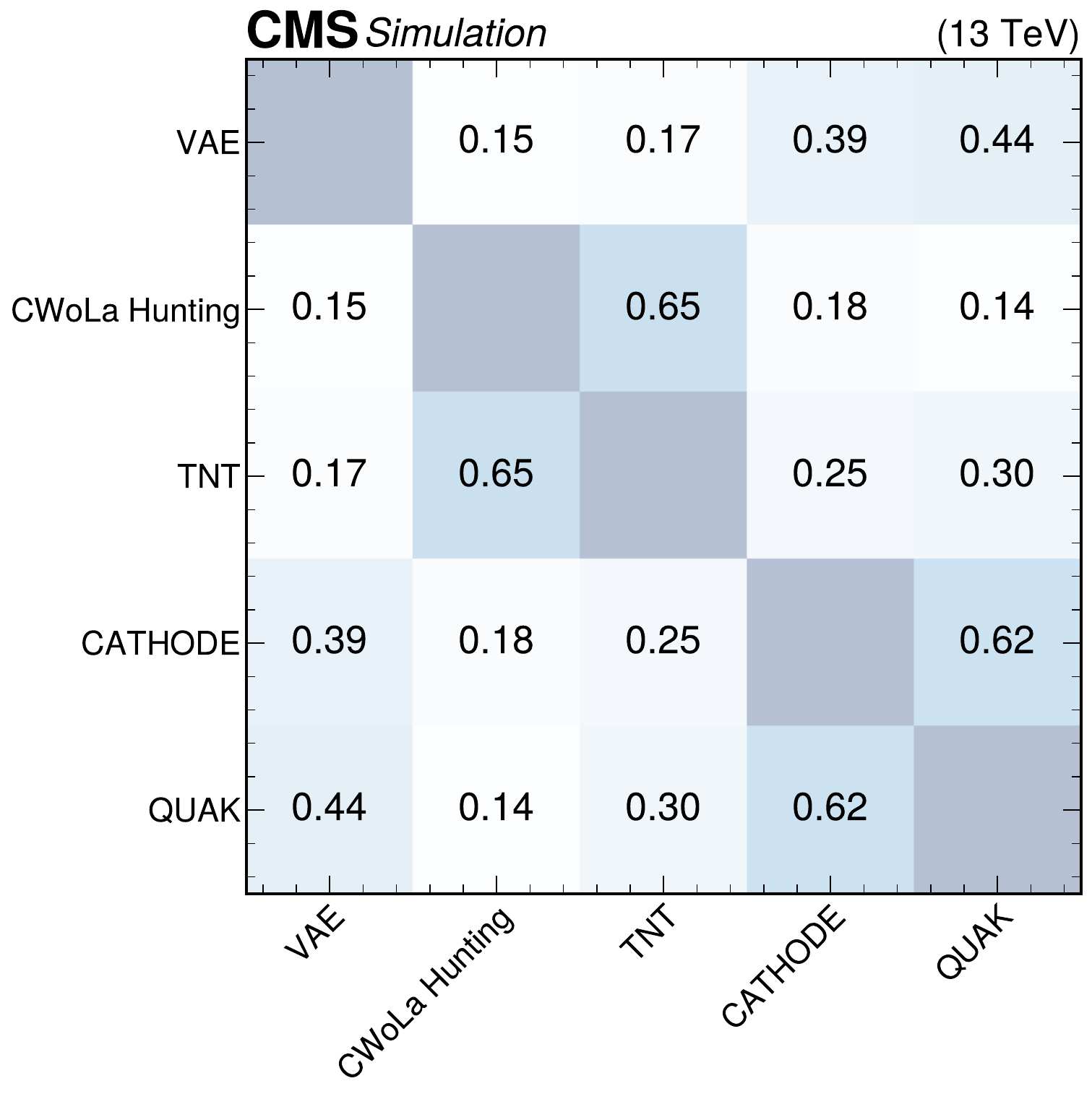}%
    \includegraphics[width=0.5\textwidth]{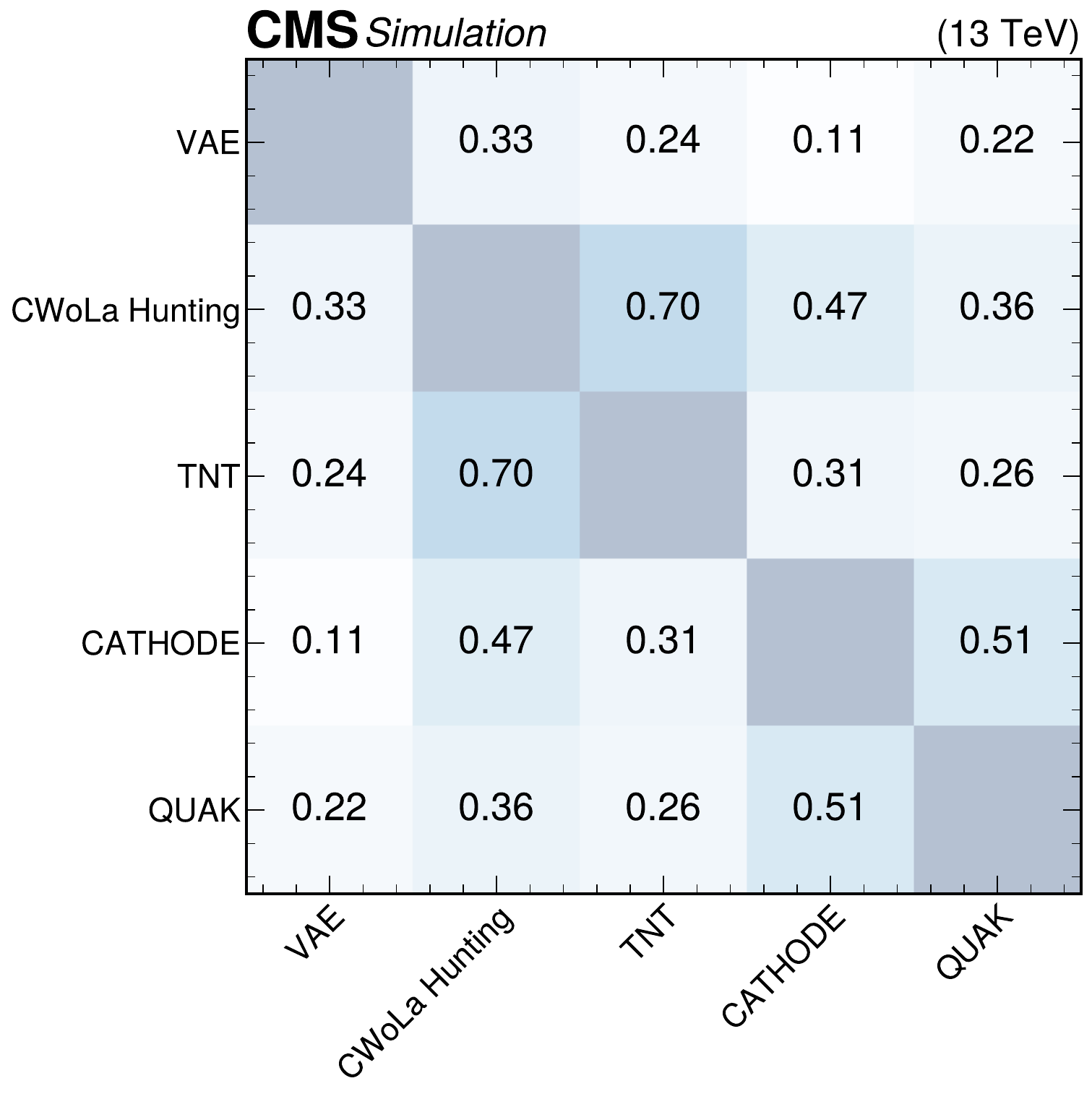}
    \caption{
      Summary plots showing the Pearson correlation coefficient for each pair
      of anomaly detection algorithms as evaluated on simulated events from the
      SM backgrounds (upper), \XtoYY signal (lower left), and \Wp signal (lower
      right).
      In many cases, the correlations are weak, indicating complementarity
      between the different approaches.
    }
    \label{fig:corr_summary}
\end{figure*}

Because the anomaly scores of the different methods have varying ranges and
distributions, we apply a preprocessing to standardize them before making a
comparison.
From the distribution of anomaly scores of a given method on a given sample, we
use a quantile transformation to convert the distribution to a Gaussian of mean
zero and variance one.
For example, the median anomaly score in the original distribution is mapped
to a value of 0, and a score at the 16th percentile is mapped to a value of
$-1$.
We then investigate the correlations between the scores assigned by the
different methods for background, \XtoYY, and \Wp events.
Direct comparisons are shown as scatter plots in
\cmsAppendix\ref{sec:score-comparisons}.
In each case
we compute the Pearson correlation coefficient, which measures the linear
correlation between the two distributions, as
well as the distance correlation (DisCo~\cite{Kasieczka:2020yyl}), which also
captures nonlinear relationships.
The Pearson correlation matrices between the different methods are shown in
Fig.~\ref{fig:corr_summary}.
We assume that \cathode and \cathodeb are strongly correlated and only include
\cathode.

In general, we find the correlations between the different methods to be
relatively low.
The two methods with the strongest correlations are \cwola and \TNT, which have
Pearson correlations of ${\sim}60\%$.
This is to be expected because they are the only two methods that use exactly the
same input features, and they have similar training procedures.
A similar correlation also exists between \cathode and \quak, which both use
information from two jets during training.
The low correlations indicate that the different approaches are complementary:
they identify anomalous events based on different criteria, meaning each method
captures anomalies that others might miss.
This also indicates that more powerful search strategies could
be obtained by combining the strengths of the ones presented here.

We stress that because the normalization is done separately for each sample, one cannot compare scores as shown in these plots across the different samples.
Additionally, because the comparison is done separately with ``pure'' samples of
signal or background, this is an investigation of which of the signal events
or background events look the most anomalous to each method.
This means that there can be low correlation between two performant methods.
For example, they may disagree about which is the ``more anomalous'' of two
signal events but still agree that both are much more anomalous than background
events.

\begin{figure*}
    \centering
    \includegraphics[width=0.49\textwidth]{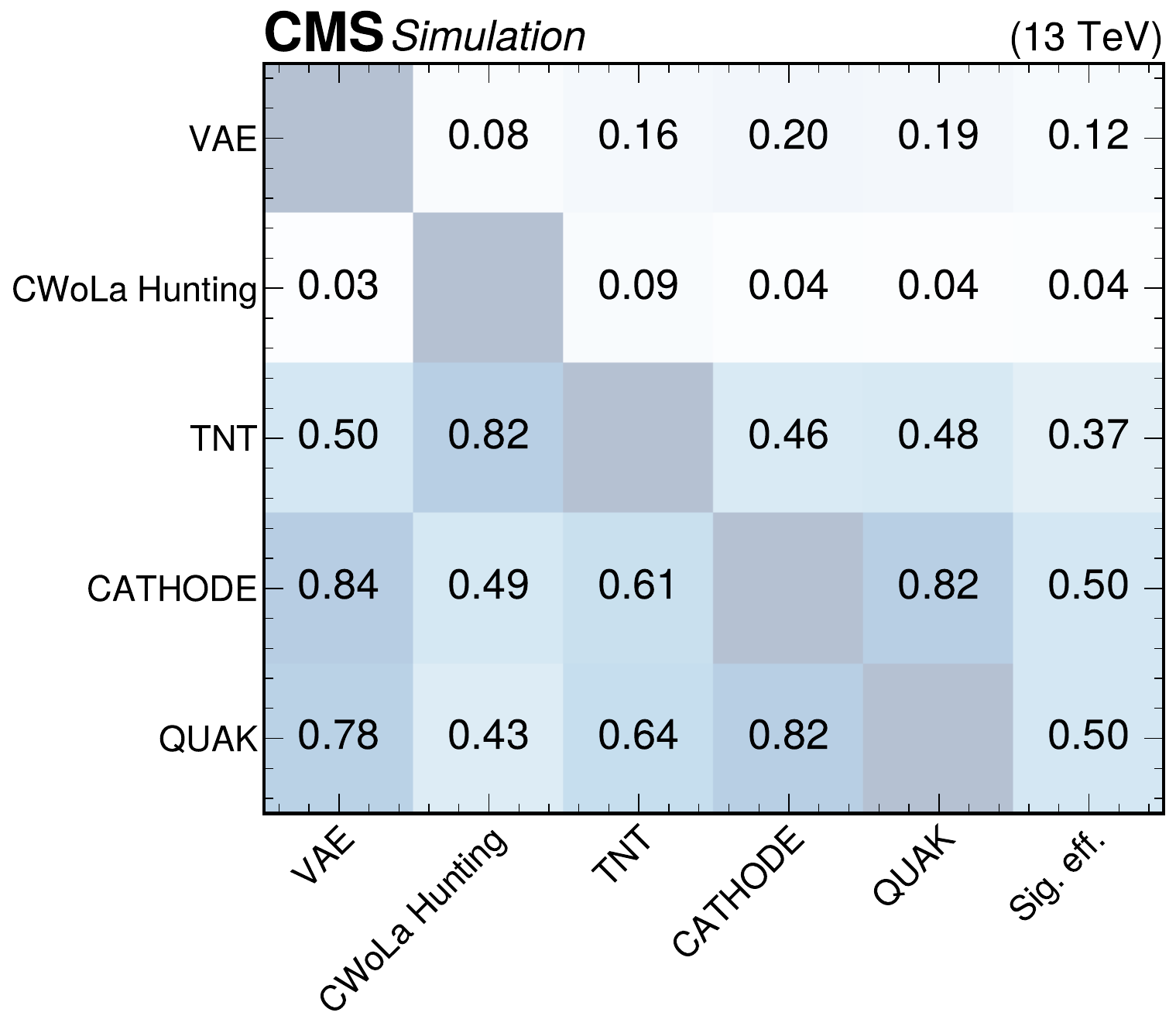}
    \includegraphics[width=0.49\textwidth]{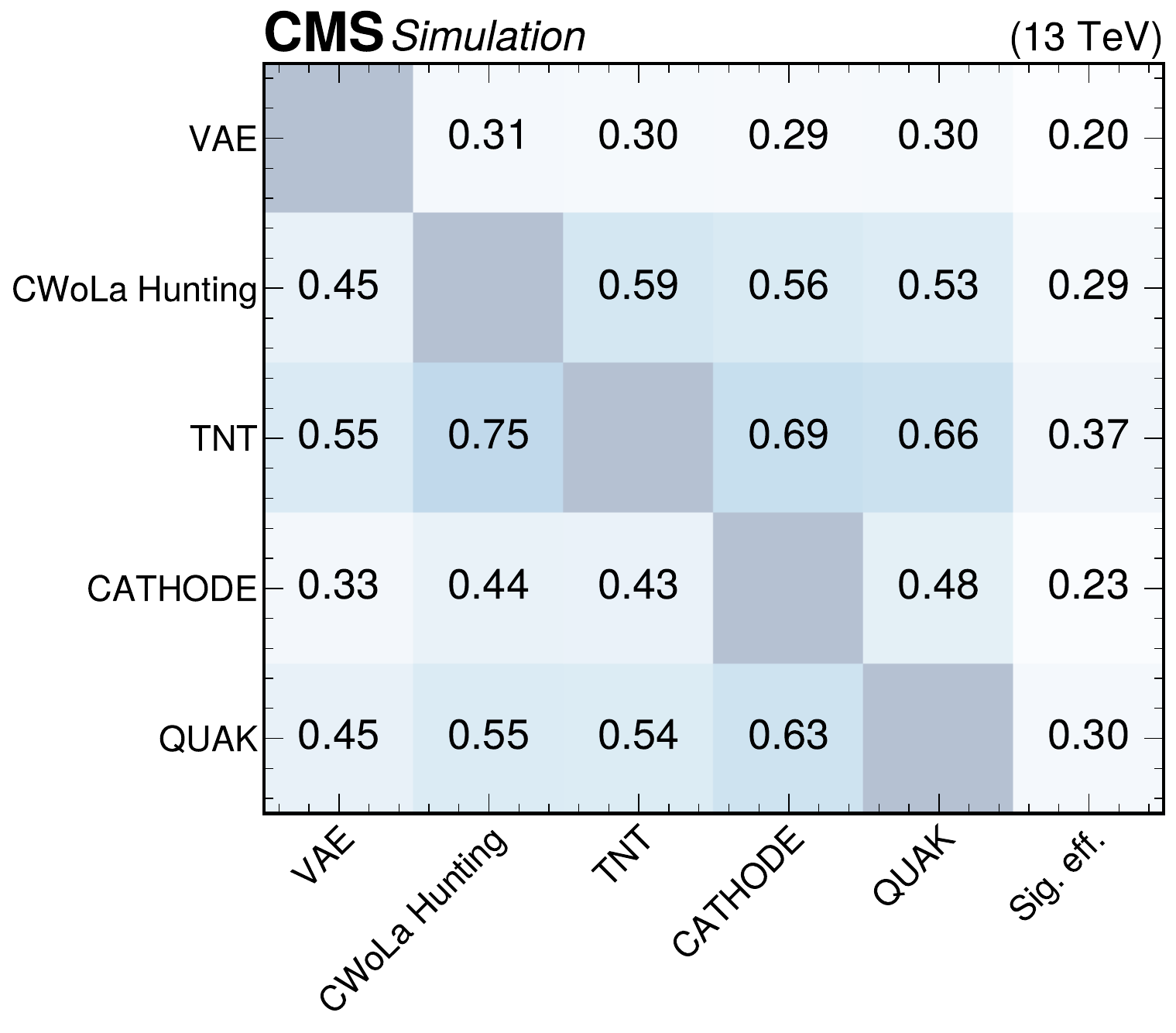}
    \caption{
      Overlap between the events selected using the anomaly detection methods,
      for simulated \XtoYY (left) and \Wp (right) signal events, with a false
      positive rate of 1\%.
      Each cell corresponds to the fraction of the events selected by the method
      shown on the horizontal axis that are also selected by the method on the
      vertical axis.
      The last column denotes the overall efficiency of selecting signal events for each method.
      For instance, the \cathode method selects 50\% of all \XtoYY events.
      Of those events selected by \cathode, 46\% are also selected by \TNT,
      but only 4\% are found by \cwola.
    }
    \label{fig:confusion}
\end{figure*}

A different perspective is obtained by considering the selected events directly.
For each method, we choose the threshold leading to a 1\% background efficiency,
and we compare the selected signal events.
The results of this study are shown in Fig.~\ref{fig:confusion}.
For the \XtoYY signal, \cathode and \quak select half of the events and reach
discovery-level significance.
\TNT follows with close to 40\% of the signal events correctly selected.
The \vae only finds 12\% of the signal and \cwola is not sensitive.
Unsurprisingly, methods with correlated anomaly scores also tend to have the
largest overlap, with up to 82\% between \cathode and \quak.
Most events found with \vae are also identified by \cathode and \quak; the
overlap with \TNT is generally lower.
For the more complex \Wp signal, all methods achieve comparable sensitivity by
selecting 20--40\% of the signal events, without strong overlap.
This confirms our earlier observation that the different methods are
complementary.

\subsection{Impact of input features}
\label{sec:supervised}

Except for \TNT and \cwola, the five methods described in this \doc all use different input features.
It was already seen in Section~\ref{sec:tagging-performance} that this has an
impact on their performance.
Furthermore, the most correlated methods share part of their input features.
This motivates a more complete study of the impact of the choice of inputs.
In this section, we perform an experiment to investigate the extent to which differences in sensitivity can be attributed to the set of input features, by training supervised classifiers to distinguish between signal and background.

Ideally, a supervised classifier trained with the binary cross-entropy loss learns the likelihood ratio between signal and background, resulting in optimal classification performance.
Thus, we can use such classifiers as proxies to study the separation power in a given feature set.
When possible, we reuse architectures already used in the anomaly detection
methods: this is the case for \cwola, \TNT, \cathode, and \cathodeb.
The \cwola and \TNT classifiers operate at the level of individual jets.
The scores from both jets are combined differently, as described in
Sections~\ref{subsec:cwola} and \ref{subsec:tagntrain}.
We reflect this difference by training one set of jet-level supervised
classifiers and combining their scores following the respective method.
For \quak, we use a fully connected network with 64, 64, and 32 nodes with ReLU activation in the three hidden layers, and a single node with sigmoid activation for the classification output.
Owing to the difficulty of training a supervised classifier on the low-level input features used by the \vae, it is not included in this study.

Supervised classifiers were trained for all signal models described in Section~\ref{sec:datasets}, using data in the corresponding SR (Table~\ref{tab:weakly_supervised_massbins}) as a background proxy.
Using data instead of simulation guarantees the absence of modeling issues.
A contamination of data by a signal strong enough to affect classifier trainings significantly is ruled out by the negative result of the weakly supervised searches~\cite{CMS:2024nsz}.

We choose the SIC at a false positive rate of 1\% as the main metric for comparisons.
Indeed, this corresponds most closely to the gain in sensitivity achieved after selecting events based on the classifier score.
The working point of 1\% corresponds to the one used by all weakly supervised
methods for $\mA=3\TeV$, and by \cathode and \cathodeb for $\mA=5\TeV$.

\begin{figure*}
\centering
\includegraphics[width=0.5\textwidth]{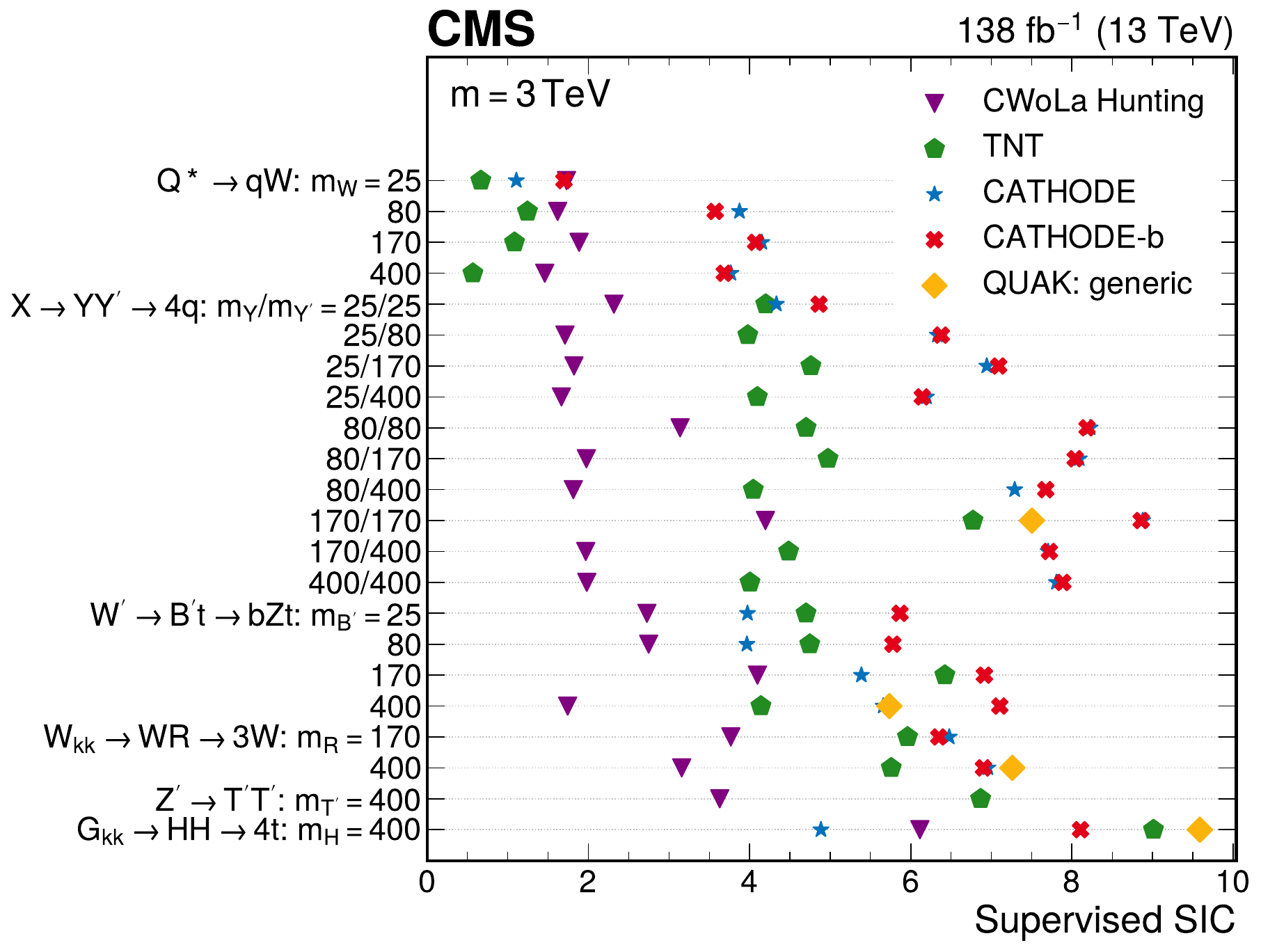}%
\includegraphics[width=0.5\textwidth]{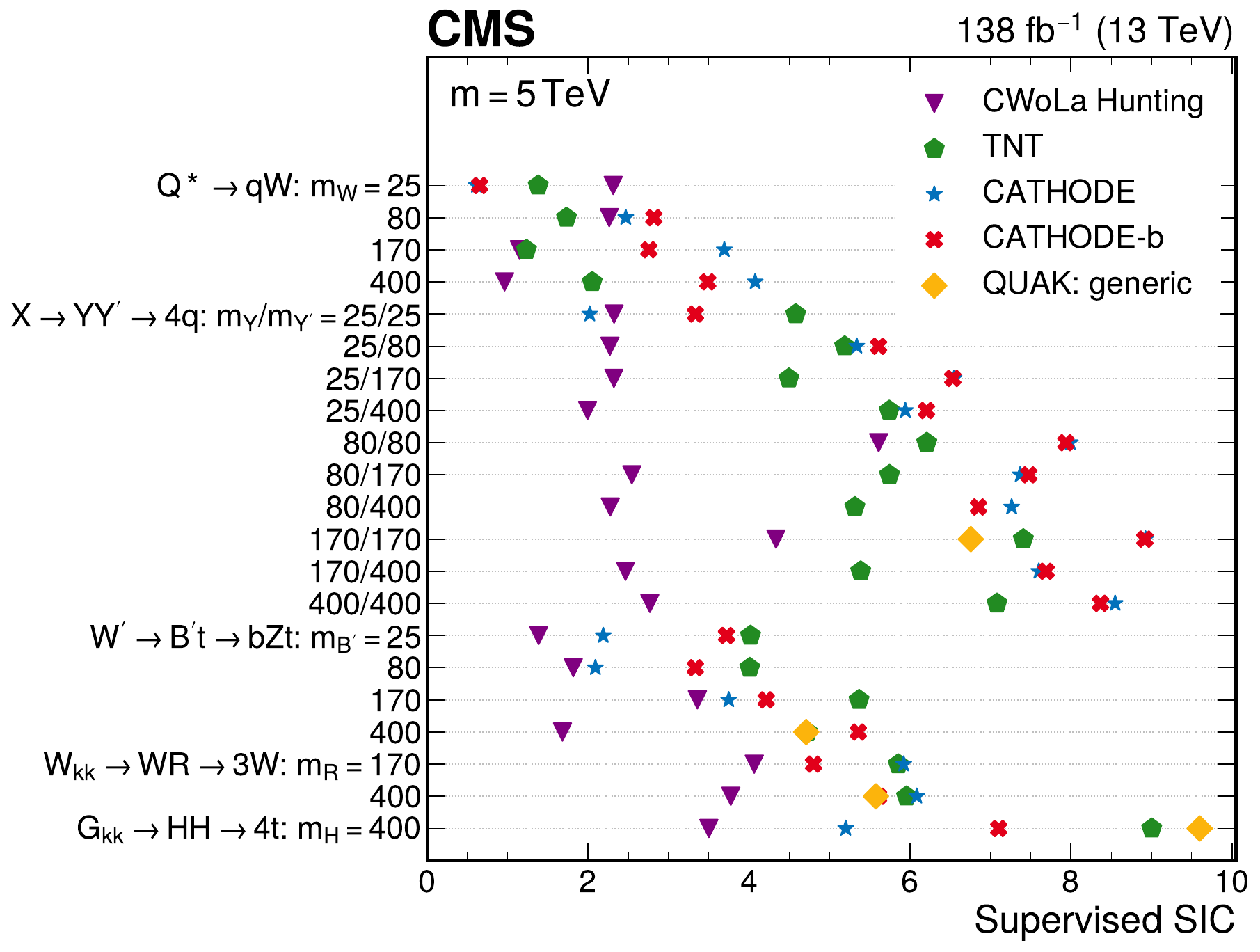}
\caption{%
    Significance improvement obtained from supervised classifiers trained using
    the same inputs as the anomaly detection methods to distinguish between
    simulated signal and data.
    The left (right) panel shows the 3\,(5)\TeV mass point for all signal models.
    The masses of intermediate decay particles are listed in \GeVns.
    The \quak method was not tested on all models.
  }
  \label{fig:supervised-sic}
\end{figure*}

The SIC obtained by the supervised classifiers is shown in Fig.~\ref{fig:supervised-sic}.
The \cwola-like supervised classifiers achieve a SIC of two for most signals.
Combining the jet scores as in \TNT leads to a higher SIC (about four at 3\TeV
and six at 5\TeV) for all signals with two anomalous jets, but performs worse
for the \Qstar signal.
This is expected since \TNT uses the product of the scores, enhancing events
with two anomalous jets, unlike in this signal model.
For most signals, the \cathode, \cathodeb, and \quak feature sets yield similar
results.
They outperform the \cwola- and \TNT-like classifiers for 3\TeV signals, while
the difference with \TNT is less marked at 5\TeV.
A comparison of \cathode and \cathodeb shows very similar results except for
signals involving \PQb quark jets, to which the additional \DeepB inputs of
\cathodeb add extra sensitivity.

\begin{figure}
  \centering
  \includegraphics[width=0.5\textwidth]{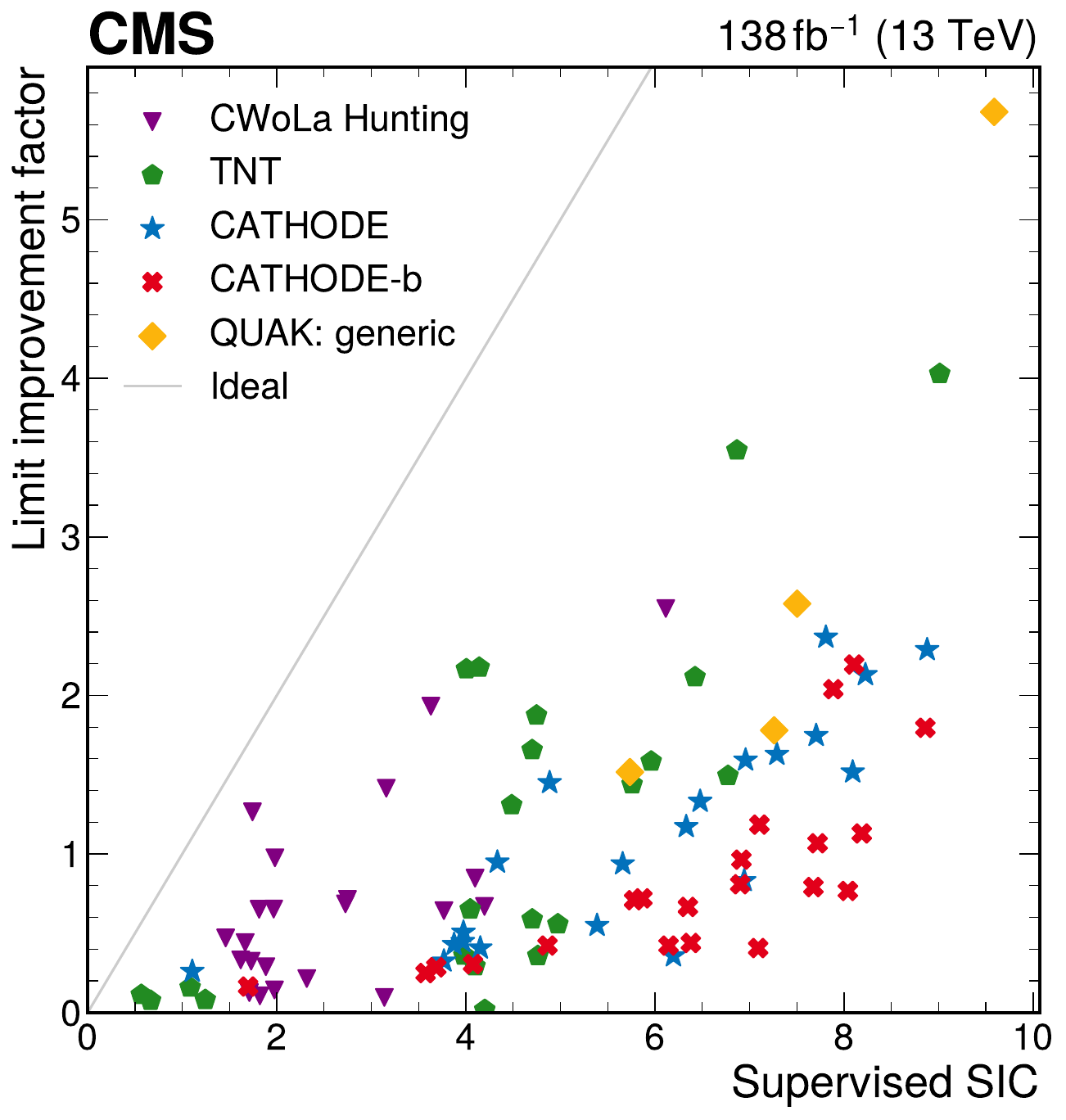}
  \caption{%
    Comparison of anomaly detection and supervised classification performance.
    The factors by which the anomaly detection methods improve the upper limits
    on the signal cross sections are compared to
    the SIC of classifiers trained with the same inputs.
    The 3\TeV mass point is shown for all models.
  }
  \label{fig:supervised-vs-ad}
\end{figure}

We now turn to investigating to what extent the differences in the input features
can explain differences between the sensitivity of the anomaly detection methods.
We measure the improvement in sensitivity by comparing the expected upper limits
on signal cross sections obtained in Ref.~\cite{CMS:2024nsz} in data, without
and with anomaly detection.
We calculate the limit improvement factor by taking the ratio of the limit
without, to the limit with, anomaly detection.
A value larger than one for this metric indicates that using anomaly detection
improves the expected limit by the same factor with respect to the inclusive
search.

Figure~\ref{fig:supervised-vs-ad} compares the SIC and the limit improvement
factor.
We observe a clear correlation between the two variables.
In particular, for each method we observe a clear correlation between the
separation power of the input features for a given signal and the performance on
the same signal.
Depending on the method, the improvements in the limits from the anomaly detection methods are between two and eight times smaller than the SIC obtained from the supervised classifiers.
This performance gap is mainly caused by the inferior tagging of anomaly
detection methods owing to their model-agnostic approaches.
A difference of at most $\sim$30\% can also be attributed to effects accounted
for in the limit-setting procedure but not in the SIC: systematic uncertainties
(Section~\ref{sec:limits_sys_uncs}) and the uncertainty in the expected
background yield in the signal region.

The size of the performance gap is seen to vary significantly between different anomaly detection methods.
The best performing methods in this comparison are \cwola and \TNT, which reach
half of the improvement of the corresponding supervised classifier for several signals.
Despite its use of information about the signals during training, \quak does not
reach the same performance.
\cathode and \cathodeb follow, being the methods least successful at extracting
the separation power present in their datasets.
This is partially compensated by stronger input features for \cathode and
\cathodeb, especially for the \XtoYY signal family at 3\TeV.

The direct comparison of \cathode and \cathodeb, for which only the input
features are different, also highlights an effect already described in
Ref.~\cite{Finke:2023ltw}:
input features that carry no information about the signal can reduce the performance
of weakly supervised methods.
Indeed, it can be seen in Fig.~\ref{fig:supervised-vs-ad} that \cathode almost
always outperforms \cathodeb.
A careful investigation unveils that \cathodeb surpasses \cathode for all signal
models that contain b jets, but is otherwise less performant.
Using classifiers based on large ensembles of boosted decision trees, rather than neural networks
as employed in this study, has been seen to mitigate the performance loss due to uninformative features~\cite{Finke:2023ltw},
and may be employed in future searches.

\section{Validation in data: identification of \texorpdfstring{\ttbar}{ttbar} events}
\label{sec:ttbar_val}

Despite the tests on simulation, it is useful to demonstrate that the anomaly detection methods can successfully find rare anomalies in data.
However, no SM particles fit the topology of the original search
strategy: the decay of a heavy resonance to daughters producing large-radius jets
with substructure differing from the SM backgrounds.
Therefore, a variation of the search strategy is developed and employed on data to identify \ttbar events.
This modified search strategy, based on weak supervision, targets nonresonant pair
production of two particles producing large-radius jets.
It assumes that the two jets have identical masses and substructure
characteristics.
This strategy is able to successfully identify Lorentz-boosted \ttbar production,
initially a ${\sim}1\%$ fraction of the sample, from the SM backgrounds.

Identification of \ttbar using anomaly detection methods has already been
demonstrated in the semileptonic decay channel~\cite{Knapp:2020dde}, using open
data released by the CMS Collaboration.
The algorithm used~\cite{ALAD} was based on unsupervised learning.
In contrast, we target the more challenging fully hadronic decay channel and
use weak supervision.

\subsection{Anomaly detection setup}

The study begins from the same basic selection criteria as the resonant anomaly search outlined in Section~\ref{sec:datasets}.
The two leading jets in the event are additionally required to have $\pt > 400\GeV$.
The requirement on the pseudorapidity difference between the two jets,
$\abs{\DeltaEta} < 1.3$, is kept unchanged.
At this $\pt$, fully hadronic top quark decays are partially merged: sometimes
all the decay products from the top quark are contained in the jet, but in other
events only the quarks from the \PW decay are contained and the \PQb quark forms
a separate jet.
These selection criteria were not optimized and a true nonresonant pair
production search would potentially perform better.
They are, however, sufficient for the desired validation on Lorentz-boosted \ttbar events.

The natural variable to use to extract the signal events
is the soft-drop mass of one of the jets,
which should peak at the resonance mass ($\sim$175\GeV for top quark jets),
but is smoothly falling for the QCD multijet background.
This contrasts with the rest of this paper, where the invariant mass of the
dijet system is used.
A split strategy is employed to circumvent sculpting of the \mSD distribution by
the selection based on the anomaly score:
the two jets in the event are randomly assigned to the two categories J1 and J2
and anomaly detection is applied to J2, while
the signal contribution is extracted using J1,
ensuring that there is no correlation between the anomaly score of J2 and the
mass of J1 for background events.
This enables the estimation of the contribution of QCD multijet events in the
final SR based on sidebands in data.

\begin{figure}
    \centering
    \includegraphics[width=0.49\textwidth]{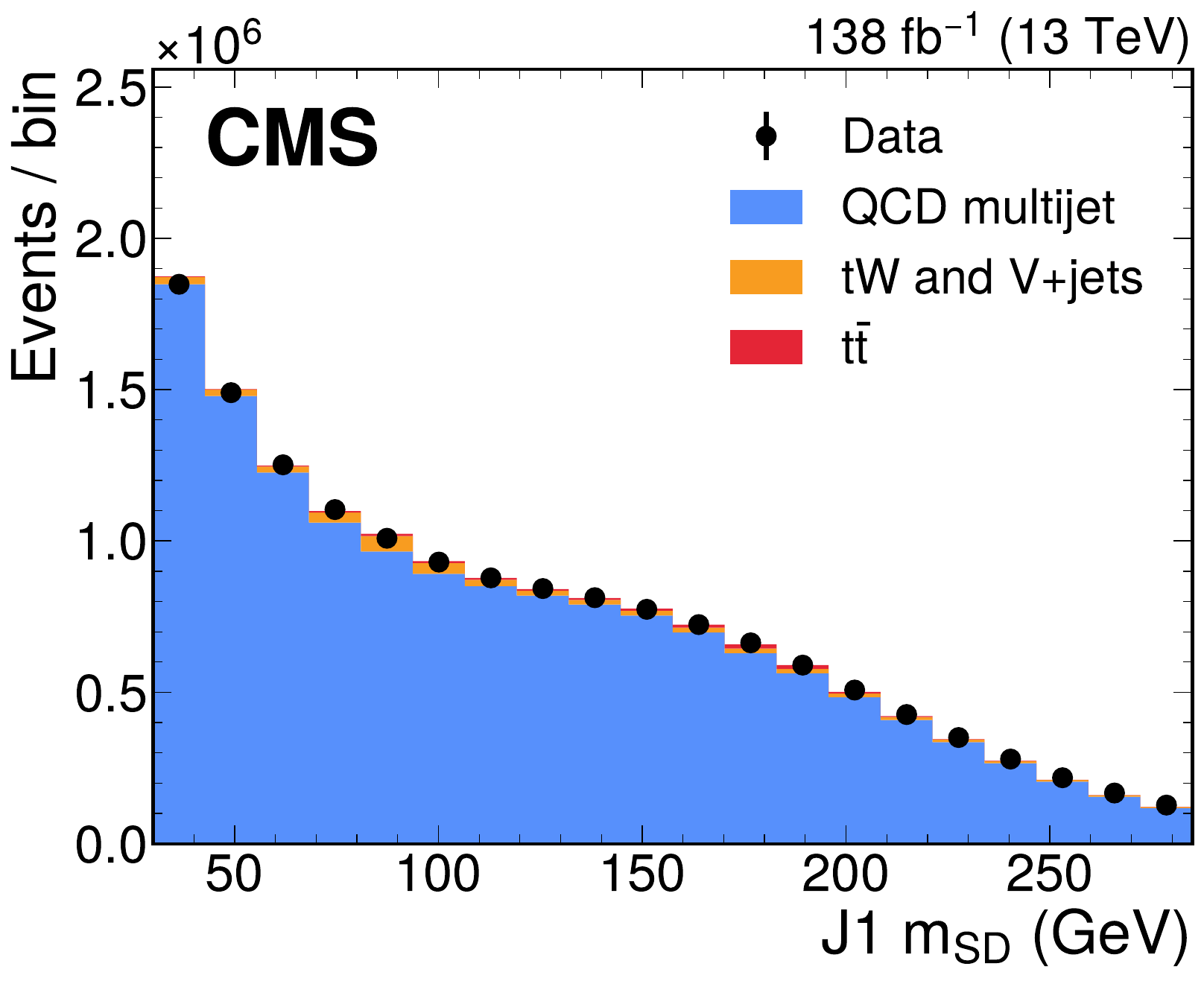}
    \includegraphics[width=0.49\textwidth]{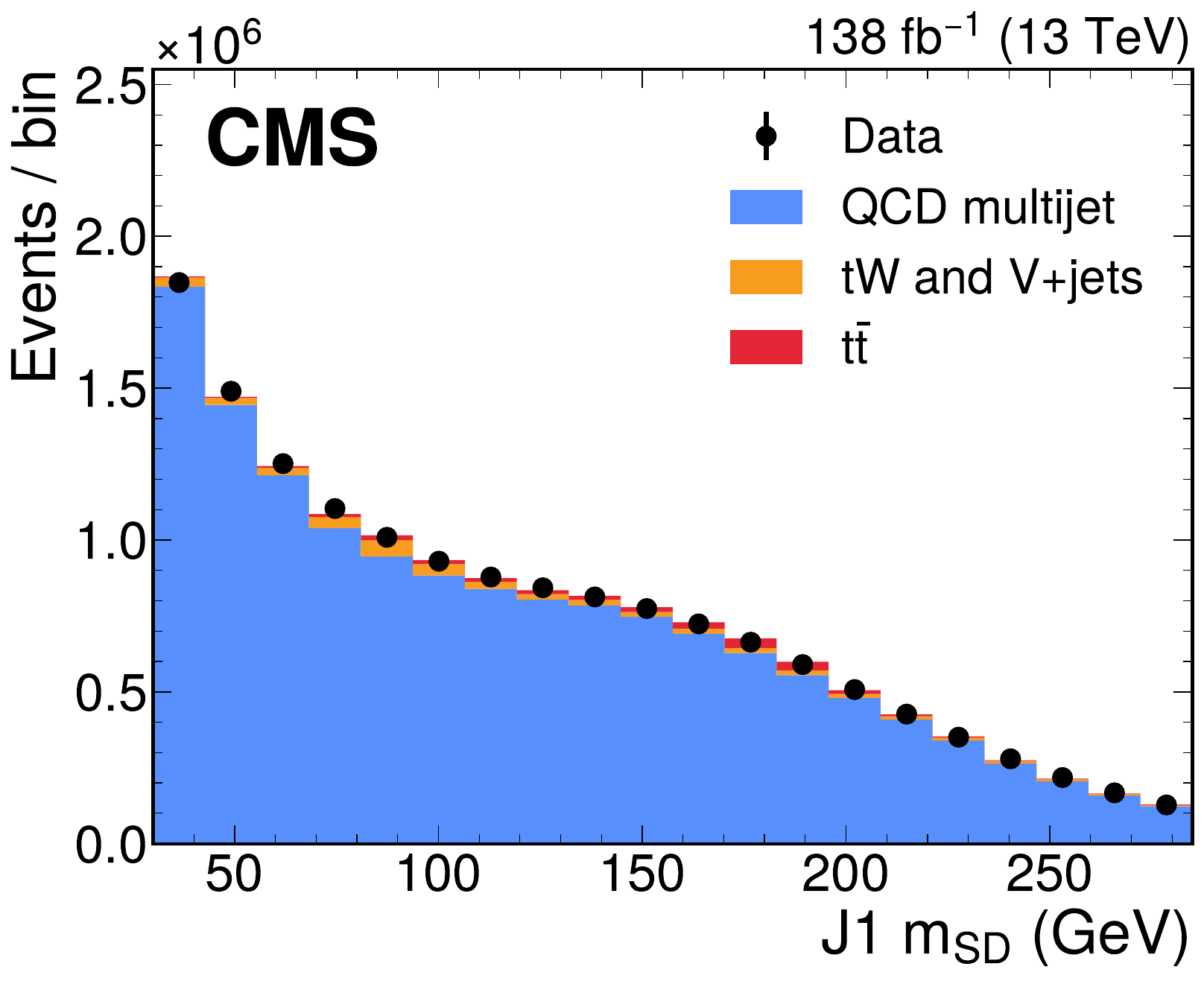}
    \caption{
      Distribution of the J1 soft-drop mass after the basic (\cmsLeft) and b-tagged
      (\cmsRight) preselections.
      The black dots represent the data and the colored histograms correspond
      to simulated events.
      In both cases the sample is dominated by the QCD multijet background
      (blue).
      While the b tagging preselection increases the relative contribution of
      \ttbar (red) from 1 to 2.5\% around the top quark mass (105--220\GeV),
      its contribution remains small.
      Other background processes are shown in yellow.
    }
    \label{fig:ttbar_presel_jet_mass}
\end{figure}

After the preselection described above, about 1\% of the events with J1
\mSD around the top quark mass (105--220\GeV) are expected to originate from \ttbar
production.
As discussed in Section~\ref{sec:limits_sys_uncs}, the performance of weakly
supervised anomaly detection methods depends on the fraction of \ttbar events in
the base dataset.
In order to explore this effect in data, we also consider a more restrictive
preselection in which J1 is required to contain a b-tagged subjet, using the
``loose'' working point of DeepCSV~\cite{BTV-16-002}.
This additional selection increases the expected \ttbar fraction in events with
J1 \mSD around the top quark mass to about 2.5\%.
The jet mass distributions after the two versions of the preselection are shown
in Fig.~\ref{fig:ttbar_presel_jet_mass}.

We chose to use the same input variables for this study as in the \cwola and
\TNT methods, since they are well-suited to discriminate between top quark jets
and the background.
There is a strong correlation between \mSD and the other substructure features
for the background jets, which could cause problems for the weakly supervised
training if the network can use the correlation to distinguish between the SR
and the sidebands.
Inspired by the \TNT approach, the signal-rich and background-rich regions are defined using the J1 jet mass, but the J2 jet features are used as inputs to the anomaly tagger.
For the QCD multijet background, the masses and substructure features are
generally uncorrelated between the two jets.
By defining signal-rich and background-rich regions based on the mass of J1, we
obtain samples of J2 that exhibit identical fractions of QCD multijet background
events in both regions.

In a true model-independent search, where the mass of the pair produced resonance is unknown, one would want to repeat the procedure
with different signal windows defined on the J1 jet mass, similar to the different \mjj windows used in the dijet search.
For this demonstration we take a simplified approach, employing the procedure on three partially overlapping signal windows defined on the J1 \mSD:
65--150, 105--220, and 145--250\GeV.
The corresponding background-rich regions is defined as all events outside the signal window.

The weakly supervised training is performed using the same model architecture
and hyperparameters as the \TNT method employed in the main search.
For each considered \mSD window, the training is performed separately for both
the baseline and b-tagged preselections.
As a simplification, a two-fold cross validation scheme is used rather than
five-fold in the dijet search.
The dataset is split into two halves, the first of which is used for training
the anomaly tagger.
After training, the tagger is applied to the second half of the data to select
the anomalous jets.

The anomaly score of the J2 jets is evaluated, and the 0.1\% most anomalous jets
are selected to define the ``pass'' region.
The events which fail this selection define the ``fail'' region.
For presentation purposes, the selected data are separated into background and
signal contributions.
Their normalization is extracted from a simultaneous maximum likelihood fit of
the two regions performed using the \textsc{Combine} package~\cite{CMS:2024onh}.
The \mSD distribution of the QCD multijet background in the pass region is
estimated from data, using the \mSD distribution of events in the fail region
multiplied by a freely floating transfer factor.
It was verified in simulation that the shapes of the QCD background in the pass
and fail regions are sufficiently similar for the transfer factor to be
parameterized by a single constant, without any dependence on the jet mass.
The shapes of resonant backgrounds from \PW and \PZ boson production in
association with jets, as well as the \ttbar signal shape, are estimated using
simulation.
The normalization of all contributions is determined entirely from the fit,
without any prior constraint.

\subsection{Results}

\begin{figure*}
    \centering
    \includegraphics[width=0.49\textwidth]{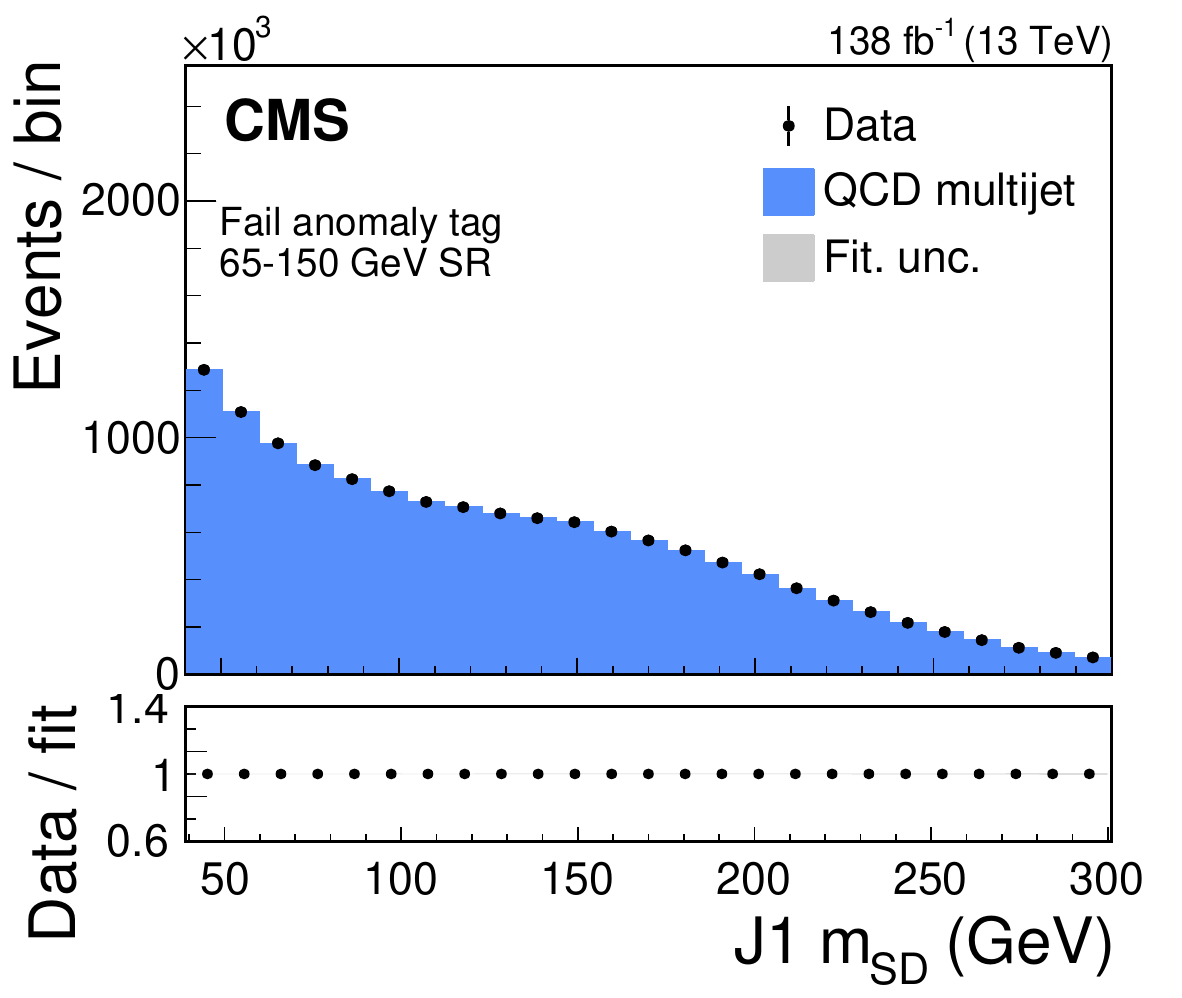}
    \includegraphics[width=0.49\textwidth]{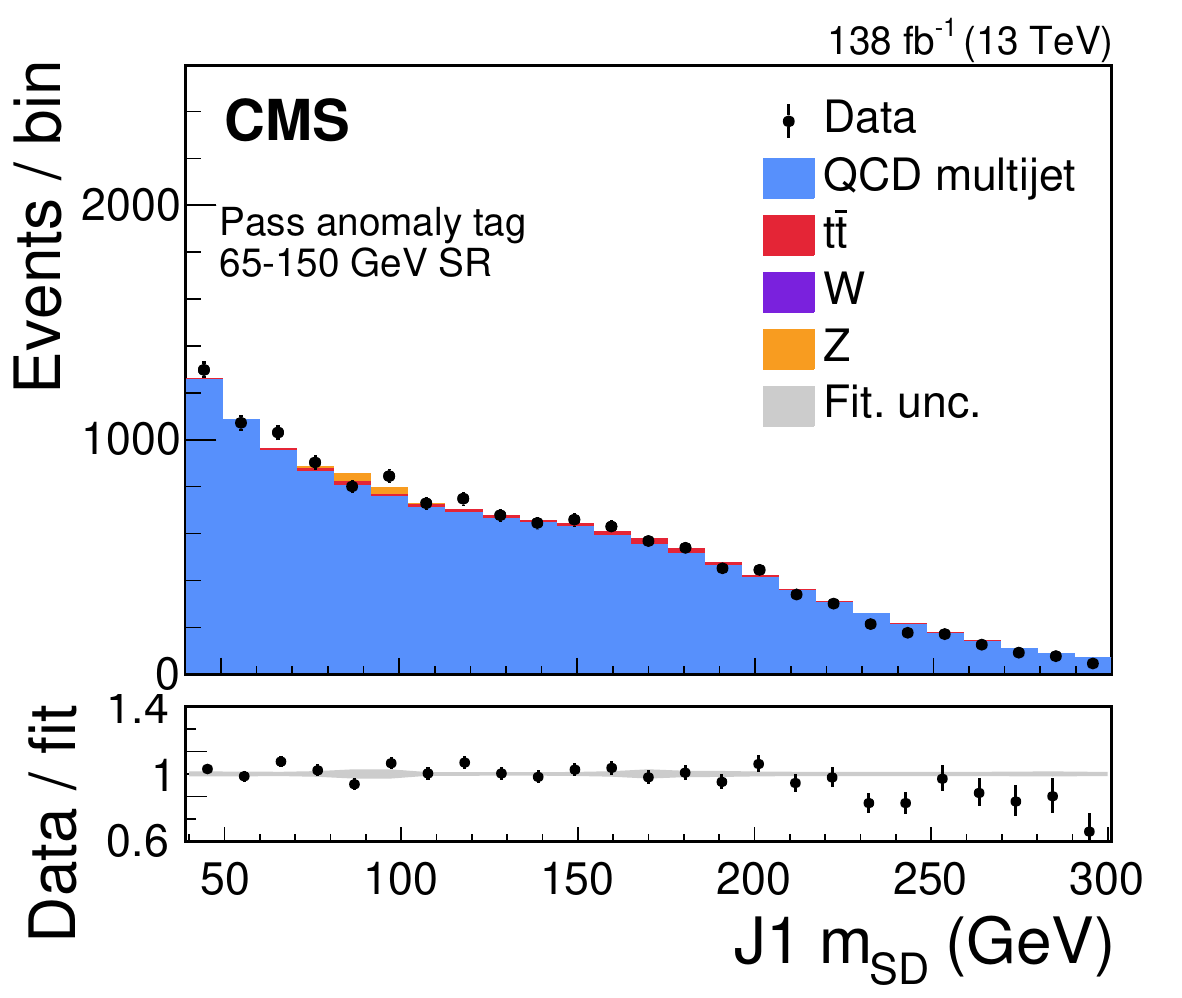} \\
    \includegraphics[width=0.49\textwidth]{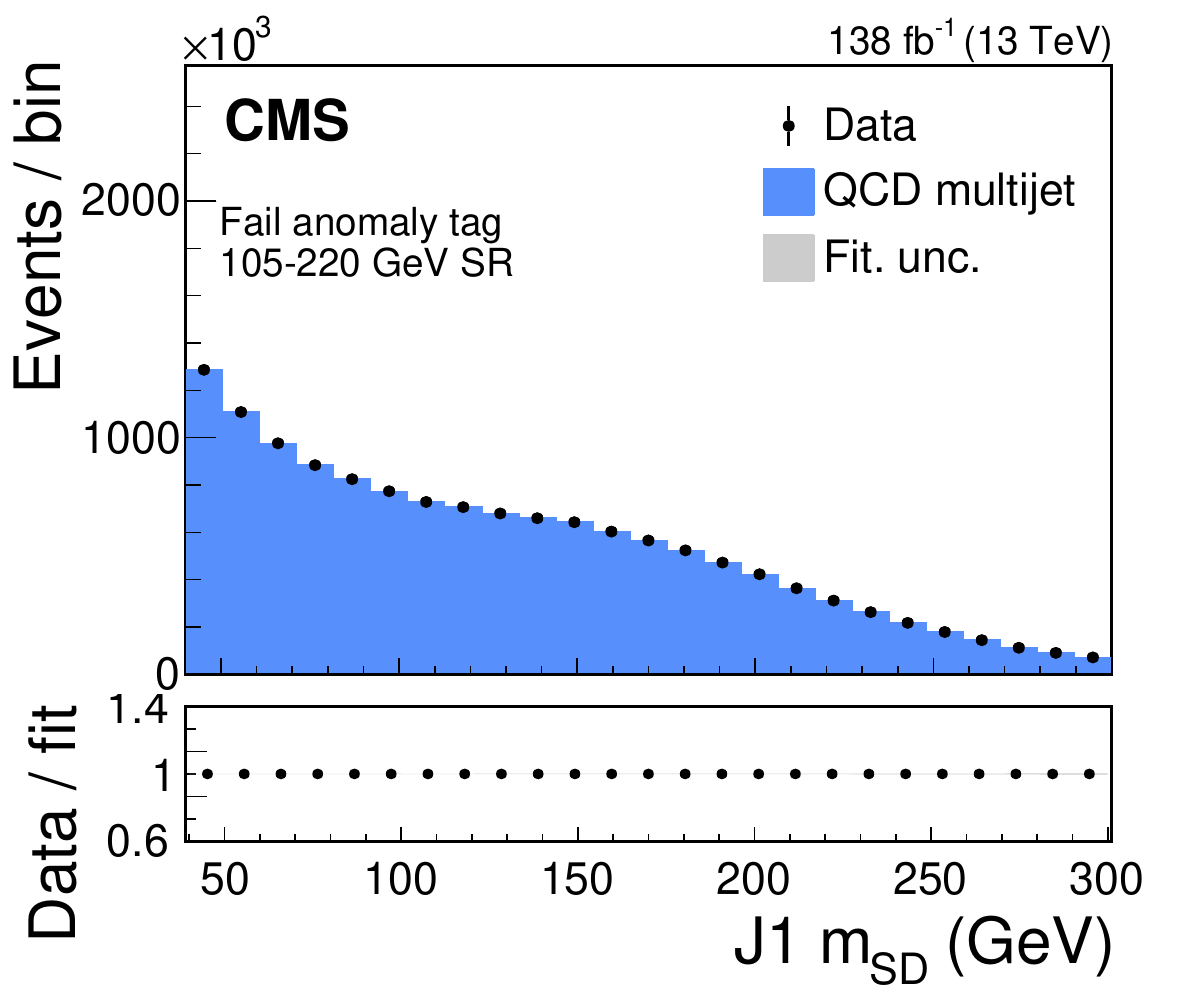}
    \includegraphics[width=0.49\textwidth]{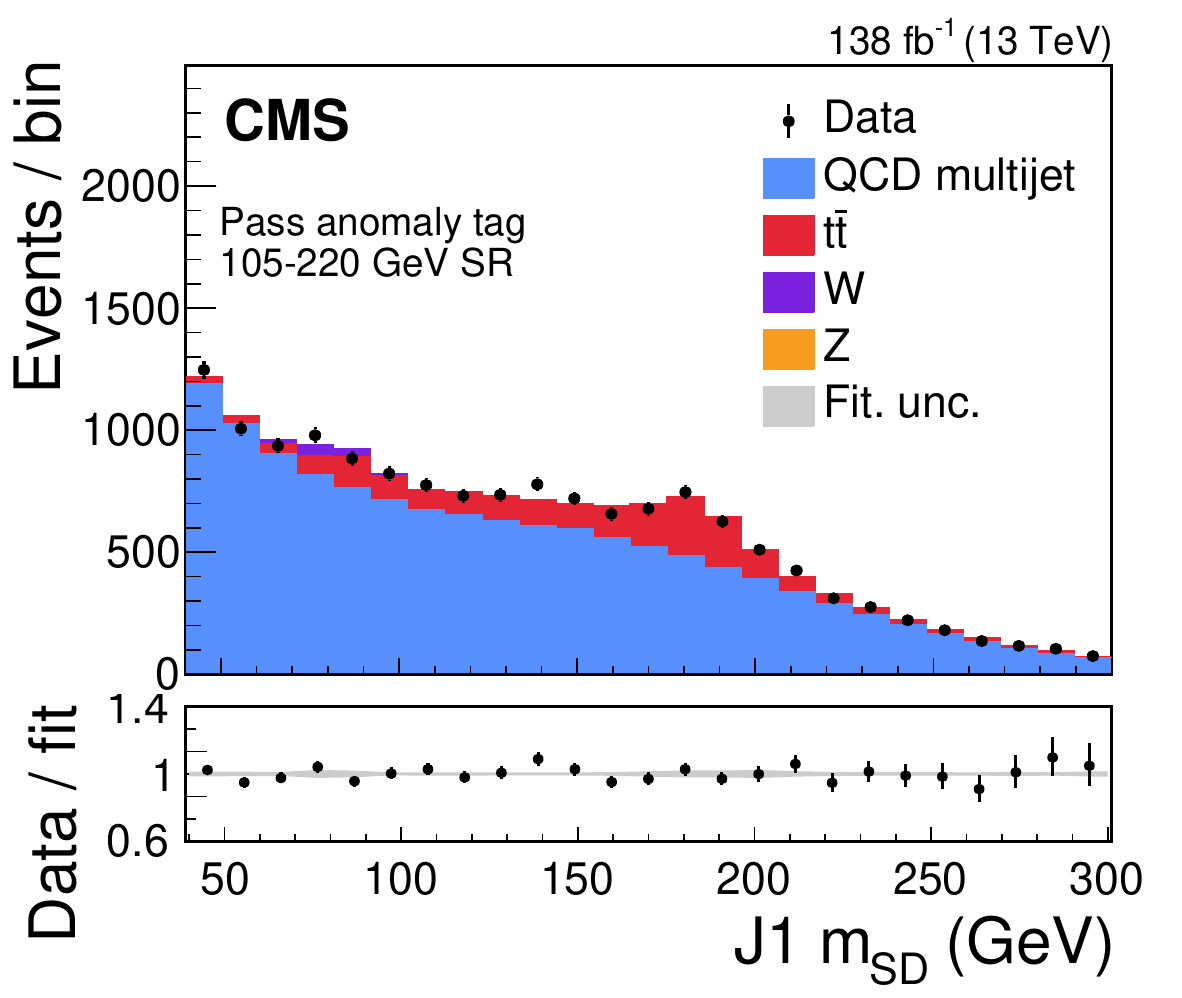} \\
    \includegraphics[width=0.49\textwidth]{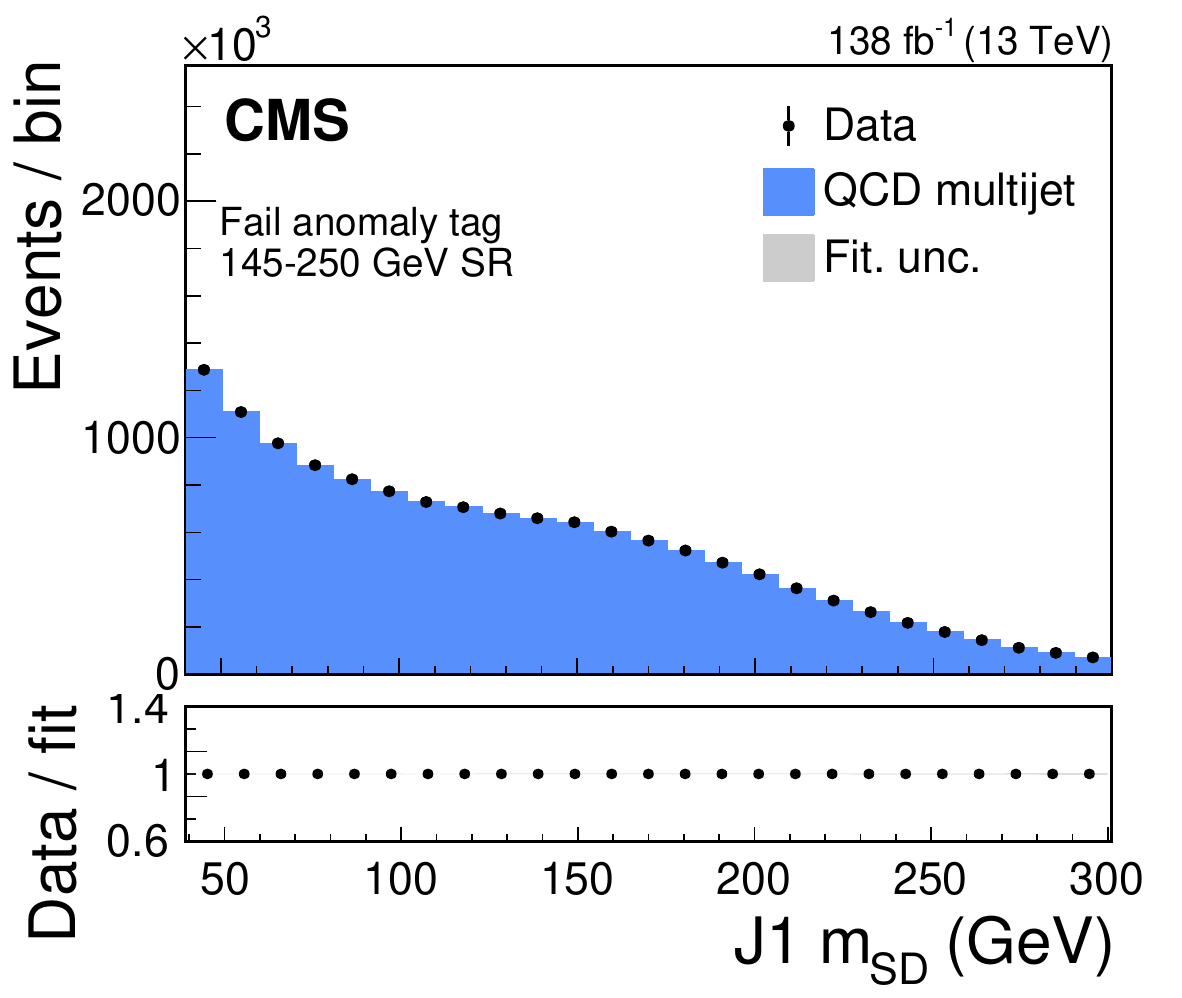}
    \includegraphics[width=0.49\textwidth]{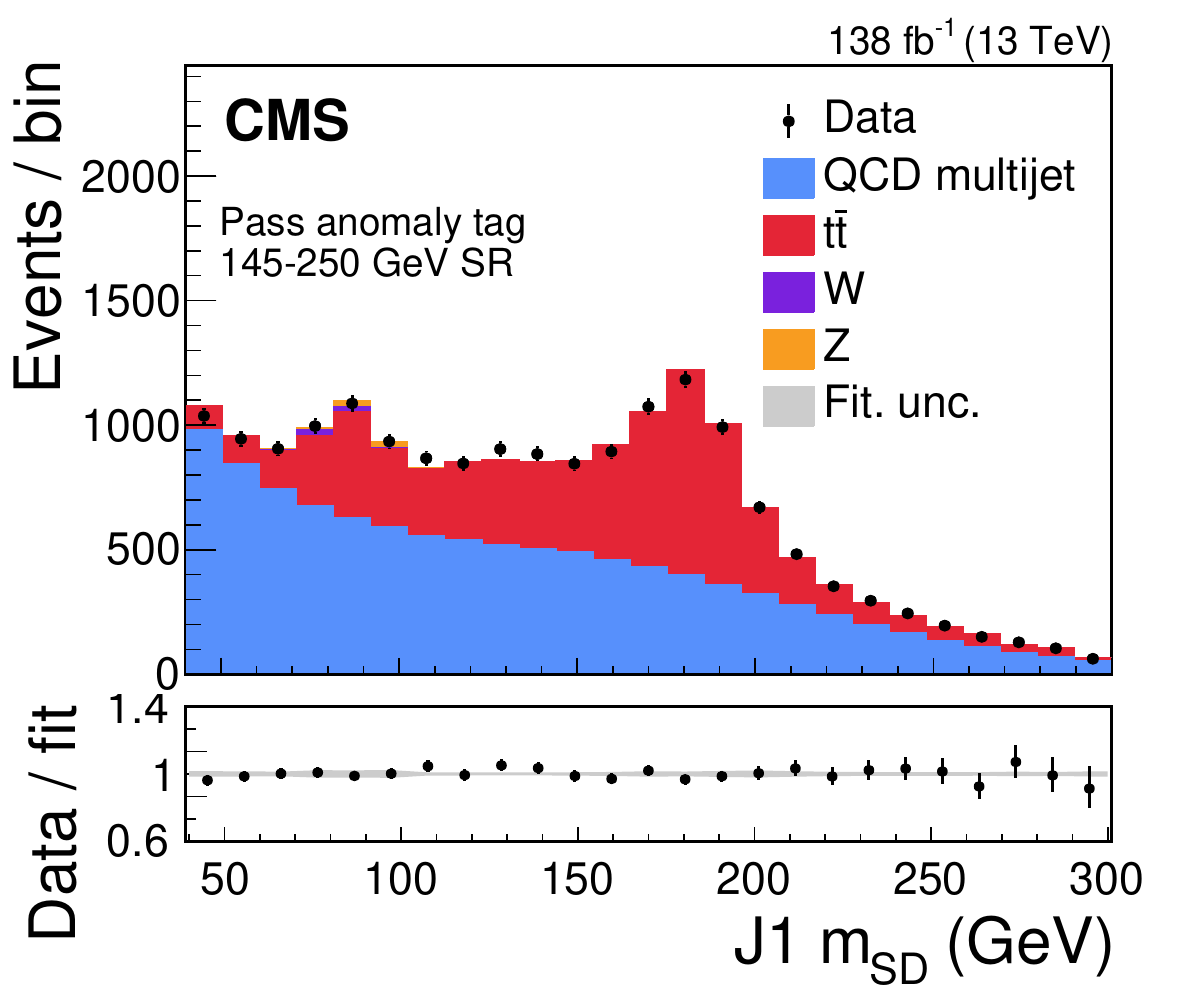} \\
    \caption{ Post-fit plots of the fail (left) and pass (right) regions for the
      \ttbar extraction procedure performed for the
      65--150 (upper), 105--220 (middle), and 145--250\GeV (lower) signal windows.
      Data (black points with error bars) is compared to the fitted
      estimates of the QCD multijet (blue), $\PZ$+jets (orange), $\PW$+jets
      (purple), and \ttbar (red) processes.
      The lower panel shows the ratio between the observed data points and the fitted estimates. The gray shading denotes the systematic uncertainty.
      The contribution from \ttbar is clearly visible in the pass region of the 105--220 and 145--250\GeV signal windows.
    }
    \label{fig:ttbar_fit_baseline}
\end{figure*}

The procedure is performed separately for the two versions of the preselection
and for the three considered signal windows.
The results using the baseline preselection are shown in
Fig.~\ref{fig:ttbar_fit_baseline}.
For the 65--150\GeV mass window, which is not well aligned with the top quark mass
peak, the weakly supervised training does not learn to identify top
quark jets, and the pass region contains no significant \ttbar contribution.
In contrast, for the 105--220 and 145--250\GeV mass windows, the procedure is
successful and large contributions from \ttbar, significantly enhanced with
respect to Fig.~\ref{fig:ttbar_presel_jet_mass}, are clearly visible in the pass
region.
In the \ttbar signal, separate contributions from large-radius jets containing
merged \PW or \PQt decay products can be seen at masses of ${\sim}80$ and ${\sim}175\GeV$, respectively.
The former can occur if the \ttbar pair is produced with moderate \pt such that
only the \PW decay, and not the additional \PQb quark, is merged into a single large-radius jet.
In these mass windows, naive estimates of the statistical significance, based on
an asymptotic approximation~\cite{Cowan:2010js}, are very large
(${>}10$ standard deviations).

\begin{figure}[htbp]
    \centering
    \includegraphics[width=\cmsFigWidthvii]{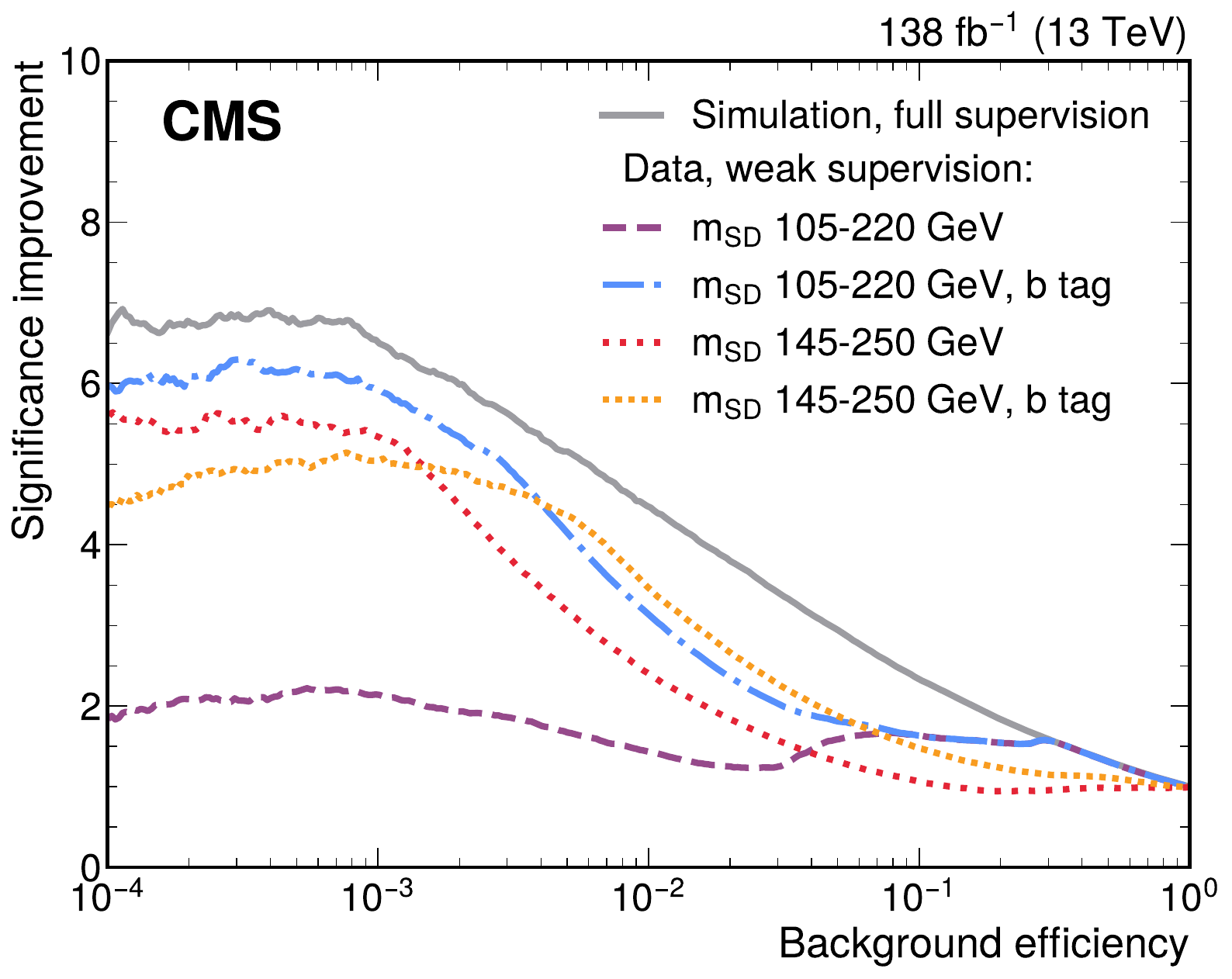}
    \caption{
      A comparison of the top quark identification performance of classifiers
      trained in different ways, evaluated in simulation.
      The two models trained in the 145--250\GeV mass window (red and yellow),
      as well as the one trained using the b tagging preselection in the in the
      105--220\GeV mass window (blue),
      nearly match the performance of a supervised classifier (gray).
      The classifier trained with the baseline preselection in the 105--220\GeV
      mass window (purple) exhibits an improvement that is smaller than the
      others, yet larger than one.
   }
    \label{fig:ttbar_sic}
\end{figure}

To further study the performance at identifying \ttbar, each trained model is
tested in simulation.
In addition, we compare them to a fully supervised model trained with labeled
simulations of \ttbar and QCD multijet events, and using the same input
features and network architecture as the weakly supervised models.
The SIC curves for the 105--220 and 145--250\GeV mass windows are shown in
Fig.~\ref{fig:ttbar_sic}.
The models trained from the 65--150\GeV region are found to be unsuccessful top taggers with significance improvement smaller than one and therefore not shown.

For the 105--220\GeV mass window, the model trained with the b tagging
preselection outperforms the one trained without, which is explained by the
different signal fractions.
However, even this worse model is capable of identifying \ttbar events,
reaching a significance improvement larger than two at the background efficiency
of $10^{-3}$ employed in this section.
Both models trained in the 145--250\GeV mass window achieve a performance close
to the supervised classifier.
We emphasize that even though the weakly supervised models were trained solely
on data and used no truth-labeled examples, the best performing model achieves a
significance improvement only $\sim$15\% worse than that of the supervised
classifier.
This illustrates the power of anomaly detection approaches.

\subsection{Signal characterization}
\label{sec:interp}

In the case of any significant excess of events found by an anomaly
detection algorithm,
one would like to infer its features, both to
validate that the models are picking up a genuine physics-based signal and as
input for further analysis.
In the following, we employ two strategies to determine the reasons that cause
jets to be classified as anomalous and demonstrate them in the context of
the \ttbar validation study.
They are discussed in Appendix~B of Ref.~\cite{CMS:2024nsz}, and a brief summary
is provided here for completeness.
We apply these techniques to the model trained with the \PQb tagging
preselection in the 105--220\GeV mass window.

\begin{figure}[htbp]
  \centering
  \includegraphics[width=\cmsFigWidthvi]{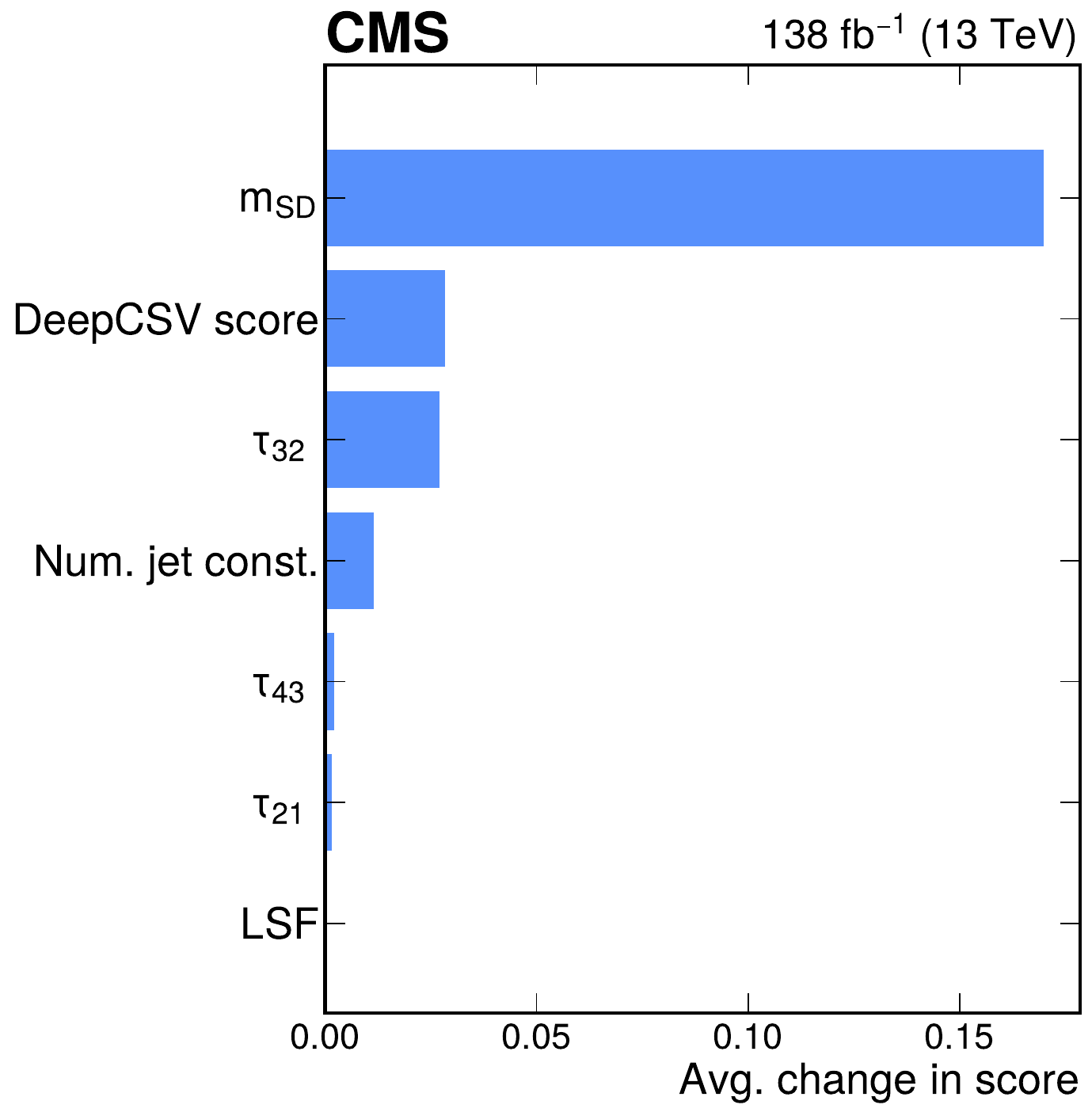}
  \caption{
    Excess characterization for the weakly supervised anomaly detection strategy applied to the \ttbar region with the b tagging preselection.
    The sensitivity of the anomaly score to the different input observables is assessed to aid in the determination of the properties of the excess.
    The jet mass, b tagging score, and $\tau_{32}$ are seen to be the most important observables, consistent with the properties of large-radius top quark jets.
  }
  \label{fig:ttbar_excess_feat_importance}
\end{figure}

The first strategy is used to interpret which input features are most important
for the computed anomaly score.
An interpretability technique called the permutation feature
importance~\cite{perm_feat_importance},
suitable for interpretation without labeled examples, is used.
To assess the importance of a feature, we randomly swap its value for the events
with the 1\% largest anomaly scores with the values of other events chosen at
random.
We then compute the average absolute change of the classifier output with
respect to the original anomalous events; a large change indicates that the
value of the feature was important for the classifier score.
The results are stabilized by taking the average of 100 repetitions of the
procedure with different random replacements.
The results of this study are shown in Fig.~\ref{fig:ttbar_excess_feat_importance}.
They indicate that the three most important features are \mSD, the \textsc{DeepCSV}
score, and $\tau_{32}$. The number of jet constituents also seems to play a role
in the network's decision.

The second strategy we use is based on investigating the features that differ
between the excess events and typical events.
The distributions of features of events with high anomaly scores are compared to the distributions for all events in
the region of the excess, without any anomaly score selection.
Differences between these two distributions give insight into how anomalous events and background events differ.

In the case of weakly supervised anomaly detection, caution must be used when
reporting properties of the signal from the distributions of the high-anomaly-score
sample.
Because the weakly supervised methods learn to identify the specific characteristics of the signal, they preferentially select
background events that have the same characteristics as the signal.
Thus, the high-anomaly-score sample has a biased selection of background events,
which would have to be accounted for in a true measurement of signal properties.
However, the qualitative features of the distributions are still useful in characterizing the basic phenomenological properties of the excess.

\begin{figure*}
  \centering
  \includegraphics[width=0.5\textwidth]{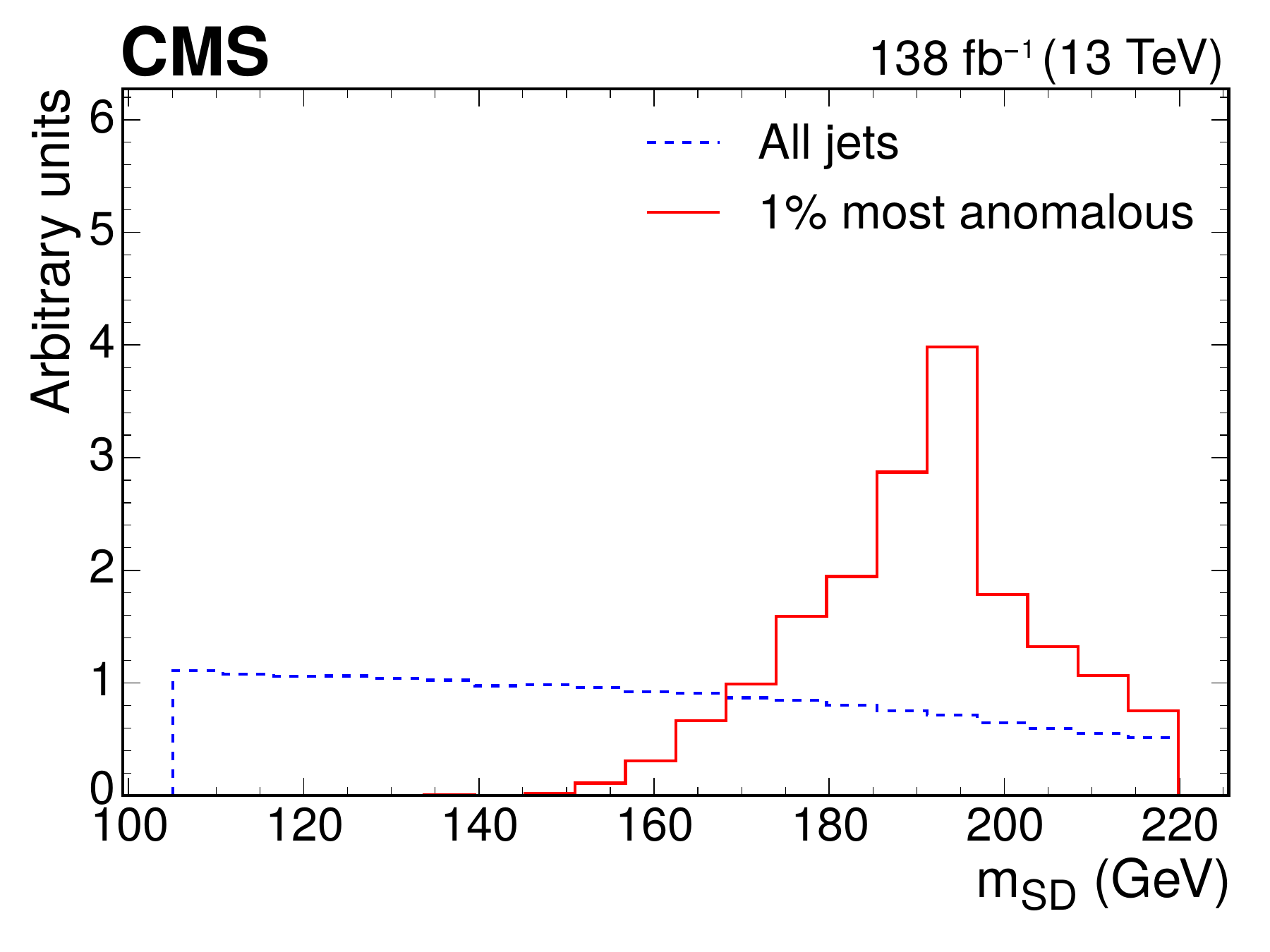}%
  \includegraphics[width=0.5\textwidth]{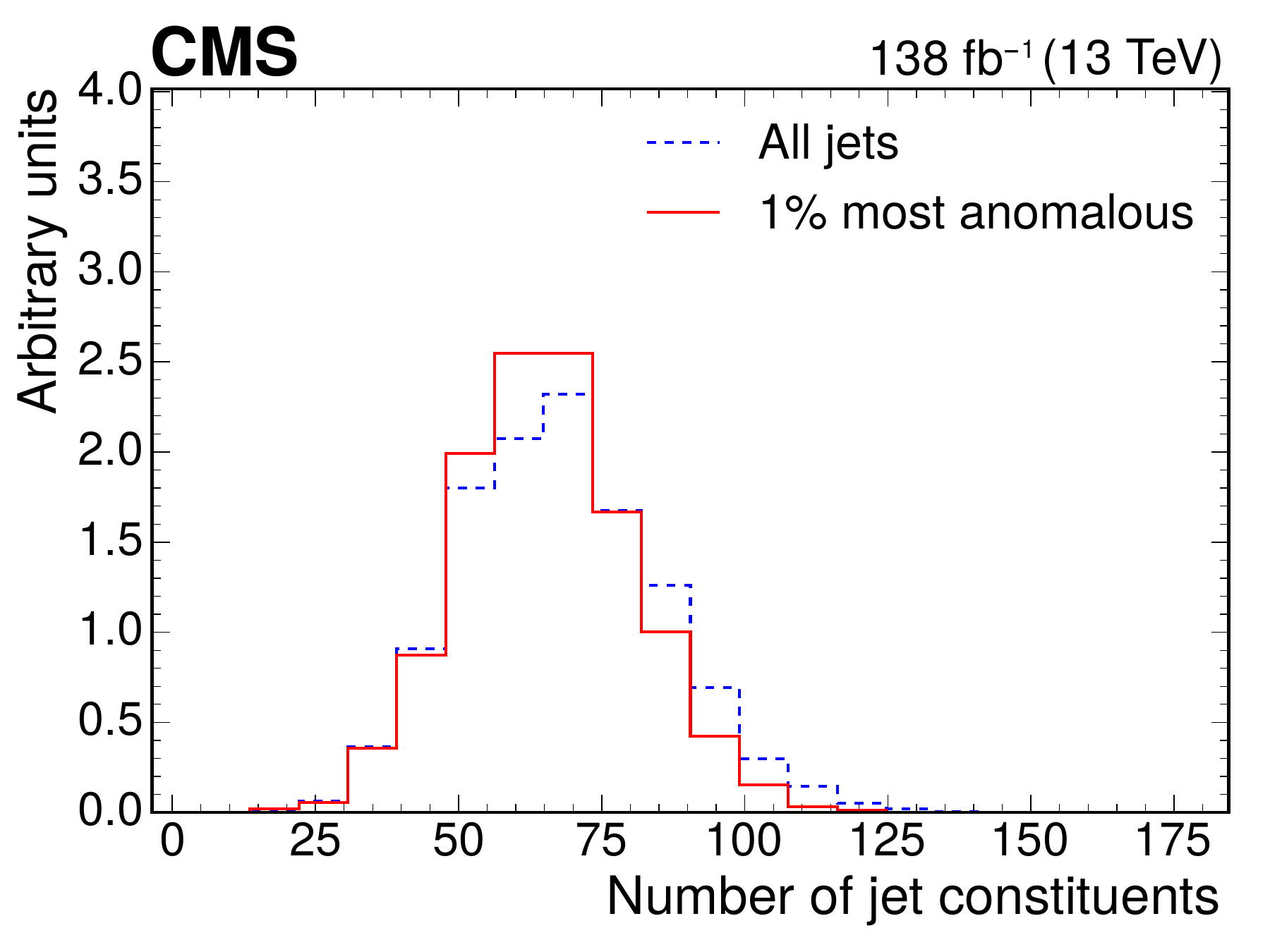}\\
  \includegraphics[width=0.5\textwidth]{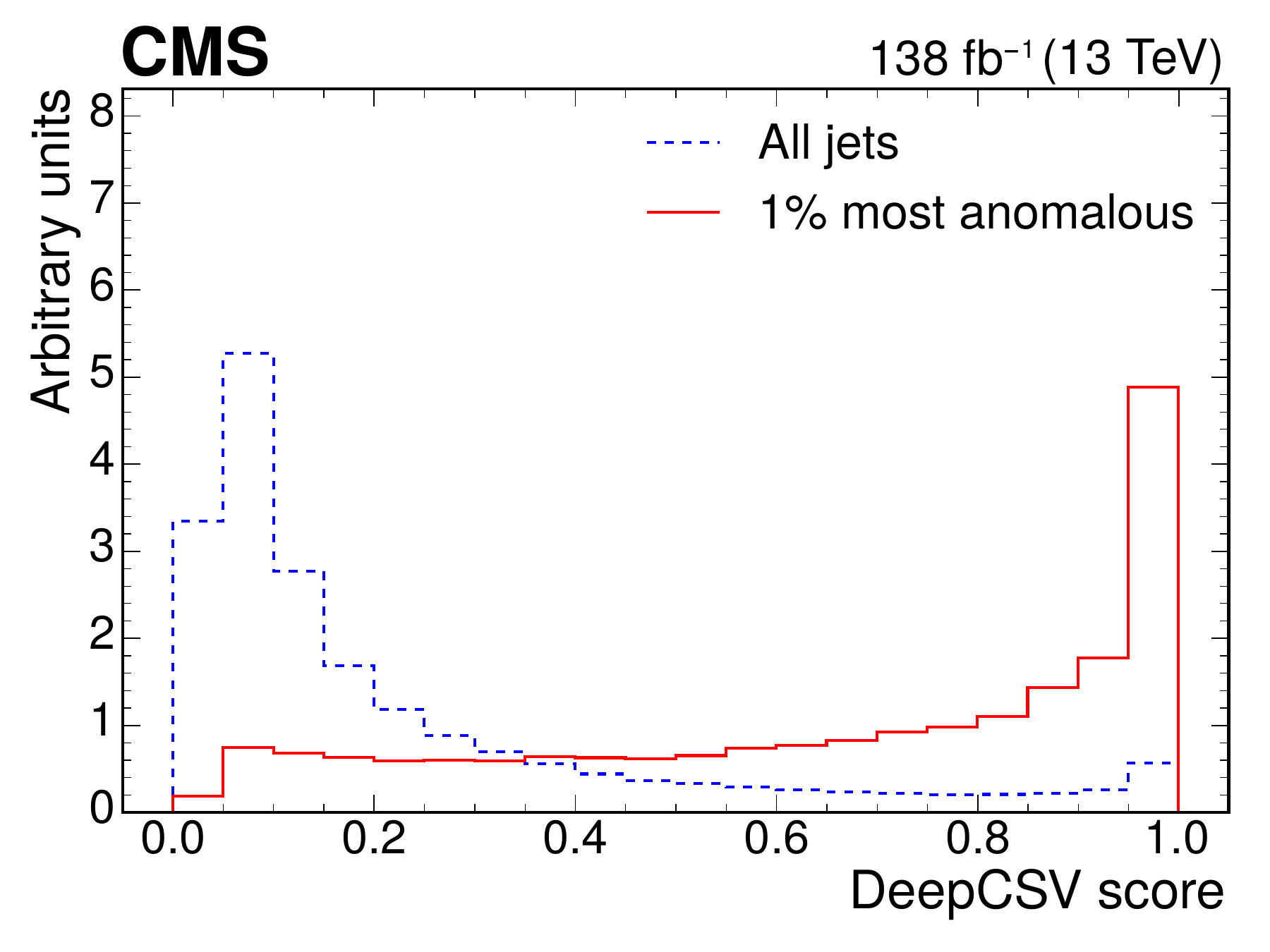}%
  \includegraphics[width=0.5\textwidth]{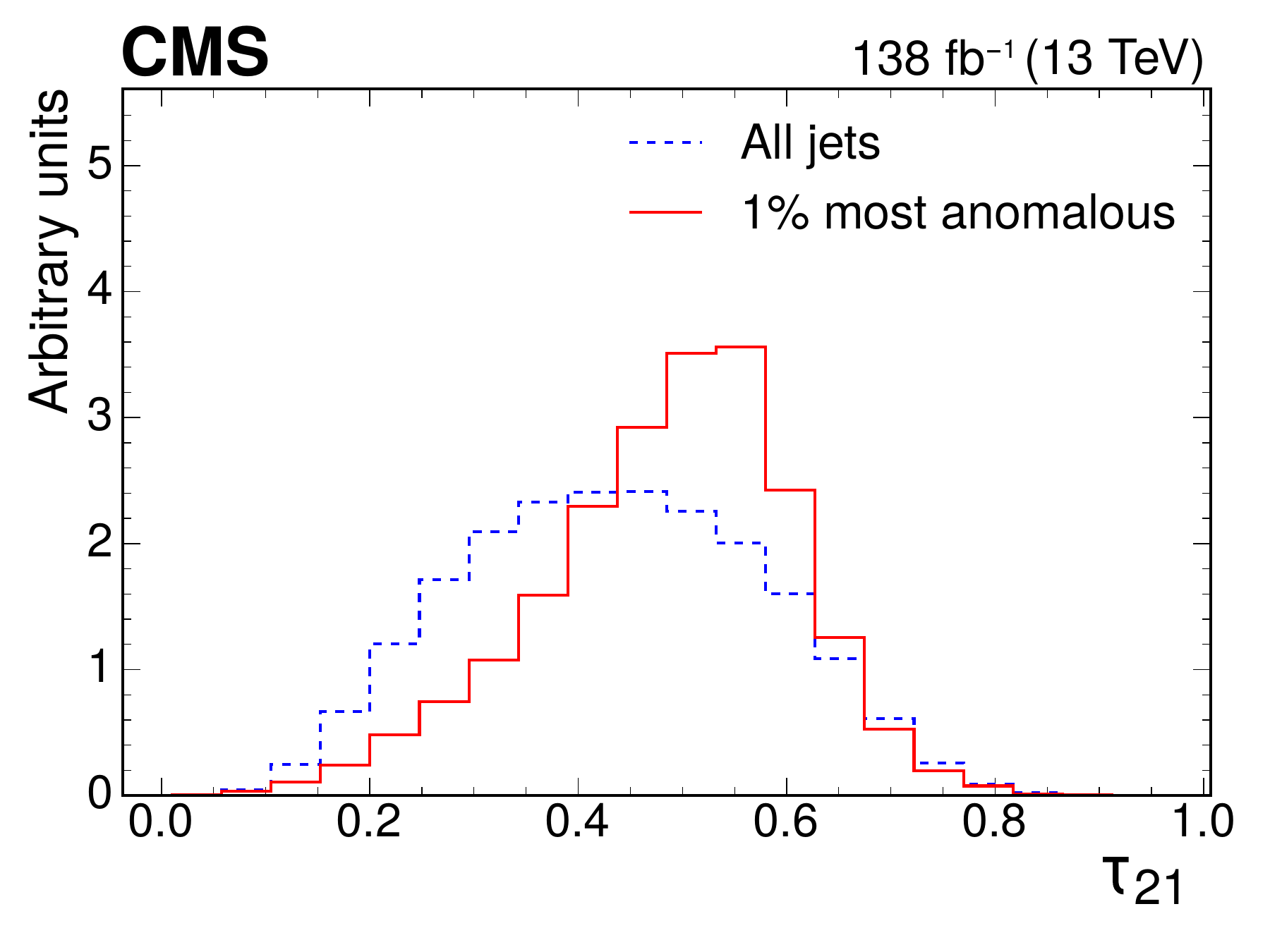}\\
  \includegraphics[width=0.5\textwidth]{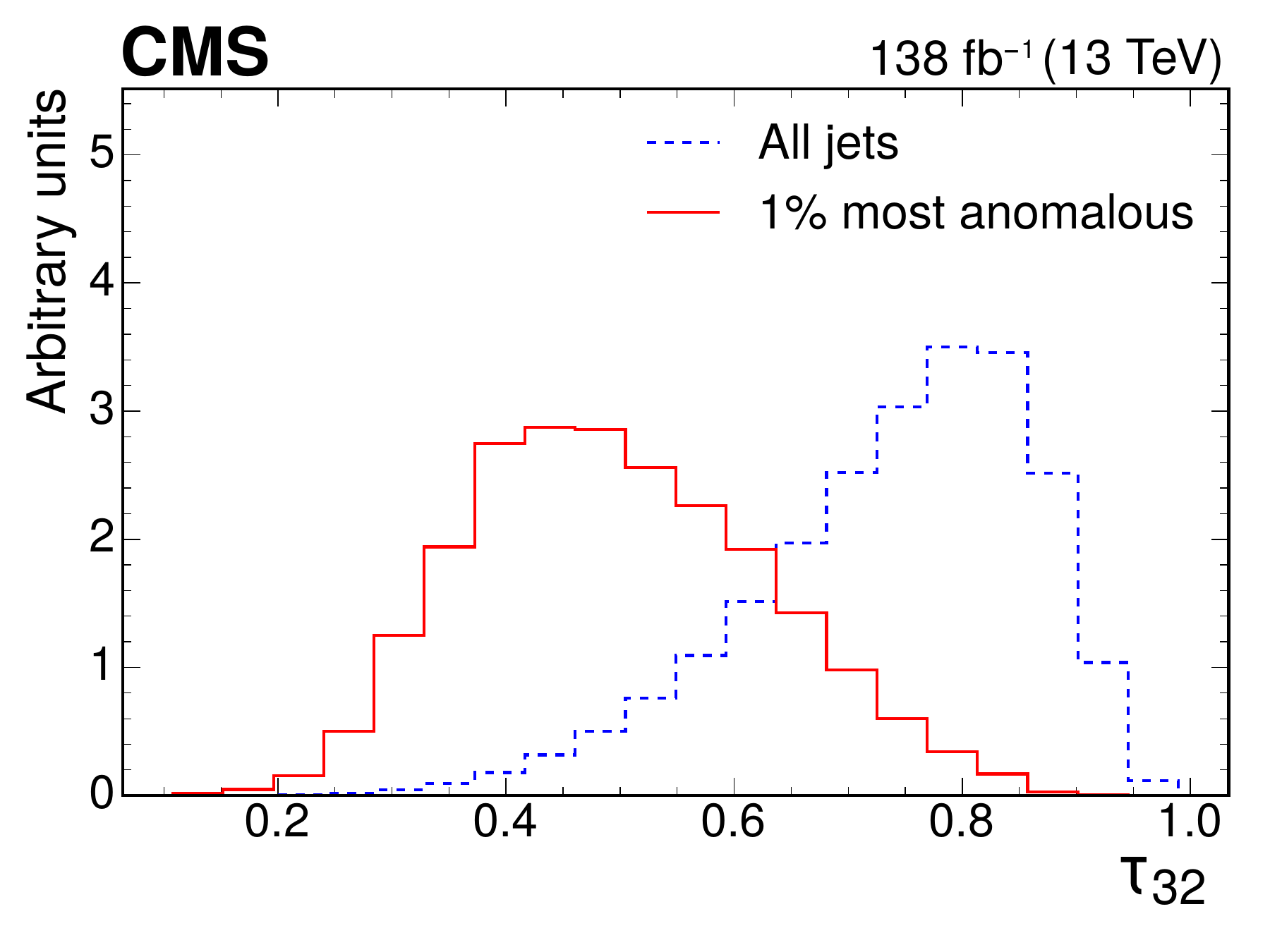}%
  \includegraphics[width=0.5\textwidth]{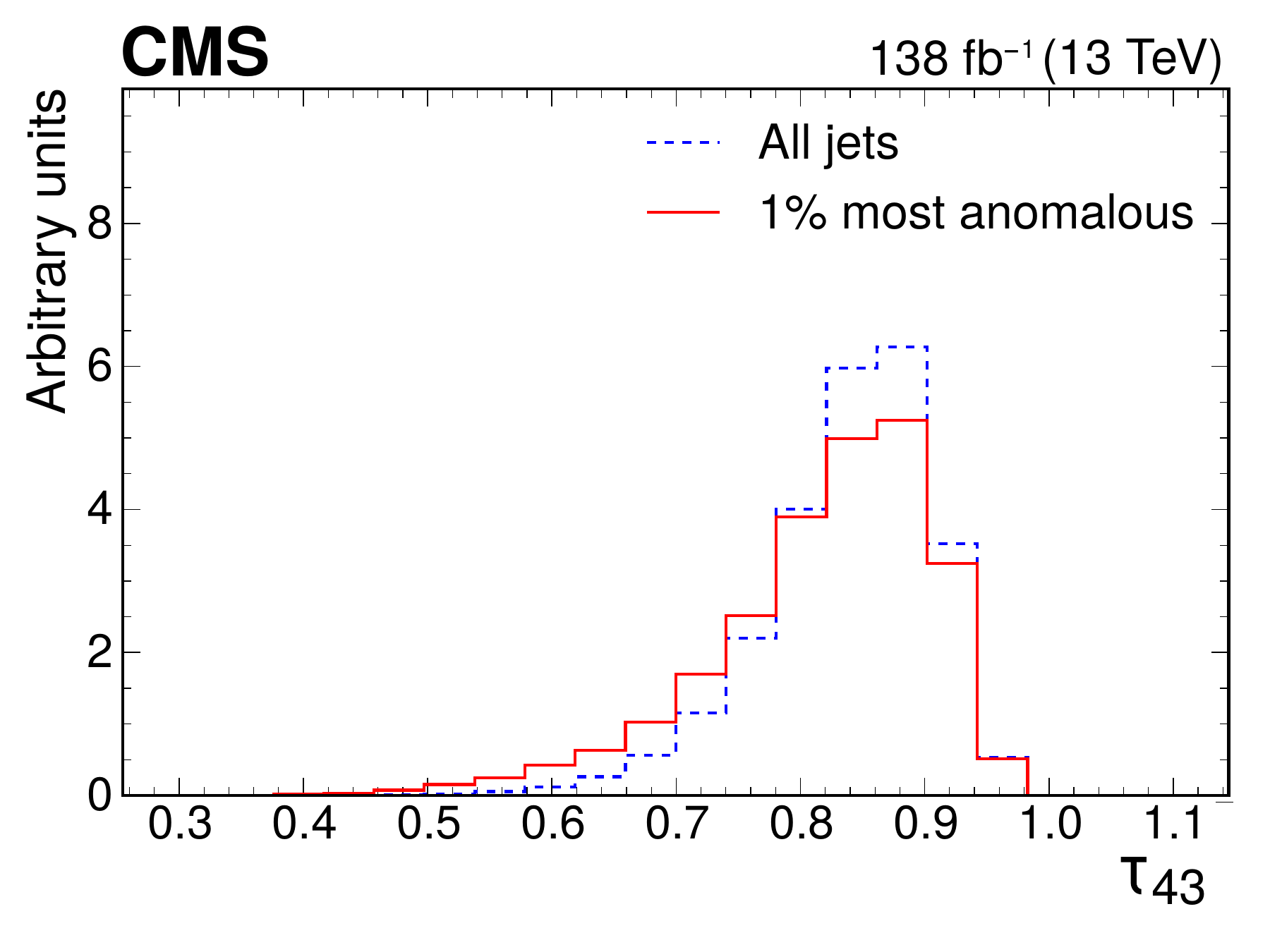}
  \caption{
    Excess characterization for the weakly supervised anomaly detection
    strategy applied to the \ttbar region with the b tagging preselection.
    The plots compare the properties of the jets with the highest anomaly score
    (red) to those for all jets in the region of the excess (blue).
    The variables shown are the soft-drop mass \mSD (upper left),
    the number of jet constituents \nPF (upper right),
    the \textsc{DeepCSV} score (middle left), and the three subjettiness ratios
    $\tau_{21}$ (middle right), $\tau_{32}$ (lower left), and $\tau_{43}$
    (lower right).
    The three-pronged nature of the signal is clear from the low $\tau_{32}$
    scores, the presence of b tags from the high \textsc{DeepCSV} score,
    and the jet mass (\mSD) distribution peaks close to the top quark mass.
}
  \label{fig:ttbar_excess_interp}
\end{figure*}

The results of this comparison are shown in Fig.~\ref{fig:ttbar_excess_interp}.
Significant differences between the two samples are visible for features already
flagged by the importance study.
A jet mass of ${\sim}190\GeV$ is visible in the \mSD distribution (which was not
explicitly calibrated),
the \textsc{DeepCSV} score indicates the presence of \PQb quarks,
and the $\tau_{32}$ distribution clearly highlights the three-pronged nature
of large-radius top quark jets.
No significant difference is observed for the number of jet constituents, which the network may
only be using in combination with another feature.
Recovering these properties of the top quark shows how an anomaly search could
be interpreted if it yields positive results.

\ifthenelse{\boolean{cms@external}}{}{\clearpage}
\section{Summary}
\label{sec:summary}

We have presented a detailed description of the five anomaly detection
methods used to search for new particles decaying to two anomalous
jets in Ref.~\cite{CMS:2024nsz}, using data collected by CMS between 2016 and
2018.
Approaches based on weakly supervised, unsupervised, and semi-supervised
training paradigms have been explored.
All methods were successfully able to identify some classes of anomalous jets as
distinct from standard model backgrounds, and were therefore able to enhance the
sensitivity to the signatures of new particles in CMS data in a model-agnostic
fashion.

The sensitivity of the methods to the presence of an anomalous signal in the
data has been shown to be higher than an inclusive search or simple selections
based on jet substructure.
The performance of the five methods has been found to depend on the considered
signal model, with no single method outperforming all others.
Further investigation has shown that correlations between the anomaly scores are
low, indicating that methods identify anomalous signal events in different ways.
Combining the different approaches could lead to more powerful methods.
Furthermore, the impact of differences in the input features has been explored
and shown to explain performance differences between signal models, but not
between methods.

Finally, a weakly supervised anomaly detection method
has been used to separate top quark jets from jets produced in other standard
model processes.
The achieved separation power comes very close to that of a fully supervised
classifier
trained with the same input variables.
This constitutes a validation of resonant anomaly detection in collider data.
In addition, interpretation techniques have been used to successfully retrieve
some of the main properties of the top quark.
This would enable further investigation and confirmation of the signal in the event of
a positive result from an anomaly search.

\clearpage

\begin{acknowledgments}
We congratulate our colleagues in the CERN accelerator departments for the excellent performance of the LHC and thank the technical and administrative staffs at CERN and at other CMS institutes for their contributions to the success of the CMS effort. In addition, we gratefully acknowledge the computing centers and personnel of the Worldwide LHC Computing Grid and other centers for delivering so effectively the computing infrastructure essential to our analyses. Finally, we acknowledge the enduring support for the construction and operation of the LHC, the CMS detector, and the supporting computing infrastructure provided by the following funding agencies: SC (Armenia), BMBWF and FWF (Austria); FNRS and FWO (Belgium); CNPq, CAPES, FAPERJ, FAPERGS, and FAPESP (Brazil); MES and BNSF (Bulgaria); CERN; CAS, MoST, and NSFC (China); MINCIENCIAS (Colombia); MSES and CSF (Croatia); RIF (Cyprus); SENESCYT (Ecuador); ERC PRG, TARISTU24-TK10 and MoER TK202 (Estonia); Academy of Finland, MEC, and HIP (Finland); CEA and CNRS/IN2P3 (France); SRNSF (Georgia); BMFTR, DFG, and HGF (Germany); GSRI (Greece); NKFIH (Hungary); DAE and DST (India); IPM (Iran); SFI (Ireland); INFN (Italy); MSIT and NRF (Republic of Korea); MES (Latvia); LMTLT (Lithuania); MOE and UM (Malaysia); BUAP, CINVESTAV, CONACYT, LNS, SEP, and UASLP-FAI (Mexico); MOS (Montenegro); MBIE (New Zealand); PAEC (Pakistan); MES, NSC, and NAWA (Poland); FCT (Portugal); MESTD (Serbia); MICIU/AEI and PCTI (Spain); MOSTR (Sri Lanka); Swiss Funding Agencies (Switzerland); MST (Taipei); MHESI (Thailand); TUBITAK and TENMAK (T\"{u}rkiye); NASU (Ukraine); STFC (United Kingdom); DOE and NSF (USA).

\hyphenation{Rachada-pisek} Individuals have received support from the Marie-Curie program and the European Research Council and Horizon 2020 Grant, contract Nos.\ 675440, 724704, 752730, 758316, 765710, 824093, 101115353, 101002207, 101001205, and COST Action CA16108 (European Union); the Leventis Foundation; the Alfred P.\ Sloan Foundation; the Alexander von Humboldt Foundation; the Science Committee, project no. 22rl-037 (Armenia); the Fonds pour la Formation \`a la Recherche dans l'Industrie et dans l'Agriculture (FRIA) and Fonds voor Wetenschappelijk Onderzoek contract No. 1228724N (Belgium); the Beijing Municipal Science \& Technology Commission, No. Z191100007219010, the Fundamental Research Funds for the Central Universities, the Ministry of Science and Technology of China under Grant No. 2023YFA1605804, the Natural Science Foundation of China under Grant No. 12061141002, 12535004, and USTC Research Funds of the Double First-Class Initiative No.\ YD2030002017 (China); the Ministry of Education, Youth and Sports (MEYS) of the Czech Republic; the Shota Rustaveli National Science Foundation, grant FR-22-985 (Georgia); the Deutsche Forschungsgemeinschaft (DFG), among others, under Germany's Excellence Strategy -- EXC 2121 ``Quantum Universe" -- 390833306, and under project number 400140256 - GRK2497; the Hellenic Foundation for Research and Innovation (HFRI), Project Number 2288 (Greece); the Hungarian Academy of Sciences, the New National Excellence Program - \'UNKP, the NKFIH research grants K 131991, K 133046, K 138136, K 143460, K 143477, K 146913, K 146914, K 147048, 2020-2.2.1-ED-2021-00181, TKP2021-NKTA-64, and 2025-1.1.5-NEMZ\_KI-2025-00004 (Hungary); the Council of Science and Industrial Research, India; ICSC -- National Research Center for High Performance Computing, Big Data and Quantum Computing, FAIR -- Future Artificial Intelligence Research, and CUP I53D23001070006 (Mission 4 Component 1), funded by the NextGenerationEU program (Italy); the Latvian Council of Science; the Ministry of Education and Science, project no. 2022/WK/14, and the National Science Center, contracts Opus 2021/41/B/ST2/01369, 2021/43/B/ST2/01552, 2023/49/B/ST2/03273, and the NAWA contract BPN/PPO/2021/1/00011 (Poland); the Funda\c{c}\~ao para a Ci\^encia e a Tecnologia, grant CEECIND/01334/2018 (Portugal); the National Priorities Research Program by Qatar National Research Fund;  MICIU/AEI/10.13039/501100011033, ERDF/EU, ``European Union NextGenerationEU/PRTR", and Programa Severo Ochoa del Principado de Asturias (Spain); the Chulalongkorn Academic into Its 2nd Century Project Advancement Project, the National Science, Research and Innovation Fund program IND\_FF\_68\_369\_2300\_097, and the Program Management Unit for Human Resources \& Institutional Development, Research and Innovation, grant B39G680009 (Thailand); the Kavli Foundation; the Nvidia Corporation; the SuperMicro Corporation; the Welch Foundation, contract C-1845; and the Weston Havens Foundation (USA).  
\end{acknowledgments}

\bibliography{auto_generated}

\clearpage
\appendix

\ifthenelse{\boolean{cms@external}}{
}{
\numberwithin{figure}{section}
\numberwithin{table}{section}
}

\section{The CMS detector and event reconstruction}
\label{sec:cms}

The CMS apparatus~\cite{Chatrchyan:2008zzk,CMS:2023gfb} is a multipurpose, nearly hermetic detector, designed to trigger on~\cite{CMS:2020cmk,CMS:2016ngn,CMS:2024aqx} and identify electrons, muons, photons, and (charged and neutral) hadrons~\cite{CMS:2020uim,CMS:2018rym,CMS:2014pgm}. Its central feature is a superconducting solenoid of 6\unit{m} internal diameter, providing a magnetic field of 3.8\unit{T}. Within the solenoid volume are a silicon pixel and strip tracker, a lead tungstate crystal electromagnetic calorimeter (ECAL), and a brass and scintillator hadron calorimeter (HCAL), each composed of a barrel and two endcap sections. Forward calorimeters extend the pseudorapidity coverage provided by the barrel and endcap detectors. Muons are reconstructed using gas-ionization detectors embedded in the steel flux-return yoke outside the solenoid. More detailed descriptions of the CMS detector, together with a definition of the coordinate system used and the relevant kinematic variables, can be found in Refs.~\cite{Chatrchyan:2008zzk,CMS:2023gfb}.

Events of interest are selected using a two-tiered trigger system. The first level, composed of custom hardware processors, uses information from the calorimeters and muon detectors to select events at a rate of around 100\unit{kHz} within a fixed latency of 4\mus~\cite{CMS:2020cmk}. The second level, known as the high-level trigger, consists of a farm of processors running a version of the full event reconstruction software optimized for fast processing, and reduces the event rate to a few kHz before data storage~\cite{CMS:2016ngn}.

The primary vertex is taken to be the vertex corresponding to the hardest scattering in the event, evaluated using tracking information alone, as described in Section 9.4.1 of Ref.~\cite{CMS-TDR-15-02}.
A particle-flow (PF) algorithm~\cite{CMS:2017yfk} aims to reconstruct and identify each individual particle in an event, with an optimized combination of information from the various elements of the CMS detector.
Jets are clustered from the PF candidates in an event using the anti-\kt jet finding algorithm~\cite{Cacciari:2008gp,Cacciari:2011ma}.
In this analysis, large-radius jets with a distance parameter of $R = 0.8$ are used.
Jet momentum is determined as the vectorial sum of all particle momenta in the jet, and is found from simulation to be, on average, within 5 to 10\% of the true momentum over the whole \pt spectrum and detector acceptance.
Additional proton-proton interactions within the same or nearby bunch crossings (pileup) can contribute additional tracks and calorimetric energy depositions, increasing the apparent jet momentum.
The pileup per particle identification algorithm~\cite{CMS:2020ebo,Bertolini:2014bba} is used to mitigate the effect of pileup at the reconstructed particle level, making use of local shape information, event pileup properties, and tracking information.
Jet energy corrections are derived from simulation studies and control samples in
data so that the average measured energy of jets becomes identical to that of
particle level jets~\cite{CMS:2016lmd, CMS:2020ebo}.

The missing transverse momentum vector \ptvecmiss is defined as the projection onto the plane perpendicular to the beam axis of the negative vector sum of the momenta of all reconstructed particle-flow objects in an event. Its magnitude is referred to as \ptmiss.

\section{Companion dataset}
\label{sec:dataset-release}

This Appendix describes the companion dataset~\cite{zenodo} of this paper made
available on
Zenodo to support the development of new anomaly detection algorithms.
It consists of an unweighted dataset of 24 million simulated SM background
events and 52 different BSM signal samples.
The considered background and signal processes, as well as the event generators
used, are the ones described in Section~\ref{sec:datasets}.
In the following, we focus on file contents.

Events included in the datasets undergo a loose preselection.
They are required to pass the trigger, contain two jets with $\pt>300\GeV$ and
$\abs{\eta}<2.5$, and have $\mjj>1200\GeV$.
For signal samples, the efficiency of this selection is recorded in the
\texttt{preselection\_eff} variable.
Two systematic variations, corresponding to the uncertainties in the jet energy
scale (JES) and resolution (JER), are also stored as up/down pairs.

Basic event information is stored in the \texttt{event\_info} array, which
contains for every event the event number, \ptmiss, its angle $\phi$, the
original generator weight, whether the event contains a leptonic decay, an
empty placeholder field, the year of the simulated data-taking conditions, and the total
number of jets in the event.
The generated physics process is encoded in the \texttt{truth\_label} column,
with signal events having positive labels.
The QCD multijet events have label 0; $-1$ is used for background processes
involving top quarks, $-2$ for processes involving \PW bosons, and $-3$ for
\PZ boson production.
The kinematic information of the two leading jets is stored as follows in the
\texttt{jet\_kinematics} array: the first two entries correspond to \mjj and
\DeltaEta.
This is followed by the four-momenta of the two jets saved as \pt, $\eta$,
$\phi$, and \mSD.
All masses and momenta use \GeVns as the default unit.

Substructure information is provided in three tables per jet.
The first, \texttt{extraInfo}, contains the high-level variables $\tau_1$,
$\tau_2$, $\tau_3$, $\tau_4$, \LSF, \DeepB, and \nPF.
The second table stores the four-momenta of the 100 highest-\pt particles in
each jet sorted by descending \pt and encoded as $p_x$, $p_y$, $p_z$, and $E$.
Finally, the properties of up to six secondary vertices (SVs) contained in the jet
are saved in the \texttt{SVs} table as the mass, \pt, the number of tracks, the
normalized $\chi^2$ of the SV fit, and the two- and three-dimensional
displacement significances.

For signal samples, we provide limited generator-level information, experimental
corrections, and variables needed to evaluate systematic uncertainties.
The \texttt{gen\_info} table contains the \pt, $\eta$, $\phi$, and particle type
ID~\cite{PDG2024} of the quarks produced in the decay of the intermediate BSM
particles, the number of which depends on the signal topology.

An important correction to the event weights is stored in the
\texttt{sys\_weights} array.
The first entry contains the product of experimental correction factors.
The rest of the vector contains 10 up/down pairs of systematic variations that
apply multiplicatively to the event weights, corresponding to experimental
and theory uncertainties.

A second weight aiming to correct the description of jets in boosted
topologies~\cite{Lund_PAS} is provided in the \texttt{lund\_weights} table.
Statistical uncertainties in this correction are provided as twice 100 replicas,
\texttt{pt\_var} and \texttt{stat\_var}, covering \pt extrapolation and
statistical uncertainties, respectively.
Each uncertainty is given by the standard deviation of the replicas.
In addition, five sources of systematic uncertainty are encoded in the
\texttt{sys\_var} list as up/down pairs.

Finally, uncertainties in jet kinematic variables are also included in one table
per jet (\texttt{JME\_vars}).
Each table contains the jet \pt and \mSD varied up and down under the JES and
JER uncertainties, as well as variations of the jet mass scale (JMS) and
resolution (JMR)~\cite{CMS:2017wyc}.
These uncertainties should be applied together with the corresponding variations
of the preselection efficiency.
The table layout is as follows:
$\pt^\text{JES up}$,
$\mSD^\text{JES up}$,
$\pt^\text{JES down}$,
$\mSD^\text{JES down}$,
$\pt^\text{JER up}$,
$\mSD^\text{JER up}$,
$\pt^\text{JER down}$,
$\mSD^\text{JER down}$,
$\mSD^\text{JMS up}$,
$\mSD^\text{JMS down}$,
$\mSD^\text{JMR up}$,
$\mSD^\text{JMR down}$.

\section{Anomaly score comparisons}
\label{sec:score-comparisons}

\begin{figure*}
  \centering
  \includegraphics[width=0.45\textwidth]{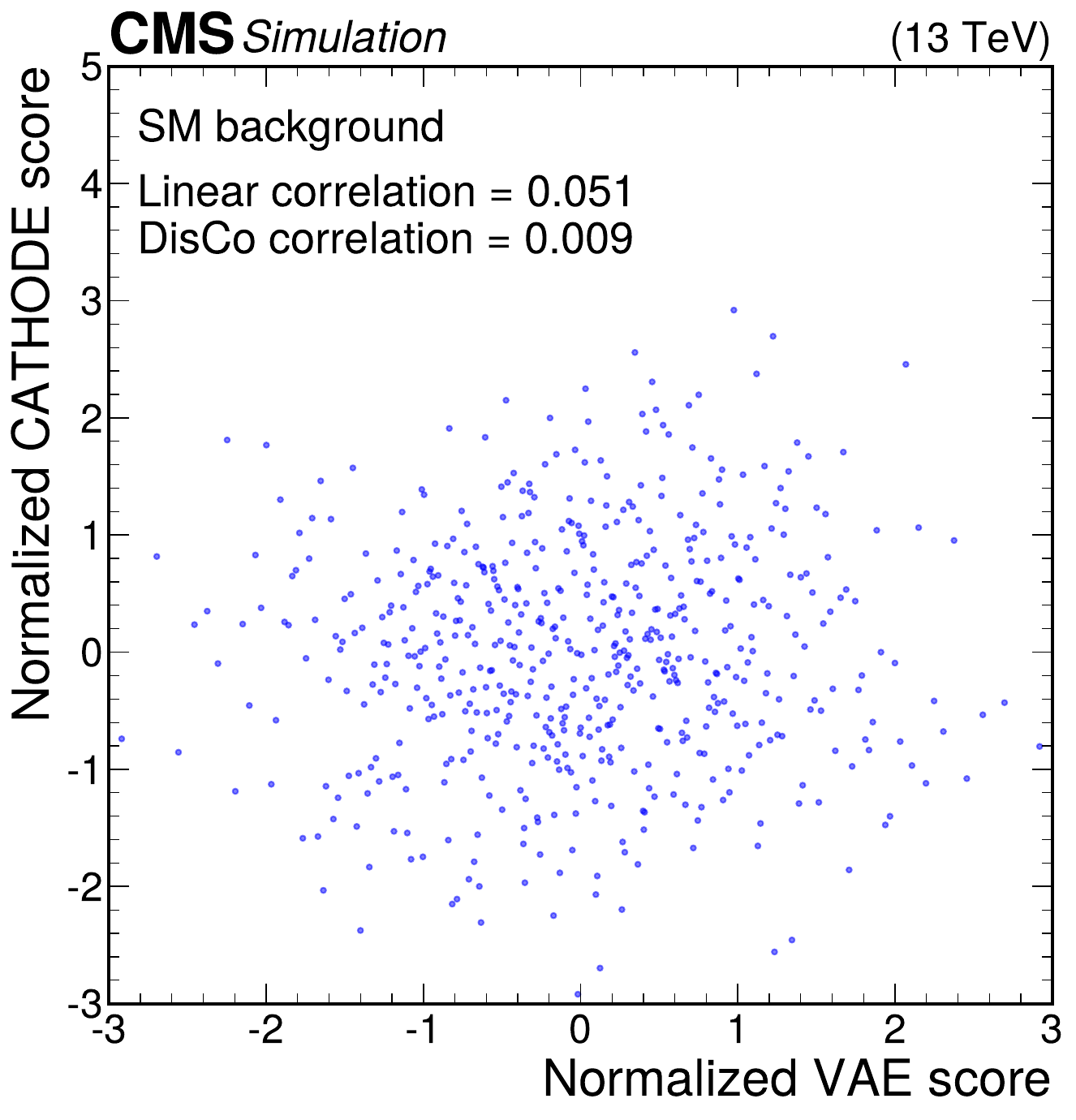}%
  \includegraphics[width=0.45\textwidth]{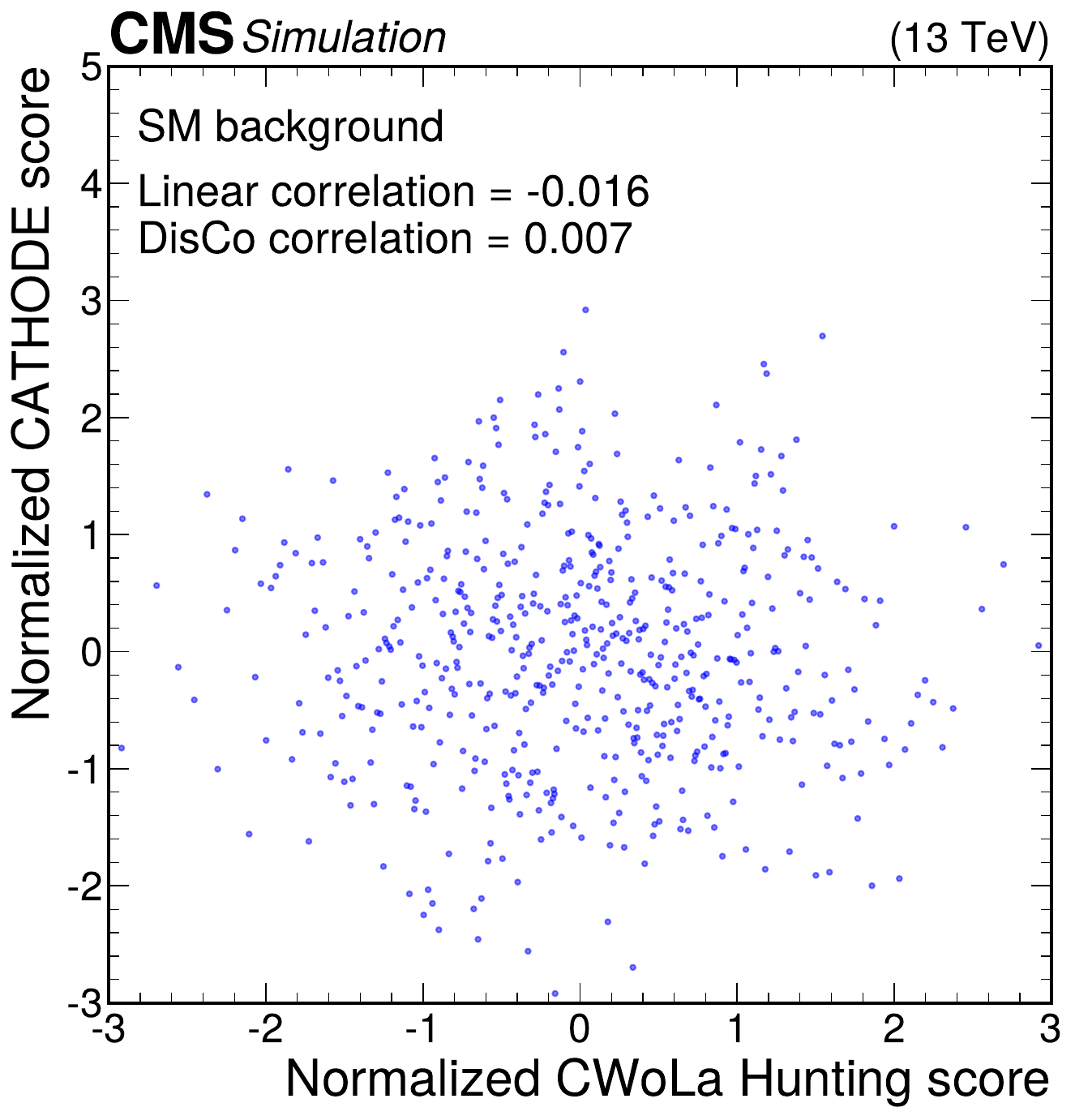}\\
  \includegraphics[width=0.45\textwidth]{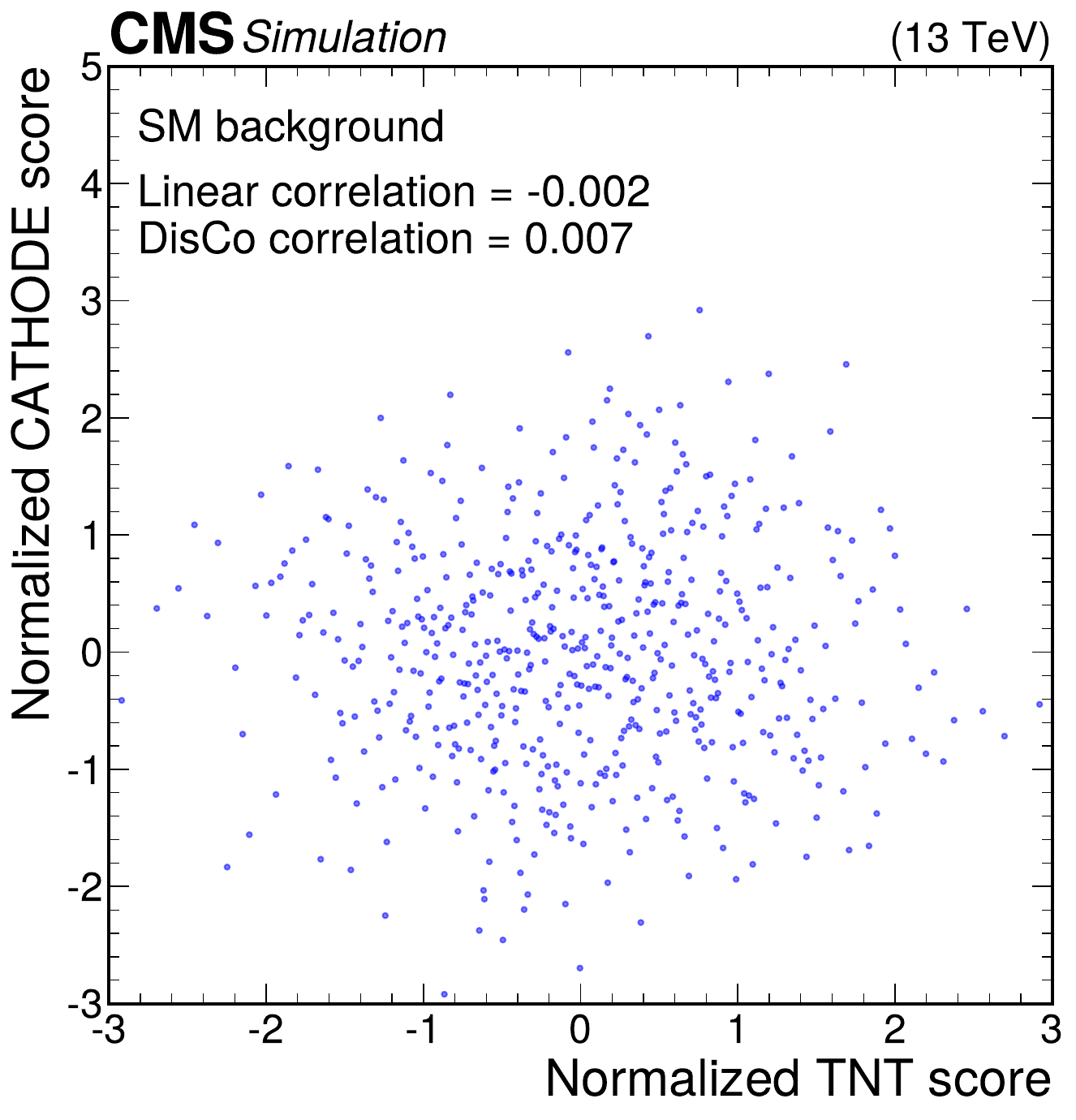}%
  \includegraphics[width=0.45\textwidth]{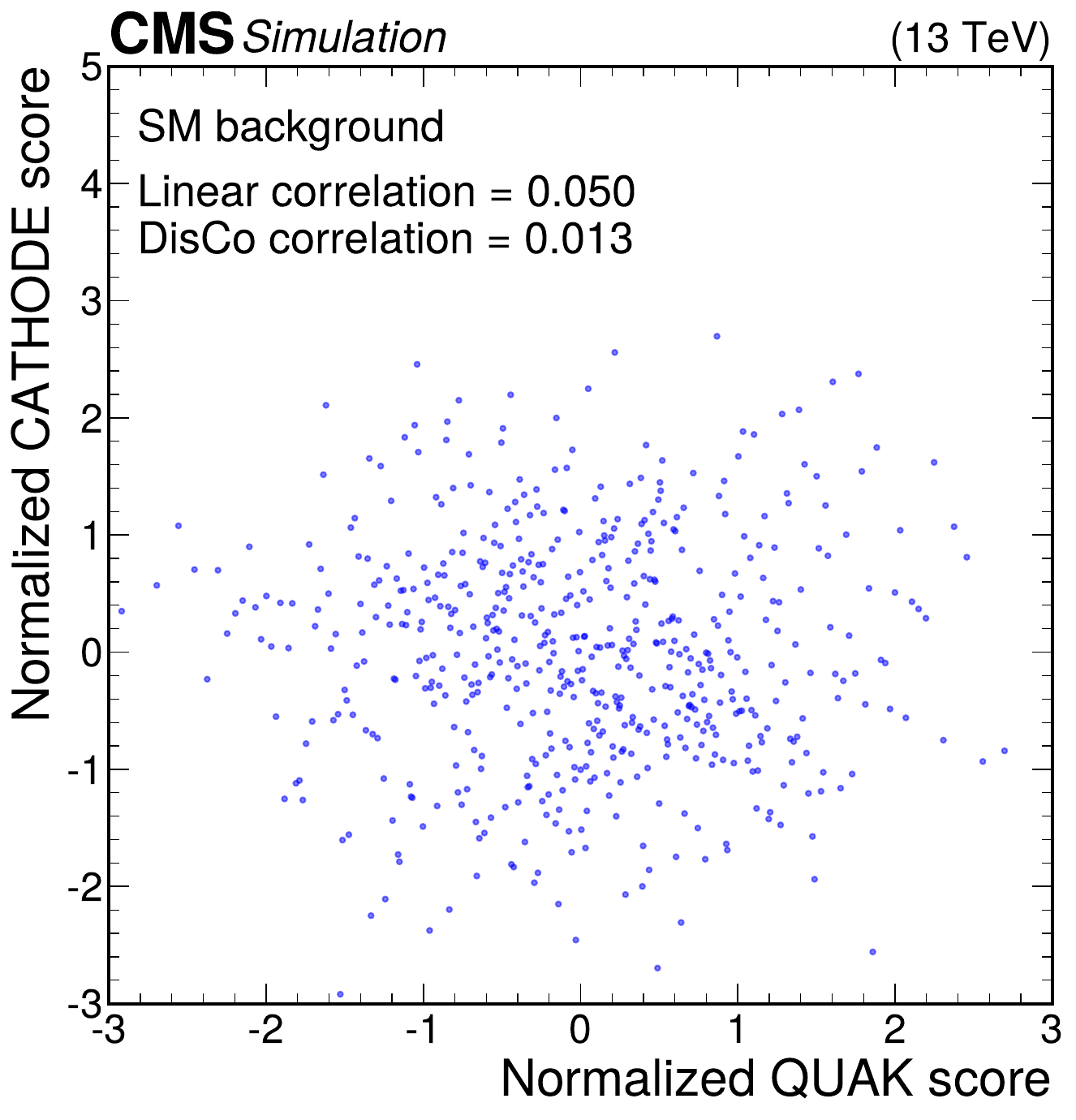}
  \caption{
    Anomaly score correlations of different methods on the simulated background
    sample.
    Scores are transformed to follow a normal distribution.
    The Pearson linear correlation coefficient and distance correlation (DisCo)
    are listed for each pairing.
    Details are given in the text.
  }
  \label{fig:bkg_corr}
\end{figure*}

\begin{figure*}
  \centering
  \includegraphics[width=0.45\textwidth]{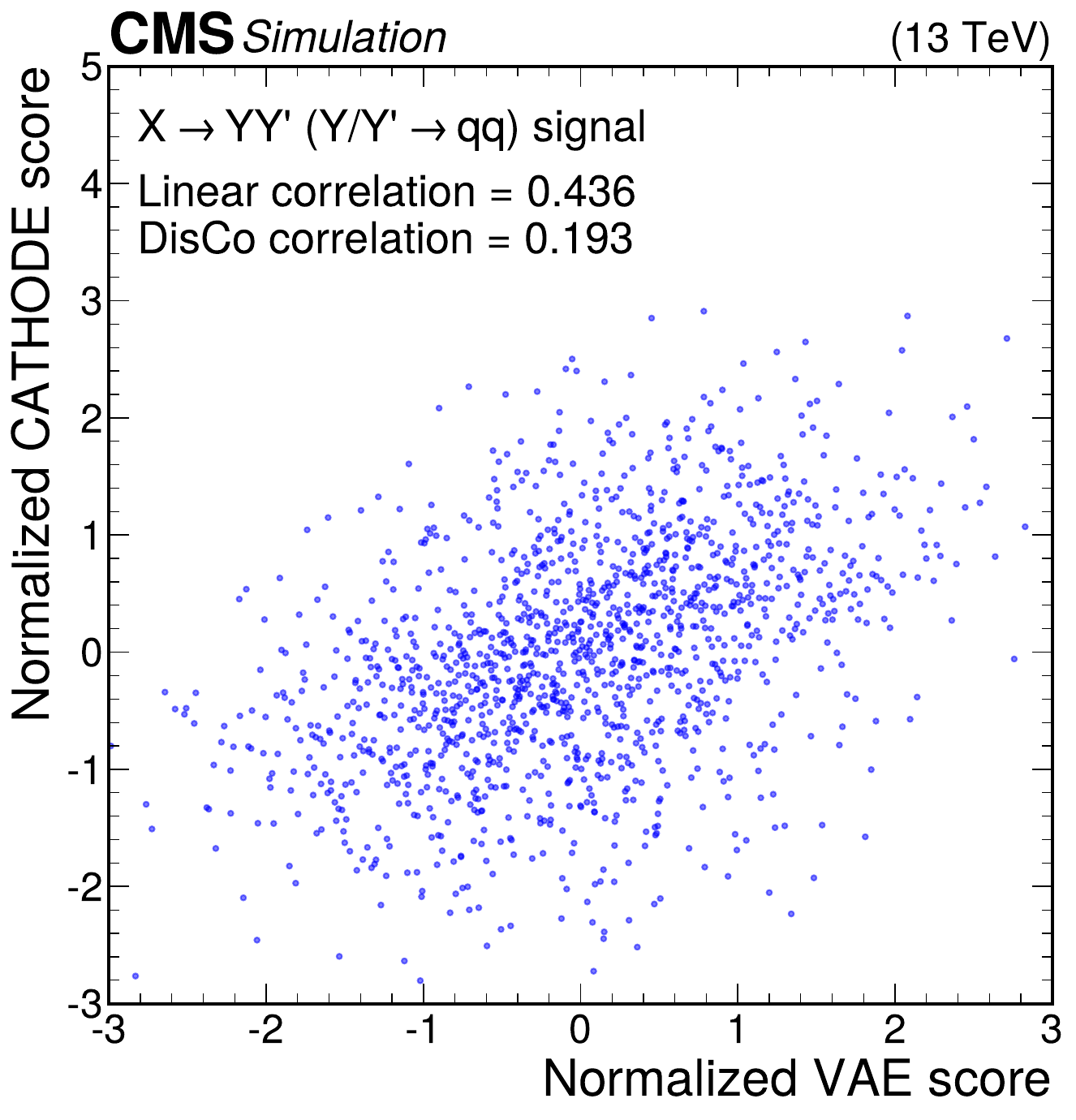}%
  \includegraphics[width=0.45\textwidth]{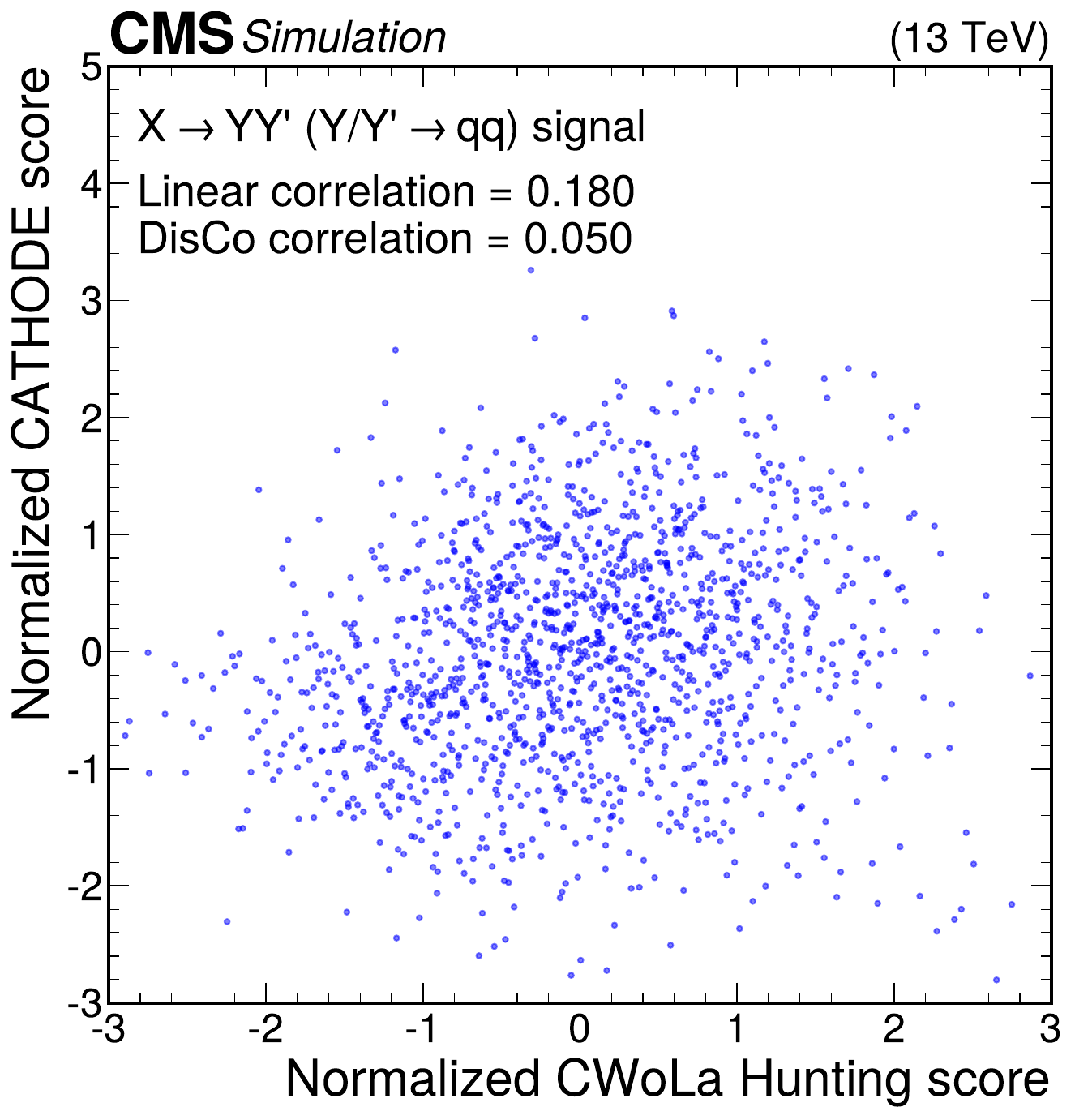}\\
  \includegraphics[width=0.45\textwidth]{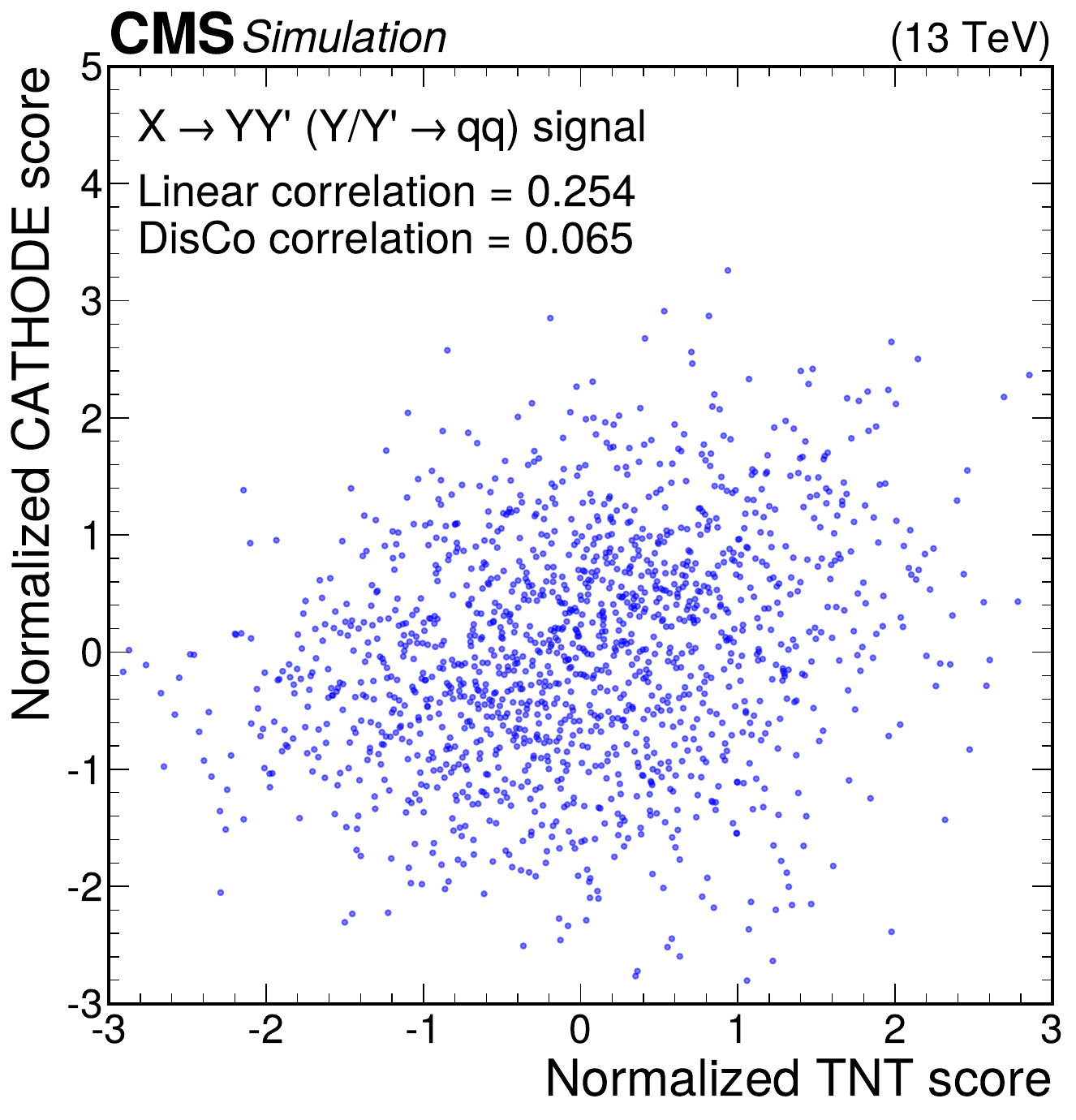}%
  \includegraphics[width=0.45\textwidth]{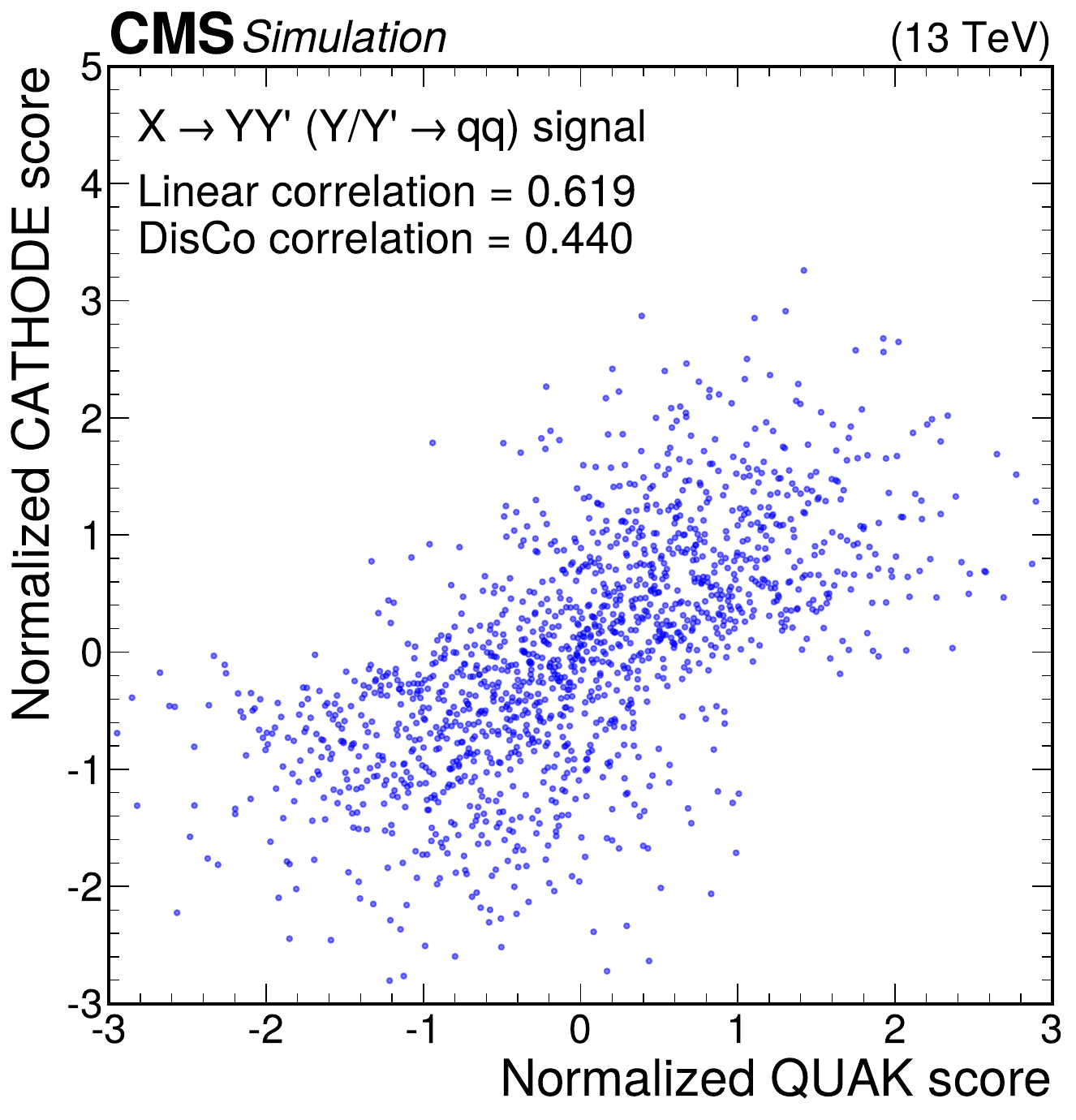}
  \caption{
    Anomaly score correlations of different methods on the \XtoYY signal model.
    Scores are transformed to follow a normal distribution.
    The Pearson linear correlation coefficient and distance correlation (DisCo)
    are listed for each pairing.
    Details are given in the text.
  }
  \label{fig:XYY_corr}
\end{figure*}

\begin{figure*}
  \centering
  \includegraphics[width=0.45\textwidth]{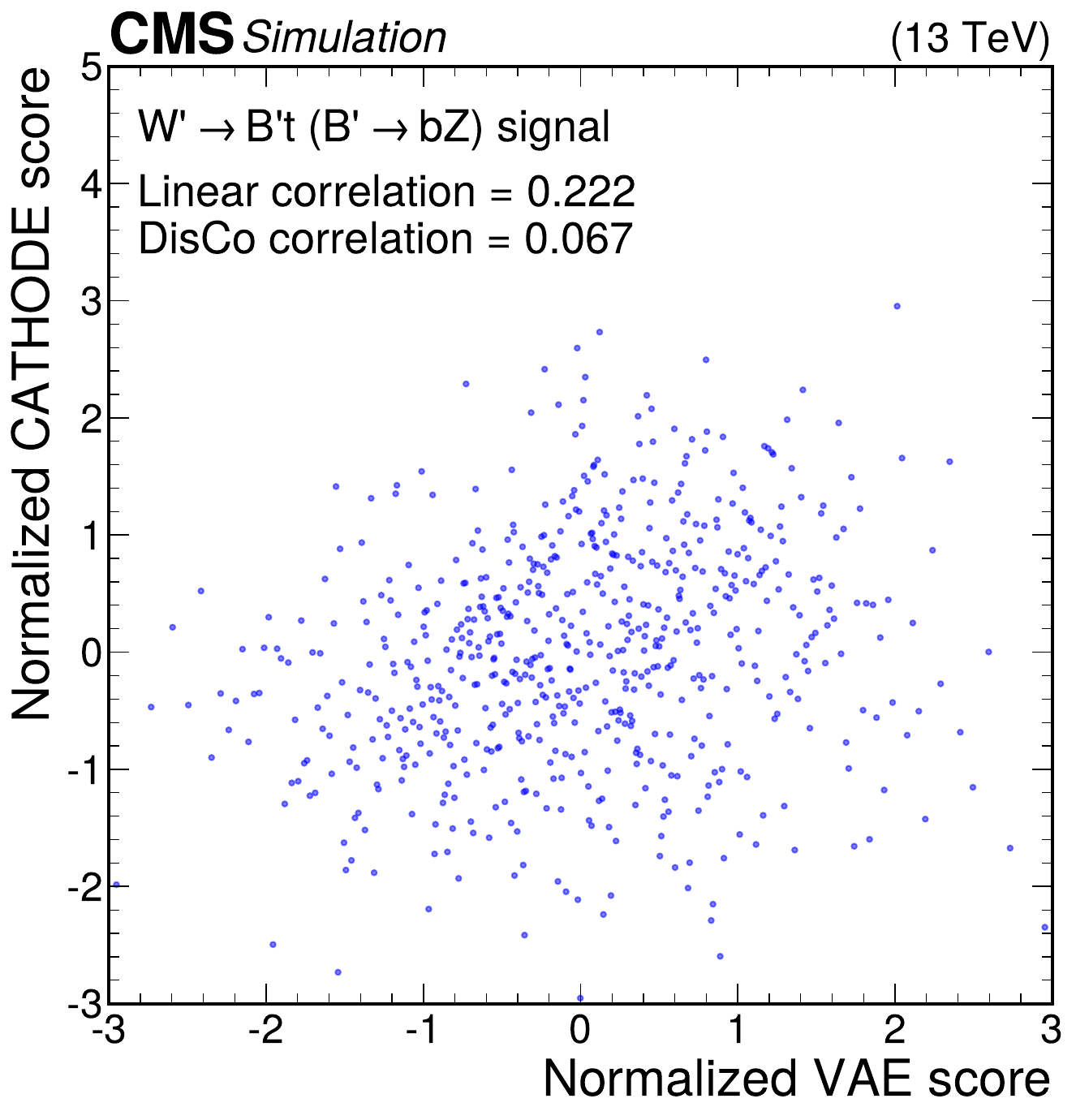}%
  \includegraphics[width=0.45\textwidth]{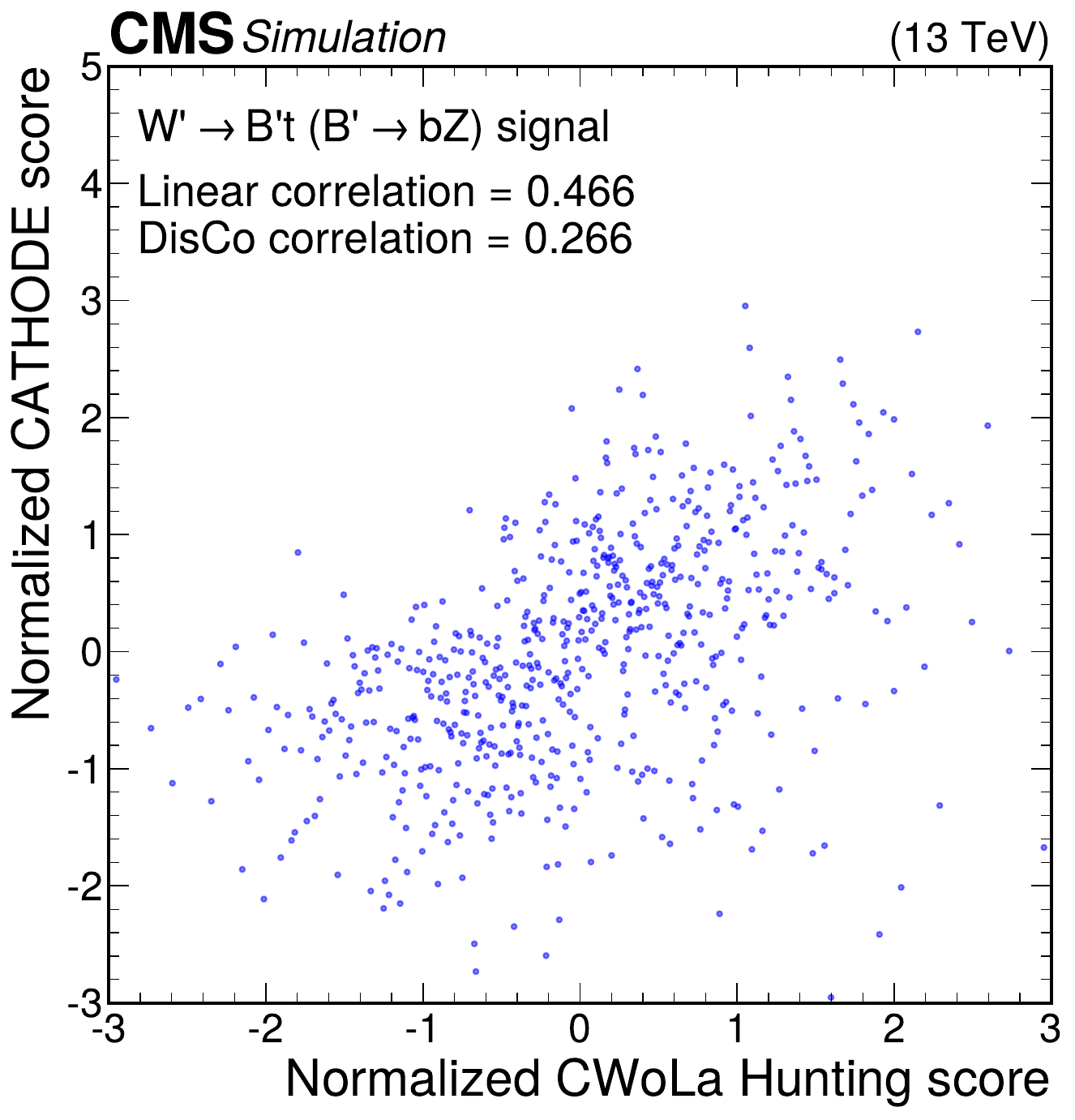}\\
  \includegraphics[width=0.45\textwidth]{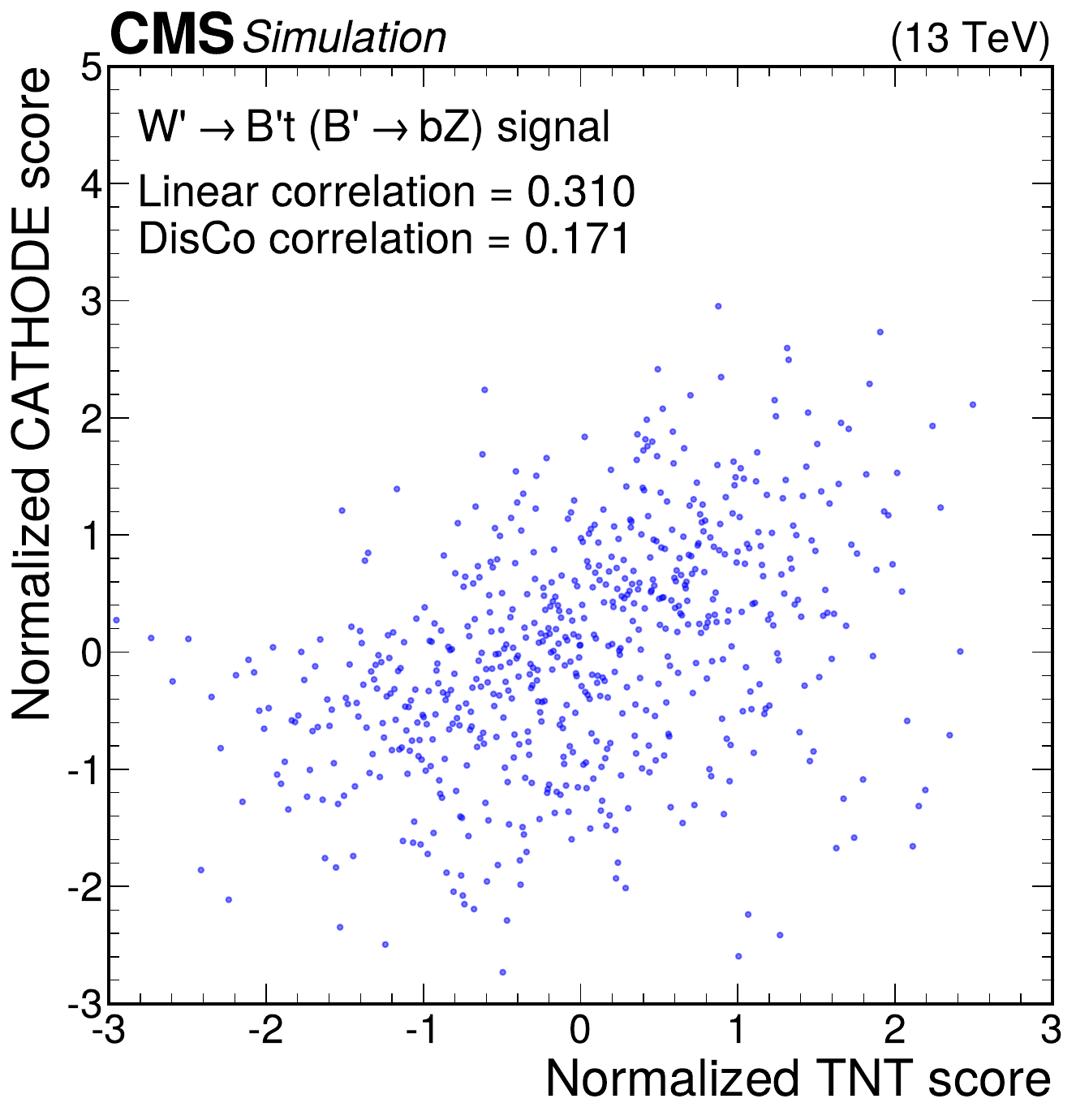}%
  \includegraphics[width=0.45\textwidth]{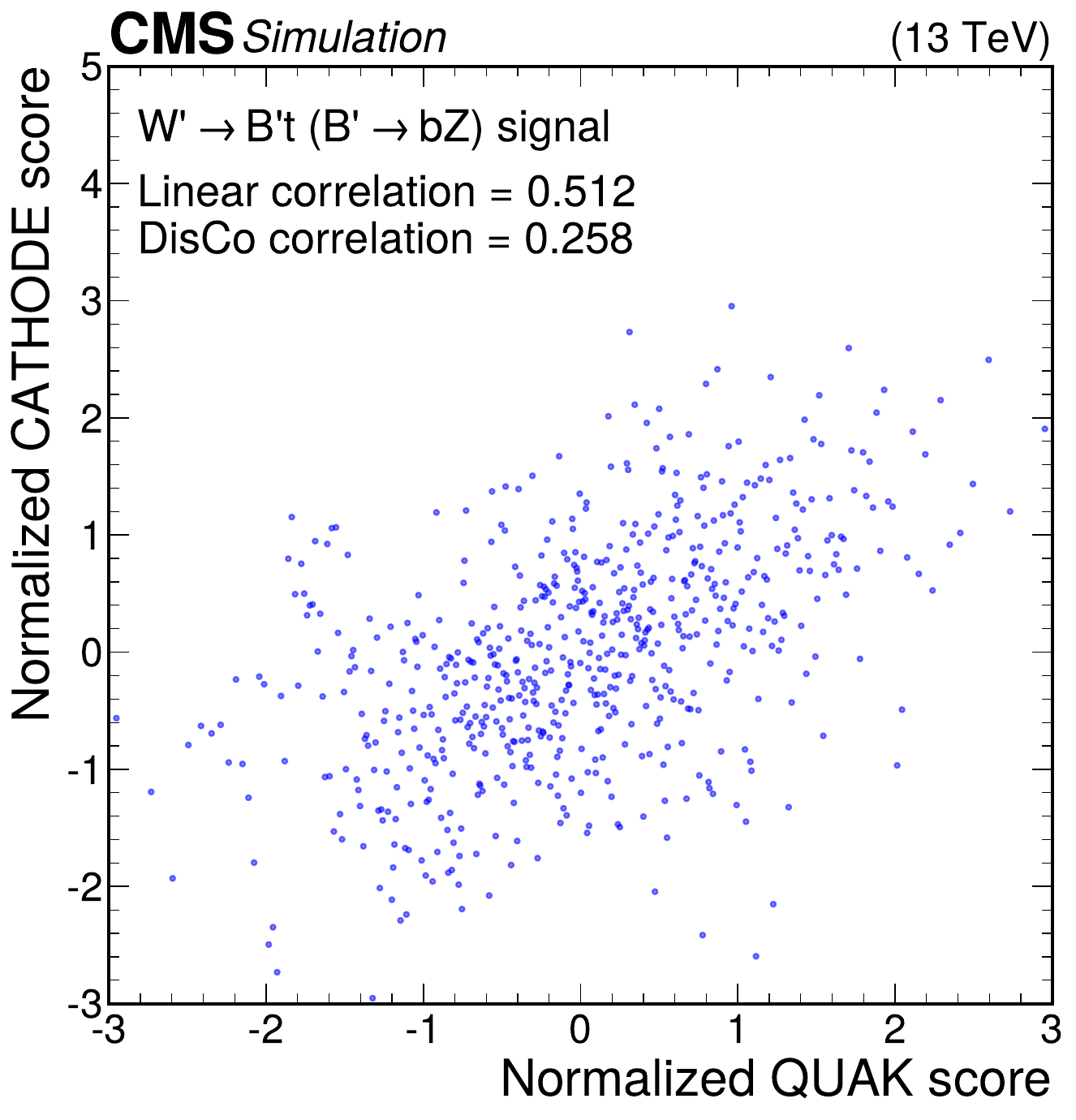}
  \caption{
    Anomaly score correlations of different methods on the \Wp signal model.
    Scores are transformed to follow a normal distribution.
    The Pearson linear correlation coefficient and distance correlation (DisCo)
    are listed for each pairing.
    Details are given in the text.
  }
  \label{fig:Wp_corr}
\end{figure*}

In this Appendix, we show comparisons of the anomaly scores assigned by
different methods to background (Fig.~\ref{fig:bkg_corr}), \XtoYY
(Fig.~\ref{fig:XYY_corr}), and \Wp (Fig.~\ref{fig:Wp_corr}) events.
Following the procedure in Section~\ref{sec:correlations}, each method is
applied to a simulated dataset and the scores are transformed to follow a
standard Gaussian distribution.
We then create a series of 2D scatter plots comparing the normalized
distributions of one method with another.
These plots are used as input to compute the correlations shown in
Section~\ref{sec:correlations}.

\cleardoublepage \section{The CMS Collaboration \label{app:collab}}\begin{sloppypar}\hyphenpenalty=5000\widowpenalty=500\clubpenalty=5000\cmsinstitute{Yerevan Physics Institute, Yerevan, Armenia}
{\tolerance=6000
A.~Hayrapetyan, V.~Makarenko\cmsorcid{0000-0002-8406-8605}, A.~Tumasyan\cmsAuthorMark{1}\cmsorcid{0009-0000-0684-6742}
\par}
\cmsinstitute{Institut f\"{u}r Hochenergiephysik, Vienna, Austria}
{\tolerance=6000
W.~Adam\cmsorcid{0000-0001-9099-4341}, L.~Benato\cmsorcid{0000-0001-5135-7489}, T.~Bergauer\cmsorcid{0000-0002-5786-0293}, M.~Dragicevic\cmsorcid{0000-0003-1967-6783}, C.~Giordano, P.S.~Hussain\cmsorcid{0000-0002-4825-5278}, M.~Jeitler\cmsAuthorMark{2}\cmsorcid{0000-0002-5141-9560}, N.~Krammer\cmsorcid{0000-0002-0548-0985}, A.~Li\cmsorcid{0000-0002-4547-116X}, D.~Liko\cmsorcid{0000-0002-3380-473X}, M.~Matthewman, J.~Schieck\cmsAuthorMark{2}\cmsorcid{0000-0002-1058-8093}, R.~Sch\"{o}fbeck\cmsAuthorMark{2}\cmsorcid{0000-0002-2332-8784}, M.~Shooshtari\cmsorcid{0009-0004-8882-4887}, M.~Sonawane\cmsorcid{0000-0003-0510-7010}, W.~Waltenberger\cmsorcid{0000-0002-6215-7228}, C.-E.~Wulz\cmsAuthorMark{2}\cmsorcid{0000-0001-9226-5812}
% ...
\par}
\vskip\cmsinstskip
\dag:~Deceased\\
$^{1}$Also at Yerevan State University, Yerevan, Armenia\\
$^{2}$Also at TU Wien, Vienna, Austria\\
% ...
\end{sloppypar}
\end{document}